\documentstyle[12pt]{article}

\newcommand{\gsim}{\mbox{\raisebox{-.3em}{$\stackrel{>}{\sim}$}}}
\newcommand{\lsim}{\mbox{\raisebox{-.3em}{$\stackrel{<}{\sim}$}}}
\renewcommand{\H}{I\!\!H}
\newcommand{\R}{I\!\!R}
\begin{document}
\begin{titlepage}

\title{On the interpretation of quantum cosmology 
         \thanks{
         Work supported by the Austrian Academy of Sciences
         in the framework of the ''Austrian Programme for
         Advanced Research and Technology''.}}

\author{Franz Embacher\\
        Institut f{\"u}r Theoretische Physik\\
        Universit{\"a}t Wien\\
        Boltzmanngasse 5\\
        A-1090 Wien\\
        \\
        E-mail: fe@pap.univie.ac.at\\
        \\
        UWThPh-1996-22\\
        gr-qc/9605019}
\date{}

\maketitle

\begin{abstract}
We formulate a ''minimal'' interpretational scheme for
fairly general (minisuperspace) quantum cosmological
models. Admitting as few exact mathematical structure as 
is reasonably possible at the fundamental level, we apply
approximate WKB-techniques locally in minisuperspace in 
order to make contact with the realm of predictions, and 
propose how to deal with the problems of mode decomposition 
and almost-classicality without introducing further principles.
In order to emphasize the general nature of approximate local 
quantum mechanical structures, we modify the standard 
WKB-expansion method so as to rely on exact congruences 
of classical paths, rather than a division of variables 
into classical and quantum. 

The only exact mathematical structures our interpretation
needs are the space of
solutions of the Wheeler-DeWitt equation and the
Klein-Gordon type indefinite scalar product. The latter
boils down to plus or minus the ordinary quantum
mechanical scalar product in the local quantum structures.
According to our approach all further structures,
in particular the concepts encountered in 
conventional physics, such as observables, time and unitarity, 
are approximate. Our interpretation coincides to some extent 
with the standard WKB-oriented view, but the way in which
the conventional concepts emerge, and the accuracy at which they 
are defined at all, are more transparent. A simpler book-keeping 
of normalization issues is automatically achieved. 
Applying our scheme to the Hawking model, we find hints
that the no-boundary wave function
predicts a cosmic catastrophe with some non-zero
probability.
\medskip

\end{abstract}

{\small \tableofcontents  }

\end{titlepage}

\section{Introduction}
\setcounter{equation}{0}

The purpose of this paper is to formulate an interpretational
scheme for quantum cosmology that relies on a minimal
amount of pre-supposed mathematical structure. 
The fundamental level is provided by the space $\H$ of 
solutions to the Wheeler-DeWitt equation and the 
Klein-Gordon-type scalar product, denoted by 
$Q(\psi_1,\psi_2)$, 
whose power has probably been understimated 
so far. In our formulation, we restrict ourselves to the 
minisuperspace approximation, but part of our program is 
of general conceptual nature, so that it might
carry over to the full theory.
(For a review of the field see e.g. Ref.
\cite{Halliwell3}).
As an example for illustration we use the ''Hawking model'', 
i.e. a closed Friedmann-Robertson-Walker universe filled with 
a homogeneous minimally coupled massive scalar field $\phi$. 
(Containing several basic features that have become paradigms in 
describing the evolution of the universe  ---
''birth from nothing''  by tunnelling, inflation,
matter dominated expansion to a maximum size, and recontraction
 --- this model is rich enough to 
provide a testing ground for conceptual questions and even for 
observational predictions). However, much of what will be said 
in this article applies to simpler and to more sophisticated 
minisuperspace models as well. 
\medskip

We begin, in Section 2, to sketch the basics necessary
to talk about the interpretational issue, in particular the
Wheeler-DeWitt equation in minisuperspace $\cal M$. 
Sections 3 -- 5 are devoted to two of the prominent 
interpretations, the WKB- and the Bohm-type approach,
respectively. In Section 3, we review how the nature 
of rapid oscillations in wave functions of the WKB-type gives
rise to the structure of usual physics, but only at an 
approximate level and locally in $\cal M$. 
We put some emphasis on the question how conventional quantum 
physics (i.e. an approximate unitary Schr{\"o}dinger-type
evolution) may be reconstructed. 
In the standard WKB-expansion approach one
divides the variables into classical and quantum ones by
means of a limit of the type $\ell_P\rightarrow 0$.
In order to obtain a general notation of approximate 
local quantum structures (Hilbert space and unitary evolution), 
it is convenient to have a modified
WKB-expansion at hand, based on geometric quantities
in $\cal M$. The ''(quasi)classical background'', provided by the
evolution of the classical (''macroscopic'') variables 
as considered in the standard WKB-expansion is
replaced by a congruence of exact classical trajectories. 
Section 3 contains an outline of this approach, whereas the
details are described at the end of the
article, in Section 12 (which serves as an appendix). 
Section 4 deals with the possibility that a given wave function,
when written down within the local quantum framework of some
WKB-domain, may exhibit catastrophic exponential
behavior that prevents interpretation. We introduce the 
distinction between {\it true} and {\it false local quantum 
mechanics}. Section 5 comments on the Bohm-interpretation. 
\medskip

Our conceptual point of view in this article will be 
that these WKB-techniques are the {\it only} means to
make contact between solutions of the Wheeler-DeWitt
equation and physical observations. In Section 6, we begin
to present the idea for an alternative conceptual 
scheme which is not new from the point of view of mathematical 
structures involved, but which seems to have attained few 
attention so far, at least as far as minisuperspace 
quantum cosmology is concerned. The scalar product 
$Q(\psi_1,\psi_2)$ between solutions of the Wheeler-DeWitt 
equation is proposed to be the
{\it only} pre-supposed exact mathematical structure on 
$\H$. All predictions must be expressed in terms of such 
numbers. Conventional physics is approximately identified by 
recognizing its conceptual structures  as ''hidden'' in the 
fundamental framework defined by $(\H,Q)$. In the
WKB-approximation, $Q$ boils down to plus or minus the local
quantum mechanics scalar product. 
We try as radically as possible
to reject any further fundamental structure or principle,
and denote this program ''minimal'' interpretation.
In using words like ''approximate'' and ''local'', we
are well aware of the fact that there is not much
mathematical rigor in these. Regrettably,
to some extent this is unavoidable: We talk about the 
{\it approximate} identification of structures from 
some {\it exact} underlying framework. Determining
the degree of accuracy to which this is possible will
provide technical difficulties by its own
--- most of which we are not able to answer in detail --- and
will presumably depend on the particular model. 
\medskip

The most severe difficulty in our program is provided by the 
hyperbolic nature of the Wheeler-DeWitt equation or, 
likewise, by the indefiniteness of the scalar product $Q$. 
In simple minisuperspace models, it is related to the fact that
the universe may contract or expand. This is the 
decomposition problem.  In Section 7 we sketch how it can be 
dealt with on account of the ''minimal'' interpretation.
Provided certain assumptions on the model hold, the 
decomposition of wave functions into ''incoming'' and 
''outgoing'' modes can be defined approximately and
locally in certain regions 
of minisuperspace by means of the same WKB-techniques that
serve for identifying observable quantities and local quantum
structures. A deep conceptual problem arises by the fact that 
the universe at sufficiently large scales is experienced
as a classical background, and that we need some structure
of this type to state clear predictions. Thus, the possible
breakdown of physics as the possibility to describe the universe
in terms of quantitative predictions is related with the
issue of classicality. Motivated by the ''minimality'' of our 
program, we formulate a condition which ensures that some
almost-classical ''macroscopic history'' of the universe
can be experienced. The appearance of a {\it false} local
quantum mechanical situation is a hint for an ''interruption''
of such a history. The observational details of the split 
into classical and quantum is addressed to a decoherence 
mechanism acting locally in $\cal M$.  
\medskip

Having discussed the issues of the identification of
observables, decomposition and almost-classicality,
we are ready in Section 8 to assign approximate 
probabilities for alternative outcomes of observations,
with respect to local WKB-domains. The space $\H$ serves
as collection of ''reference'' states with respect to
which a given wave function of the universe can be
interpreted. As compared to the standard 
WKB-interpretation, our scheme implies automatically the 
correct book-keeping in normalization issues as soon 
as several WKB-branches are involved, and includes a 
conceptually clearer treatment of quantum gravitational 
corrections and long-range unitarity-violating effects. 
The philosophy in treating the latter is essentially
to WKB-analyze a {\it given} wave function in various
adjacent WKB-domains (rather than to evolve them by
means of a local Schr{\"o}dinger equation). It is
expected that this accounts for all quantum
gravitational effects contained in some the model, to
the degree of accuracy at which observational
situations are defined at all. 
\medskip

Section 9 is devoted to three comments. Concerning the
relation of our interpretation to the one
refering to the conserved (Klein-Gordon) current of the 
Wheeler-DeWitt equation, we recall the conceptual
differences and show that the latter
gives in fact rise to Bohm's interpretation 
of ordinary quantum mechanics in the local WKB-branches.
As a second point we observe that,
taking our proposal over to these branches, the
slightly unusual feature arises that amplitudes
should be expressed as scalar products between solutions
of the {\it time-dependent} Schr{\"o}dinger equation.
However, due to the first order nature of the latter,
this is equivalent to the usual formalism.
Lastly, we outline how our framework could possibly
be modified to yield a third-quantized theory. 
\medskip

In order to illustrate some of the features of our
scheme, we look a bit closer to the Hawking model.
The typical classical solutions in this model
undergo an inflationary phase, followed by a matter 
dominated era in which the scalar field oscillates,
until the scale factor reaches a maximum and the universe 
begins to recollapse. The matter energy $E$ is approximately 
conserved after inflation. In Section 10 we make explicit 
how approximate wave packet solutions of the 
Wheeler-DeWitt equation in the inflationary domain may be 
constructed (thereby working out an example of a local quantum 
structure) and how predictions formulated in terms of
these relate to the standard WKB-interpretation.
Also, we discuss how the limited range of the
WKB-approximation spoils the notation of a probability
density on the set of classical trajectories, and 
just allows for a probability interpretation of sufficiently
broad tubes of paths. 
\medskip

In Section 11, we present a set of wave functions in the
Hawking model 
describing families of expanding and contracting universes 
to given values $E$ of the matter energy in the 
post-inflationary domain. This amounts to an interpretation 
of the wave function of the universe in terms of observations 
of the energy. An example is provided by expanding the 
Hartle-Hawking
(no boundary) wave function in terms of these states.
Also, we display a further 
example of a local quantum structure. 
Moreover, the structure of the approximate energy 
eigenstates as a convenient 
basis for $\H$ gives some insight into the question
what happens when the universe attains its maximum size.
We encouner some hints that the no-boundary wave function
predicts a cosmic catastrophe with some non-zero probability, 
but that no logical inconsistency
is implied. A condition for a wave function to avoid such an
event may be formulated in terms of the energy states.
\medskip

Lastly, in Section 12 (which may be regarded as an appendix) 
we present our modified
WKB-expansion to extract ordinary quantum mechanics from 
the Wheeler-DeWitt equation. It is based on the geometric
structure of $\cal M$ and the equal-time hypersurfaces
of a congruence of classical trajectories.
Although the notation of a quasiclassical background history
is contained therein only implicitly and the relation
to observation is less transparent, it provides a
more general analysis of rapidly oscillating wave
functions than the standard approach
and is applicable in larger domains than the latter. 
\medskip

\section{Minisuperspace quantum cosmology}
\setcounter{equation}{0}

We will consider the approach to quantum cosmology
that is based on the Wheeler-DeWitt equation 
\cite{Wheeler}\cite{DeWitt}
for the wave function of the universe. Formally, this 
equation ${\cal H}\psi=0$ is obtained as the quantum version 
of the Hamiltonian constraint ${\cal H}=0$ of the underlying 
field theory, i.e. general relativity with some fundamental 
matter field sector. 
Despite the mathematical subtleties related to 
the transition from the classical constraint to a well-defined 
wave operator (the appearance of infinities and the operator 
ordering problem), one may state that the overall nature of the
Wheeler-DeWitt equation is that of a hyperbolic 
(Klein-Gordon-type) functional differential equation.
The background on which it is defined is the space of all 
metrics $g_{ij}(x)$ and matter field configurations $\phi(x)$ 
on a given three-manifold $\Sigma$ which fixes the spatial 
topology. (Therefore, by the way, this approach does not naturally 
accomodate for the possibility of topology fluctuations. It relies 
on one of the assumptions that such fluctuations do not exist at 
all, are negligible for quantum cosmology, or must be taken into 
account by some generalization --- see. e.g. Ref. 
\cite{HartleHawking} 
for a path integral approach). 
This space of configurations on $\Sigma$ 
is endowed with the DeWitt metric, which is defined by the 
quadratic form in which the momenta appear in the
Hamiltonian constraint, and hence in which the second (functional) 
derivatives appear in the Wheeler-DeWitt equation. 
The DeWitt metric is ultralocal and at each point $x$ of
$\Sigma$ has Lorentzian-type signature $(-,+,+,\dots)$
\cite{DeWitt}.
The wave function is a functional $\psi[g_{ij}(x),\phi(x)]$.
The classical momentum constraint ${\cal H}_i=0$ translates into the
condition that $\psi$ is invariant under diffeomorphisms
on $\Sigma$. 
This gives rise to factor out the diffeomorphism group, 
by which the space of configurations turns into the
''superspace'' (see Ref. 
\cite{Giulini}
for a recent discussion of the geometry of this construction). 
Moreover, $\psi$ does not depend on an extra ''time'' 
(or evolution) parameter, as opposed to the Schr{\"o}dinger 
wave function of ordinary quantum mechanics. Hence, there is 
no obvious Hilbert space structure, and no obvious notation of 
probabilities. The structure as well as the contents of all 
predictions of observations in the universe, including conventional 
quantum physics and the experience of an almost-classical
evolution at sufficiently large scales, must be extracted from a
functional depending only on three-geometries and matter field 
configurations. This raises the problem of how to interpret a 
given wave function in terms of observations.
\medskip

This problem is retained at a conceptual level if one truncates
the infinite number of degrees of freedom down to a finite
''minisuperspace'' (see e.g. Ref.
\cite{Halliwell3}).
Thereby one usually chooses a more or less 
symmetric ansatz for the four-metric and the matter fields
in terms of a collection of parameters, typically
$(N,y^0,y^1,\dots y^{n-1})\equiv (N,y^{\alpha})$, where $N$
represents the lapse function that accounts for the possibility
of redefining the ''time'' parameter $t$ labelling the slices
of the foliation of the four-manifold in a $3+1$-ADM-split
\cite{ADM}\cite{MTW}.
The usual type of the truncation for the four-metric is
\begin{equation}
ds^2= - N(t)^2 dt^2 + g_{ij}({\bf x},y(t))dx^i dx^j\,,
\label{trunc}
\end{equation}
where ${\bf x}\equiv (x^j)$, and the $g_{ij}$ are fixed
functions. (To be even a bit more general, one may replace
$g_{00}$ by $- b({\bf x},y(t))^2 N(t)^2 dt^2$, with $b$
a fixed function. We will admit this possibility, but
refer to $N=1$ as the ''proper time gauge'' for simplicity). 
An anologous ansatz is imposed for the matter fields,
symbolically $\phi\equiv$ $\phi({\bf x}, y(t))$.
We shall denote by $\cal M$ the parameter space labelled by 
the $y^\alpha$. (It is called minisuperspace, and is usually 
a manifold). 
If different $y\in{\cal M}$ describe non-diffeomorphic
configurations $(g_{ij},\phi)$, the omission of the
momentum constraint (by putting $g_{0i}=0$ from
the outset) is heuristically justified. 
Upon inserting the ansatz into the classical 
Einstein-Hilbert action (with appropriate
boundary term), and integrating over the surfaces $\Sigma$
of constant $t$ (which kills the appearance of ${\bf x}$), 
one obtains a Lagrangian (and from it a Hamiltonian) 
formulation of a reduced model for a finite number of
degrees of freedom. This serves as a starting point for 
(canonical) quantization {\it \`a la} Dirac
\cite{Dirac}.
The variable $N$ enters the Lagrangian only algebraically,
hence defines the reduced analogue of the Hamiltonian constraint. 
The latter is in general of the form 
\begin{equation}
{\cal H} \equiv \frac{1}{2} \left(
\gamma^{\alpha\beta} p_\alpha p_\beta + U_{\rm cl}(y) \right)
= 0\, , 
\label{hamiltonian}
\end{equation}
where $U_{\rm cl}$ is the (real valued) potential term 
originating from the curvature of $g_{ij}$ on $\Sigma$ and 
the matter couplings. 
Up to operator ordering ambiguities, one may translate this
constraint into a minisuperspace Wheeler-DeWitt equation. The 
latter is now a second order partial differential equation for the
wave function $\psi(y^0,y^1,\dots y^{n-1})\equiv$ $\psi(y)$, and 
thus provides a mathematically well-defined problem. 
The quadratic form $\gamma_{\alpha\beta}$ (the inverse of
$\gamma^{\alpha\beta}$) defines the reduced version of the DeWitt 
metric, and has Lorentzian-type signature $(-,+,+,\dots)$.
This is true at least for the case of all interesting 
minisuperspace models. 
(With respect to this metric, we will talk about ''timelike'' and
''spacelike'' directions in $\cal M$ later on). 
In general, one may choose the operator ordering such that
the minisuperspace Wheeler-DeWitt equation reads
\begin{equation}
\left( - \nabla_\alpha \nabla^\alpha + U(y) \right) \psi(y) =0,
\label{wdw1}
\end{equation}  
where $\nabla_\alpha$ is the covariant derivative with respect to 
the DeWitt metric. One may include the Ricci-scalar of the latter 
into the potential, thus slightly modifying $U_{\rm cl}$ from
(\ref{hamiltonian}), in order to make the formalism invariant 
under $y$-dependent rescalings of the lapse (conformal 
transformations of the DeWitt metric
\cite{Halliwell3}). 
\medskip

Although we admit rather general models of this type, let 
us impose {\it two} requirements. First of all,
$\cal M$ shall admit a foliation by ''spacelike''
hypersurfaces. This implies that $\cal M$ is orientable.
We suspect that this restriction still admits models that
describe a reasonable range of three-geometries of topology 
${\bf S}^3$ or $\R^3$. 
It is not entirely clear to what extent the orientability of 
$\cal M$ is just a technical issue. We leave it open whether it
may be weakened or omitted in a more general scheme. As a second
requirement we assume that the domain in $\cal M$ defined by all 
$y$ such that $U(y)>0$ is the relevant one for drawing 
quantitative predictions. This is not a very precise formulation, 
but its motivation is essentially to exclude situations that are 
too far from the structure of general relativity and realistic 
matter fields. Usually, the domain $U>0$ (which need not be 
connected) is the one in which the major oscillations in the wave 
functions take place 
\cite{Halliwell3}\cite{Halliwell1}\cite{Halliwell2}. 
This seems to be true at least for models whose geometry is close 
to Friedmann-Robertson-Walker, and whose matter contents 
is of rather realistic type (in other words: models ''close'' 
to the Hawking model). 
Regions with $U<0$ are typically entered by a classical
solution only for short periods of (proper) time.
These $U<0$ domains will be important for the problem of
almost-classicality (the issue of returning universes), but not 
for the quantitative prediction of particular measurements. 
Also, one may encounter situations --- as e.g. in the 
Hawking model; see Section 11 --- in which 
the $U<0$ domains may simply be ignored for most purposes. 
\medskip

Classically, such a truncation to minisuperspace may be 
consistent with the full dynamics. By this we mean that a 
solution $(N(t),y^\alpha(t))$ of the reduced classical system, 
when inserted into the truncation ansatz (\ref{trunc})
for the metric and the 
matter field, gives rise to a solution of the full classical 
Einstein plus matter field equations (possibly with $U_{\rm cl}$ 
replaced by $U$). This is the case for all 
reductions to homogeneous three-metrics and matter fields, and 
in particular for the most popular models of the 
Friedmann-Robertson-Walker type 
\cite{Ryan}.
In principle, we can think about more sophisticated
models, the $y^\alpha$ representing a good selection of the
information contained in the (infinite) set of 
original degrees of freedom.  
Quantum mechanically, one looses information about
the fluctuations {\it off} the minisuperspace configurations. 
Such fluctuations can be re-introduced to first 
order by means of a perturbative analysis around minisuperspace
\cite{HalliwellHawking}.
Despite their approximate nature, various minisuperspace 
models have been proposed and discussed. On the one hand, 
they may serve for a preliminary orientation about the 
quantitative nature of predictions. On the other hand, they 
are particularly useful in attacking the conceptual problems 
of quantum cosmology. From now on we would like to restrict 
ourselves to the minisuperspace approaches. 
\medskip

In order to have an example at hand, we write down the 
Wheeler-DeWitt equation of the Hawking model
\cite{Hawking2}.
The minisuperspace variables are
the scale factor $a$ of a spatially isotropic closed universe
and the spatially homogeneous value $\phi$ of a 
minimally coupled scalar field with mass $m$. The classical
Hamiltonian constraint, when expressed in terms of
canonically conjugate variables $(a,p_a;\phi,p_\phi)$
is given by
\begin{equation}
{\cal H} \equiv -\,\frac{1}{2}\left(\frac{p_a^2}{a} + a\right)
+ \frac{1}{2}
\left( \frac{p_\phi^2}{a^3} + m^2 a^3 \phi^2\right) = 0.
\label{hhm}
\end{equation}  
The DeWitt metric 
$ds^2_{\rm DW} = \gamma_{\alpha\beta}dy^\alpha dy^\beta$ is thus
\begin{equation}
ds^2_{\rm DW} = a ( - da^2 + a^2 d\phi^2 ),
\label{dwhm}
\end{equation}  
and the momenta are related to the derivatives with respect to
the classical evolution parameter $t$ by
\begin{equation}
p_a = - \,\frac{a}{N}\,\frac{da}{dt}\qquad \qquad 
p_\phi = \frac{a^3}{N}\,\frac{d\phi}{dt}\, . 
\label{momhm}
\end{equation}  
Up to an overall numerical constant (see Ref.
\cite{Hawking2} 
for units), the classical space-time metric is
\begin{equation}
ds^2 = - \,N(t)^2 dt^2 + a(t)^2 d\sigma_3^2
\label{metrichm}
\end{equation}  
with $d\sigma_3^2$ the metric on the round unit three-sphere.
In the operator ordering corresponding to (\ref{wdw1})
--- the Ricci-scalar of $ds^2_{\rm DW}$ being zero
and thus not producing any further ambiguity ---, the
Wheeler-DeWitt equation (after multiplication by $a$) reads
\cite{Hawking2}\cite{HawkingPage} 
\begin{equation}
\left(
\partial_{aa} + \frac{1}{a} \,\partial_a
- \,\frac{1}{a^2}\, \partial_{\phi\phi} + m^2 a^4 \phi^2 -a^2
\right) \psi(a,\phi) = 0.
\label{wdwhm}
\end{equation}  
The potential term as appearing in (\ref{wdw1}) is
$U(a,\phi) = $ $a(m^2 a^2\phi^2 -1)$. The minisuperspace manifold 
${\cal M}$ is given by all values $(a,\phi)$ such that $a>0$. 
\medskip

There have been made a number of proposals how physical information
should be extracted from a given wave function $\psi$, i.e. a 
solution of the Wheeler-DeWitt equation (in both full superspace and
minisuperspace). There seem to be two main ideas 
along which the problem of interpretation may be attacked, both relying 
on mathematical structures related to the Wheeler-DeWitt equation.
The first one tries to interpret $|\psi|^2$ as a 
generalized probability density, whereas the second 
one uses the conserved (Klein-Gordon-type) current 
--- to be introduced in equation (\ref{current1}) below --- for 
extracting predictions. Let us first outline the former.
\medskip

In terms of the minisuperspace version (\ref{wdw1}) of the
Wheeler-DeWitt equation, one may start with the observation that 
there is a preferred measure $d\mu =$ $d^n y\, \sqrt{-\gamma}$ 
on minisuperspace. (For the Hawking model, this is $a^2 da d\phi$).
The proposal is that with any region $G$ of 
minisuperspace one associates the positive quantity
\begin{equation}
P_G = \int_G d\mu\, \psi^* \psi.
\label{phawkingandpage}
\end{equation}  
Predictions about observations should be stated in terms of these
numbers. This interpretation has been advocated, among others 
\cite{MossWright}, 
by Hawking and Page
\cite{Hawking2}\cite{HawkingPage}.
In the words of these authors, $P_G$
is ''the probability of finding a 3-metric and matter
field configuration in a region'' $G$.
\medskip

The second of the above-mentioned ideas starts from the fact that
equation (\ref{wdw1}) admits the conserved 
(Klein-Gordon type) current 
\cite{DeWitt} 
\begin{equation}
j_\alpha = -\,\frac{i}{2} \Big(
\psi^* \partial_\alpha \psi - (\partial_\alpha \psi^*) \psi\Big)
\equiv -\,\frac{i}{2}\,\psi^*
\stackrel{\leftrightarrow}{\partial_\alpha} \psi\, . 
\label{current1}\\
\end{equation}
Using the reality of the potential $U$ one easily checks
\begin{equation}
\nabla_\alpha\, j^\alpha = 0.
\label{current2}
\end{equation}  
An analogous construction is sometimes possible if the Wheeler-DeWitt
equation is defined by a more general operator ordering than
(\ref{wdw1}). 
(In the Hawking model the contravariant components of the current
are given by $j^a=$ $-j_a/a$, $j^\phi=$ $j_\phi/a^3$, and the
conservation law (\ref{current2}) reads
$\,a\,\partial_a (a j_a) - $ $\partial_\phi j_\phi =0$).
Much like the magnetic field strength in classical
electrodynamics (satisfying $\partial_j B^j=0$), the current
(\ref{current1}) defines a flow on minisuperspace and an
associated measure on the flow lines. The flow lines are those 
curves $y^\alpha_{\rm flow}(\tau)$ 
which are tangential to the current
\begin{equation}
\frac{d}{d\tau}\, y^\alpha_{\rm flow}(\tau) =
j^\alpha(y_{\rm flow}(\tau)).
\label{flow1}
\end{equation}  
Suppose now that $j\neq 0$ in some domain of ${\cal M}$. 
If ${\cal B}$ is a ''pencil'' of flow lines, and
$\Gamma$ a hypersurface intersecting any curve 
in ${\cal B}$ only once, the surface integral
\begin{equation}
P_{\cal B} = \int_{\Gamma \cap {\cal B}} d s^\alpha \,j_\alpha
\label{probab1}
\end{equation}  
is independent of the choice of $\Gamma$. This follows from 
integrating equation (\ref{current2}) over an appropriate
$n$-dimensional domain.
If the integrand in (\ref{probab1}) is non-zero in some
domain (which, however, is not always the case) one can 
manage $P_{\cal B}$ to be positive 
by choosing the orientation of the surface element on
$\Gamma$. 
\medskip

Having made this observation, one may proceed in several ways
in order to use this structure for an interpretational scheme.
We will outline two prominent approaches relying on the
current  $j$. The first one is
the realm of WKB-techniques (Section 3), the second one is a 
generalization of the Bohm interpretation of quantum 
mechanics (Section 5). 
The former provides a powerful tool in making contact with the 
structure of conventional quantum physics, 
and in order to account for a subtlety of
interpretation that is often overlooked, some adaption to 
domains $U<0$ of $\cal M$ (which sometimes play a ''Euclidean'' 
role) will be necessary 
(Section 4). 
\medskip

\section{WKB-interpretation and local quantum structure}
\setcounter{equation}{0}

Consider a wave function whose oscillations in some domain of 
$\cal M$ are mainly of the WKB-type, i.e. which is 
of the form 
\begin{equation}
\psi(y)= A(y)\, e^{i S(y)}\, ,
\label{WKB1}
\end{equation}  
where $S$ satisfies the Hamilton-Jacobi equation
for the classical action
\begin{equation}
- (\nabla^\alpha S) \nabla_\alpha S = U
\label{HamJac}
\end{equation}  
and $A$ is a slowly changing prefactor. 
(In general, this means 
$|\nabla A|\ll |A \nabla S|$ with
$\nabla$ denoting a derivative off the hypersurfaces
$S=const$). The quantity $S$ serves 
as the classical action of a congruence ($\equiv$ 
non-intersecting $(n-1)$-parameter family) of solutions 
$(N(t),y^\alpha(t))$ to the classical equations of motion 
\cite{Halliwell3}. 
These trajectories (paths) are defined by
\begin{equation}
p_\alpha\equiv \gamma_{\alpha\beta}\,
\frac{1}{N}\,\frac{d y^{\beta}(t)}{dt} =
\partial_\alpha S\, , 
\label{classequ}
\end{equation}
where $N$ is the lapse whose choice fixes the ''time gauge'', i.e.
the evolution coordinate $t$ in space-time. (If the operator 
ordering in (\ref{wdw1}) is such 
that the Ricci-scalar of $ds^2_{\rm DW}$ is present in $U$, it is
reasonable to consider it part of the classical equations of
motion). 
In the Hawking model, the first equality in 
(\ref{classequ}) is just given by (\ref{momhm}).
In the approximation of slowly varying $A$, the current is
approximately proportional to the gradient of $S$,
\begin{equation}
j_\alpha \approx |A|^2\, \partial_\alpha S.
\label{currentWKB}
\end{equation}  
Hence, the classical trajectories 
$y(t)$ approximately coincide  --- up to a redefinition of
the time parameter $t$ --- with the flow lines
$y_{\rm flow}(\tau)$, and the construction (\ref{probab1}) 
provides (locally in $\cal M$) 
a positive definite measure on the set of these curves. It can 
be used to interprete the wave function in terms of relative 
probabilities for pencils of classical paths. This interpretational 
approach has been advocated, among others, by Vilenkin 
\cite{Vilenkininter} 
(see also Refs.
\cite{DeWitt}\cite{Hawking2}\cite{HawkingPage}), 
and it plays an important role in the formulation of
his outgoing mode proposal (tunnelling proposal)
\cite{Lindeout}--\cite{DelCampoVilenkin}.
It is an approximate concept that relies on the
WKB-form of $\psi$. When the gradient of $S$ becomes small or
when neighbouring classical trajectories intersect each other,
it will break down. 
(See e.g. Ref. 
\cite{UnruhWald} 
for a criticism of this philosophy altogether). 
Hence, the applicability of techniques related to the
WKB-form of $\psi$ is bound to particular domains of ${\cal M}$.
Sometimes the description of the wave function in
terms of classical trajectories and their action may be
extended into a region where $\psi$ is still oscillating
but (\ref{WKB1}) is no longer a good approximation and caustics 
appear (see e.g. Refs.
\cite{Page}\cite{GrishchukRozhansky}). 
In such regions the wave function usually becomes
a superposition of several WKB-branches of the type
(\ref{WKB1}), the physical reason behind such
phenomena typically being that the number of variables in
which $\psi$ rapidly oscillates (the ''classical'' variables, see 
below) depends on the location in $\cal M$. 
In contrast, domains in which $\psi$ behaves exponentially rather 
than oscillating are usually thought of as a hint for tunnelling 
phenomena which do not admit direct observation. Although a
Euclidean version of the WKB-approach 
\cite{Halliwell3}\cite{Hawking2}\cite{HartleHawking}
may formally be applied, the direct relationship 
to predictions as in (\ref{probab1}) is lost, and
no classical paths representing observable universes
can be associated with $\psi$ within these domains.
\medskip

Pushing the analysis further, one finds by means of
a WKB-expansion that the prefactor $A$ in (\ref{WKB1}) 
contains a Schr{\"o}dinger-type wave function describing
quantum fluctuations between the paths in a WKB-branch. 
We will first sketch the common approach to see this,
and afterwards impose a modification in
philosophy that fits our purposes better.
\medskip

Usually, by treating the Planck length $\ell_P$ as the small 
parameter (or, equivalently, the Planck mass $m_P$ as the
large parameter) in a WKB- (semiclassical)
expansion, one uncovers an effective 
Schr{\"o}dinger equation whose evolution parameter $t$ is 
associated with a family of classical background space-times
\cite{DeWitt}\cite{HalliwellHawking}\cite{Vilenkininter}, 
\cite{LapchinskyRubakov}--\cite{Cosandey}. 
This is not an entirely unique procedure because some
of the matter coupling constants may depend on the Planck
mass, or certain field values may be very large.
For example, in the Hawking model one has
$m/m_P\approx 10^{-6}-10^{-5}$
\cite{HalliwellHawking}\cite{LindeInfl}, 
and if one likes this ratio to 
remain intact, the matter sector must be treated on an equal
footing with gravitation in domains where $|\phi|$ is
sufficiently large, or mimicked by an effective cosmological
constant $\Lambda^{\rm eff}\approx$ $m |\phi|$. 
(In this case $\phi$ acts as an inflaton and drives
the universe to expand exponentially). 
For pure gravity,
the limits $\ell_P\rightarrow 0$ and $\hbar \rightarrow 0$ 
are equivalent
\cite{Kiefersemi}. 
The variables $y$ are divided (either by hand or by
the way how $\ell_P$ appears in the Wheeler-DeWitt equation) 
into 
''classical'' and ''quantum'' (''heavy'' and ''light'') 
ones $(y_{\rm cl},y_{\rm q})$. 
The former usually correspond to the large-scale geometric 
degrees of 
freedom, possibly in combination with some part of the matter 
sector, and provide a congruence of classical ''background'' 
configurations in the space labelled by the $y_{\rm cl}$ alone
(cf. however Ref.
\cite{Vilenkininter}, 
where it is emphasized that this division of variables into
two groups may depend on the domain in $\cal M$,
e.g. when an inflaton $\phi$ is present.
Also, one would not expect the graviton degrees of freedom 
to belong to the classical variables
\cite{Kiefersemi}). 
The background action $S_0\equiv S_0(y_{\rm cl})$ 
depends only on the classical variables 
and satisfies the Hamilton-Jacobi equation 
(\ref{HamJac}) to the first relevant order 
(which in practice is defined by a ''classical'' part of the
potential $U_0(y_{\rm cl})$). 
After a suitable choice of the background lapse 
$N(y_{\rm cl})$, these configurations are endowed with an 
evolution parameter $t$, and $\partial_t$ denotes the derivative 
along the background paths in the $y_{\rm cl}$-manifold. 
This all is mathematically somewhat analogous to the 
non-relativistic ($c\rightarrow\infty$) limit of the flat
Klein-Gordon equation (i.e. 
$\gamma_{\alpha\beta}=\eta_{\alpha\beta}={\rm diag}(-1,1,1,1)$
and $U=m^2$) by the ansatz (making $c$ and $\hbar$ explicit)
$\psi(t,\vec{x})= A(t,\vec{x}) \exp(-i m c^2\, t/\hbar)$, where
formally $y_{\rm cl}\equiv t$ and $y_{\rm q}\equiv \vec{x}$. 
The technically most important feature in both cases is the
idea that a contribution of the type $\partial_{t t} A$ is 
neglected to lowest order in the small parameter, so that the 
remaining equation is of {\it first} order in $\partial_t$. We 
will not go into the details of these
methods but just note that 
the prefactor $A$ and the phase $S$ from (\ref{WKB1}) are
usually redefined by
\begin{equation}
\psi(y) = \widetilde{\chi}(y_{\rm cl},y_{\rm q}) 
D_0(y_{\rm cl}) e^{i S_0(y_{\rm cl})} 
\label{WKB0}
\end{equation} 
for some appropriate choice of the (real valued) factor
$D_0$ which does not concern us here. Both functions $S_0$ and 
$D_0$ characterize the (''macroscopic'' or quasiclassical)
WKB-branch, 
whereas $\widetilde{\chi}$ specifies a particular (quantum) state.
To first order in the expansion (which is actually
a Born-Oppenheimer type approximation), the Wheeler-DeWitt 
equation reduces to an effective Schr{\"o}dinger equation 
\cite{DeWitt}\cite{HalliwellHawking}\cite{Vilenkininter}, 
\cite{LapchinskyRubakov}--\cite{Cosandey}  
of the type
\begin{equation}
i \,\partial_t \widetilde{\chi} = \widetilde{H}^{\rm eff} 
\widetilde{\chi}\, , 
\label{eff0}
\end{equation}   
where $\widetilde{\chi}$ may be associated with a Hilbert space 
structure (with respect to which it is symbolically denoted as
$|\widetilde{\chi}_t\rangle\,$), such that 
$\widetilde{H}^{\rm eff}$ is hermitan. (Its form depends very 
strongly on the particular appearance of the small parameter
--- usually it is just the matter Hamiltonian ---, and the 
hermiticity is achieved by a suitable choice of $D_0$).
Hence, the semiclassical methods define an approximate unitary 
evolution and 
allow for a probability interpretation.
The function $\widetilde{\chi}$ can be chosen freely,
as long as it satisfies (\ref{eff0}), is sufficiently 
concentrated inside the WKB-domain but does not undergo too 
rapid variations that would spoil the 
approximation. 
(\ref{eff0}) is a minisuperspace version of the
Tomonaga-Schwinger equation
\cite{LapchinskyRubakov}\cite{Banks}\cite{Kiefersemi}, 
and thus it also reproduces the truncated form of relativistic 
quantum ''field'' theory.
(We will not go into the details of recovering
the full quantum field theory from the full superspace
Wheeler-DeWitt equation). 
Thus, whenever we will talk about the extraction of 
''quantum physics'' from quantum cosmology in this article, we 
automatically mean quantum field theory --- in its minisuperspace 
reduction --- as well. 
\medskip

Whereas the semiclassical expansion assumes that 
conventional quantum physics essentially relies on the smallness of 
$\ell_P$ it is sometimes appropriate to have a modified 
and geometrically oriented notation of WKB-states. Let us
write some wave function according to (\ref{WKB1}) as
\begin{equation}
\psi(y) = \chi(y) D(y) e^{i S(y)} \, , 
\label{WKB2}
\end{equation} 
where $S$ is the action (\ref{HamJac}) of a congruence of 
{\it exact} classical trajectories --- as opposed to
$S_0$ ---, and define an approximation by assuming that 
$S$ is {\it rapidly varying} as compared to $D$ and $\chi$. No
{\it \`a priori} division into classical and quantum variables
is performed. Maybe this is in spirit a bit closer to the early
study of DeWitt 
\cite{DeWitt} 
than to the modern WKB-approaches. 
The advantage of putting things like this is that all types
of ''paths'' one may encounter (i.e. the exact 
classical trajectories and the flow lines of $j$) are trajectories 
in $\cal M$ --- as opposed to $y_{\rm cl}(t)$ in the standard
approaches. In practice, the split into classical and quantum
variables will {\it emerge} to some accuracy rather than being 
assumed. The classical trajectories described by $S$ serve as
the background which provides the effective quantum evolution 
parameter $t$. 
However, the reconstruction of {\it the experienced} local
Hilbert space structure (\ref{eff0}) does not seem to be
trivial. Hence, although the use of the form (\ref{WKB2}) 
seems to be more fundamental than (\ref{WKB0}), 
the relation to observation is less transparent. 
Logically, the modified formalism might turn out as an
{\em intermediate} step between the full Wheeler-DeWitt equation 
and the standard semiclassical version of local quantum mechanics, 
although closer to the latter. 
\medskip

In Section 12, we will outline this modified 
approach in some detail, and so we are rather brief here. 
By appropriately defining the (real valued) factor $D$ (which, 
together with $S$, completely defines the WKB-branch, including 
all representation and normalization issues) and setting 
$\partial_t = $ $N (\nabla^\alpha S)\nabla_\alpha\,$
--- the lapse being specified as a function $N(y)$ on 
$\cal M$ ---, one again encounters an effective 
Schr{\"o}dinger equation
\begin{equation}
i \,\partial_t \chi = H^{\rm eff} \chi 
\label{eff}
\end{equation}   
to the degree of accuracy defined by the nature of
oscillations of $\exp(i S)$. In Section 12, the effective
Hamiltonian is expressed in terms of geometric quantities
on the equal-time hypersurfaces in $\cal M$, the result being 
equation (\ref{Heff}). One may associate theses states with 
an approximate Hilbert space structure (with respect to
which the states are denoted as 
$|\chi_t\rangle\,$), just as in the semiclassical expansion
(see equations (\ref{QQQ1}) and (\ref{surff}) below).
Also, we show in Section 12 that this approximation is
but the first order of a modified WKB-expansion. 
\medskip

As before, the function $\chi$ can be chosen freely,
as long as it satisfies (\ref{eff}), is sufficiently 
concentrated inside the WKB-domain but does not undergo too 
rapid variations in directions transverse to the
classical trajectories. Large transverse variations 
give rise to large $\partial_{tt}\chi$ which would spoil the 
approximation. This also limits the width of wave packets from 
below (to be $\gg |U|^{-1/2}$ in terms of the distance 
defined by the DeWitt metric, as a rough estimate will yield 
in Section 12). 
In Section 10 we will discuss the consequences
of this fact for the issue of interpretation. 
A heuristic condition for (\ref{eff}) to be a good
approximation to the Wheeler-DeWitt equation (\ref{wdw1}) 
is that $H^{\rm eff}$ (e.g. in the sense of the expectation 
value $\varepsilon=$ $\langle \chi|H^{\rm eff}\chi\rangle\,$) 
is ''smaller'' than $N U$, as will become clearer in Section 12. 
\medskip

In many practical situations, this approach should 
not be too different from the formulation (\ref{eff0}).
Heuristically, one would expect the 
relation between $\widetilde{\chi}$ and $\chi$ to be
essentially a 
''picture changing'' transformation that interpolates between 
$D_0 \exp(i S_0)$ and $D \exp(i S)$. 
However, since in the modified formalism no division of
variables in two groups has been assumed, the classical
character of some of them might emerge only after 
an average over (superposition of) neighbouring exact 
WKB-branches and the neglection of small terms. 
Also note that in general (\ref{eff}) may be expected to come
closer to solutions of the 
full Wheeler-DeWitt equations than (\ref{eff0}), i.e. 
$H^{\rm eff}$ to be ''smaller'' than $\widetilde{H}^{\rm eff}$. From 
the point of view of our formulation, a very small value of 
$\ell_P$ will automatically allow for rapid oscillations of $\psi$ 
in just those variables that have been assumed to be the classical
ones in the standard WKB-expansion approach, although 
there may be wave functions which do not 
respect the pattern provided by the small parameters. 
\medskip

There is some ambiguity in both approaches. In the 
standard one, an $\ell_P$-dependent redefinition of the variables 
may lead to another, macroscopically equivalent (quasi)classical
background. In the modified one, two slightly different 
WKB-branches may be macroscopically indistinguishable, 
and lead effectively to the same local quantum structure.
Hence, to be precise, there is no unique notion of ''trajectories 
$y(t)$ or WKB-branches $S$ contributing'' to a wave 
function. In any such statement the trajectory or
WKB-branch must be considered as just representing a 
larger class of possible other ones. The effective equivalence 
of the according quantum structures restores the
uniqueness of the approximation to a reasonable extent. 
\medskip

This problem of ambiguity of exact WKB-branches to 
characterize macroscopic situations has a counterpart 
in the standard semiclassical formalism. 
As mentioned above, it is conceivable that the 
''picture changing'' transformation
between $\widetilde{\chi}$ and $\chi$ may be understood as
(or should be combined with) an average over macroscopically 
equivalent (and possibly also over close but non-equivalent) 
exact WKB-branches. 
Due to such an average (together with the omission of
small terms), one ''looses'' variables 
(note in particular that the semiclassical effective Hamiltonian
$\widetilde{H}^{\rm eff}$ contains no derivatives
with respect to the classical variabes $y_{\rm cl}$) and ends up
with a wave function built around the rapidly oscillating 
part $\exp(i S_0(y_{\rm cl}))$. 
In passing from the modified to the standard
formalism, one thus looses the geometric significance of the
local quantum structure with respect to
$\cal M$. This implies a typical disadvantage of the conventional 
semiclassical approach: The ''background'' (expressed as a 
family of quasiclassical evolutions $y_{\rm cl}(t)$) does not 
define paths (or even local directions) in $\cal M$. 
\medskip

Both shortcomings (ambiguity of exact WKB-branches in the
modified and semiclassical evolutions destroying geometry based 
on $\cal M$ in the standard approach) 
have the same origin, and both approaches should 
be regarded as closely related tools for recovering local 
quantum structures from the Wheeler-DeWitt equation. 
From the point of view of reconstructing the {\em actual}
variables and representations of locally observed
quantum mechanics, our framework might constitute an
intermediate step. However, the remaining steps to
recover (\ref{eff0}), once the local quantum structure 
(\ref{eff}) has been revealed, should work inside (or near)
the familiar realm of Hilbert space techniques. Hence, from 
the mere point of view of extracting a Hilbert space 
environment from the Wheeler-DeWitt equation, both approaches 
may even be regarded as equivalent. 
\medskip

One advantage for our present porposes of the modified version 
is that it is a bit more global in spirit
and can be applied to larger domains in $\cal M\,$: 
As already mentioned, it may happen that during the background
evolution some classical variables become quantum 
and {\it vice versa} (as e.g. at the end
of inflation, where the inflaton field becomes
quantum). In the conventional scheme, this 
amounts to using different collections $y_{\rm cl}$ in two
adjacent domains (or, at the end of inflation, switching 
off an effective cosmological
constant adiabatically) whereas in our approch there is a 
common structure for both, and just the numbers of variables
in $H^{\rm eff}$ that contribute to rapid and slow oscillations
in $\chi$ change. It thus applies to the {\it intermediate}
domains as well, in which the status of some variables is
just about to change (see also Ref.
\cite{Isham}, p. 245, 
where the problematic nature of the split into classical
and quantum variables was pointed out). 
Technically, our formalism 
allows us cover ${\cal M}$ by a pattern of overlapping local 
domains inside which congruences of paths exist and thus give 
rise to various local quantum structures of the type (\ref{eff})
that represent possible WKB-type oscillating behaviour there. 
The main advantage of this modified WKB-approach in the 
present context is however
its geometric formulation, which enables us to make
contact between exact and approximate issues without 
dividing variables into two groups 
(as e.g. the derivation of equation (\ref{QQ}) below). From 
now on we will include all these and related methods
to extract the structure of conventional classical and
quantum physics from the Wheeler-DeWitt equation in the
phrase ''WKB-techniques''. 
\medskip

Let us add the remark that although in minisuperspace 
there is only one lapse degree of freedom $N$, the redefinition 
of the evolution parameter along the trajectories is not a 
simple issue in the quantum context. In the full theory one has 
to deal with a lapse function $N(x)$ (and a shift vector 
$N_i(x)$) on the three-manifold
$\Sigma$, and the question whether different coordinate choices 
yield the same effective four-dimensional space-time physics 
provides a highly non-trivial problem
\cite{Isham}--\cite{GiuliniKiefer}.
Probably, this is the most significant difference
between the full and the minisuperspace framework, and it
certainly limits the straightforward application to the 
former of all ideas we sketch in this article. 
However, it is also a non-trivial question in minisuperspace
to what extent local quantum structures that are constructed from 
the same WKB-branch but with different time gauges $N(y)$ {\it and}
different choices of equal-time hypersurfaces in $\cal M$
are equivalent. 
We suspect that they are in fact equivalent to a sufficient degree 
of accuracy if the equal-time hypersurfaces do not differ too 
drastly from the equal-action hypersurfaces. 
\medskip

When going beyond the accuracy assumed so far (i.e. by looking at 
even higher orders in a WKB-expansion) one typically encounters a 
''corrected Schr{\"o}dinger equation'' with a non-hermitan part
in the Hamiltonian (see e.g. Ref. 
\cite{Kiefersemi} 
for the standard approach, and equations (\ref{wdweps1}) 
and (\ref{wdweps2}) below for our modified formalism). This 
implies violation of unitarity and the 
(conceptual) breakdown of the probabilistic interpretation 
(''loss of probability''). It is due to the fact that the 
probability flow defined by $j$ (which is approximately tangential
to the classical paths in the WKB-context) is exactly conserved 
only along the flow lines $y_{\rm flow}(\tau)$, and neither along 
the classical solutions $y(t)$ nor along the quasiclassical
trajectories $y_{\rm cl}(t)$, as appearing in 
the standard WKB-approach. We will explain in Section 7 how we
handle the appearance of unitarity-violating terms in our
approach to quantum cosmology. Let us just remark here that
we do not consider approaches that redefine the wave function
$\widetilde{\chi}\rightarrow \chi^\prime$ by means of prefactors 
and phases that depend on $t$ and on expectation values with respect 
to $\widetilde{\chi}$. In this way one may manage 
$\langle\chi^\prime_t|\chi^\prime_t\rangle$ not to depend on
$t$ (see e.g. Ref. 
\cite{Bertoni} for a recent discussion)
but it destroys the linear structure (i.e. the possibility of
superpositions) and the notation of a unitary evolution
in a Hilbert space (for which also 
$\langle\chi^\prime_{t,1}|\chi^\prime_{t,2}\rangle$ is 
a reasonable quantity and has 
to be independent of $t$). As a consequence, unitarity in the
usual sense is still an approximate concept in such 
approaches. 
\medskip 

Although the WKB-techniques proove useful in uncovering
conventional physics locally in $\cal M$, 
they do not seem to be appropriate for giving quantum cosmology 
a fundamental, mathematically firm structure. 
We rather feel that all approximate methods refering to 
classical trajectories should work {\it within} a more 
basic framework. 
The philosophy behind the WKB-procedures as we use them 
is that the  oscillations of the WKB-type wave functions 
make the identification of standard physical concepts 
possible as a kind of {\it classical optics} approximation,
and that {\it nothing} at the fundamental level tells us 
about the observational significance of a wave function 
{\it before} this identification has been performed, 
and {\it beyond} its applicability. 
The wave functions as mathematical objects (and their 
oscillations in $\cal M$) define ''the fundamental level'' 
of the theory, and reconstructing observational physics is an 
approximate issue. Also, we do not regard the WKB-techniques as
approximations to a path integral
(see e.g. Ref.
\cite{Hawking1}) 
in this article. 
\medskip

Let us as a last remark in this Section comment on the size of 
the local WKB-domains necessary for our interpretation. 
Since the local physical concepts refer to particular 
WKB-branches contained in a wave function, the basic
quantities limiting the size of such domains from
below are the classical actions $S$ associated with 
the branches. In order to identify a particular $S$, the 
domain must at least allow for several oscillations of 
the phase
$\exp(i S)$. If $t$ is the classical evolution parameter
along a trajectory, and since $dS/dt= - N U$ (which is
a consequence of the Hamilton-Jacobi equation (\ref{HamJac})) 
the minimal time scale possible (below which no physical
statement makes sense) is 
$\delta t_{\rm min}\approx$ $(N U)^{-1}$.
Since in the post-inflationary domain $U$ is of the order of
the total matter energy $E$, this is a very small scale. 
In this way the extension of a WKB-domain in the direction 
tangential to the classical paths is limited from below. This
condition may be translated into a coordinate independent statement.
The ''proper length squared'' along a piece of the trajectory 
with respect to the DeWitt metric is given by
$ds^2_{\rm DW}=$ $- dS^2/U$. Thus the minimal scale of ''proper 
length'' along the trajectory is $s_{\rm min}\approx |U|^{-1/2}$. 
The minimal size of a domain in the transverse directions
is not that strictly gouverned, but it should allow one
at least to identify the equal-action hypersurfaces. 
As already mentioned, the scale in ''proper length''
over which $\chi$ changes considerably in a direction transverse
to the classical paths must be much larger than $|U|^{-1/2}$,
thus limiting the widths of wave packets and the scale of 
transverse oscillations, and hence the transverse size of 
WKB-domains from below. 
For a given wave function, this should in general provide a
fine pattern of domains covering the relevant part of
$\cal M$, inside each of which the WKB-analysis may be performed.
On the other hand, for some purposes it may be necessary to choose 
domains as large as possible, even extending to ''infinity''
in some directions (e.g. if questions of 
normalizability or well-behavedness of $\chi$ with respect to a 
probabilistic interpretation are concerned). In the region
$U>0$, such large domains can be chosen in the vicinity of
spacelike hypersurfaces. 
\medskip

\section{''True'' and ''false'' local quantum mechanics}
\setcounter{equation}{0}

In the foregoing section we have sketched how local
quantum structures are contained in an approximate and
''hidden'' way in the solutions of the
WheelerDe-Witt equation. When
trying to give a clear meaning to this phrase, it turns
out that there are actually {\it two different}
possible motivations leading to the WKB-approximation.
The first one is to rewrite the Wheeler-DeWitt equation
under certain conditions, and to {\it use} the
effective Schr{\"o}dinger equations (\ref{eff0}) or
(\ref{eff}) in order to {\it construct} approximate
solutions. In this case, one regards the local quantum
structure (i.e. the Schr{\"o}dinger-type form of the
equation and the existence of the scalar product) 
as a tool for performing similar methods as in conventional 
quantum mechanics: to solve an initial value problem with prescribed 
$\chi_{t=0}$. One will of course put emphasis on
{\it normalizable} wave functions, i.e. one will
{\it begin with} $\widetilde{\chi}$'s or $\chi$'s that behave
reasonable under the scalar product. Imposing 
normalization on these functions (inside the WKB-domain)
leads to solutions that are more or less of a 
wave packet-like type 
(see Refs.
\cite{Zeh}--\cite{ConradiZeh} 
for wave packets in minisuperspace). As in
ordinary quantum mechanics one will also admit wave functions
of a soft non-normalizable nature (analogous to 
''distributions'', such as momentum eigenstates, associated with 
a Hilbert space). 
\medskip

However, it is a {\it different} issue if some wave 
function $\psi$ is {\it given} and shall be analyzed and
interpreted by WKB-techniques. In such a case one would
first determine the particular WKB-branches $(S,D)$ the wave
function $\psi$ can be associated with. 
If this is done, the particular 
Schr{\"o}dinger-type wave function $\chi$ belonging to some
branch may be 
{\it computed} straightforwardly from
$\psi$. For any local quantum structure we define that 
{\it true local quantum mechanics} is recovered if $\chi$
admits a probabilistic interpretation without severe problems.
Because of the local nature of all WKB-issues
this is not a precise definition, but in practice it
means that $\chi$ is normalizable with respect to the
local $\langle\,|\,\rangle$ or at least of a harmless
non-normalizable type, as mentioned above. 
One can in general expect this to be the case in WKB-domains
in which $U(y)>0$ (the classical trajectories
being ''timelike''): Consider a ''spacelike'' hypersurface 
$\Gamma$ that intersects the WKB-domain. 
Since the integral (\ref{probab1}) --- as well as the
Klein-Gordon scalar product to be considered below ---
refers to hypersurfaces of this type, one may {\it impose} a 
reasonable behaviour on $\Gamma$ for all admissible wave 
functions from the outset (in some analogy with the requirement of 
normalizability in ordinary quantum mechanics). The hyperbolic 
structure of the Wheeler-DeWitt equation allows for integration, 
once the value of a wave function and its (normal) derivative
on a ''spacelike'' hypersurface are prescribed.
Also, one may use local wave 
packet-like states (to be constructed inside the 
WKB-branch by means of {\it solving} the effective 
Schr{\"o}dinger equation, as described above) in order to
interpret $\psi$ in terms of reasonable relative probabilities
of alternative observations.
\medskip

The situation may change in a domain in which $U<0$. It is
a common feature of wave functions to exhibit exponential
rather than oscillatory character in a ''timelike'' coordinate
in the $U<0$ domains of a certain size 
\cite{Halliwell3}. 
If $\cal M$ is higher than just one-dimensional, 
some ''spacelike'' variables will take over the role of
providing the oscillations and thus the evolution parameter
in the local Schr{\"o}dinger equation (\ref{eff}) --- cf. Refs.
\cite{Zeh}--\cite{ConradiZeh}. 
The equal-time hypersurfaces inhabiting the effective
wave functions $\chi$ are no longer ''spacelike''. Still one
may {\it construct} well-localized wave packets inside such
a domain by imposing suitable initial conditions on the effective 
Schr{\"o}dinger equation (\ref{eff}), but it is by no means 
guaranteed that the effective wave function $\chi$ of some 
{\it given} state $\psi$ behaves reasonably in a global
sense. One will often 
encounter functions that are exponentially {\it de}creasing or 
{\it in}creasing in some of their variables. This provides
a particularly clear division of wave functions if the domain
extends to ''infinity'' in some direction of
$\cal M$ (as e.g. in the Hawking model the domain defined by
$|\phi|\lsim 1/(m a)$ and $a\gsim a_{\rm min}$, but $a$ 
arbitrarily large). 
Whereas the case of asymptotically vanishing $\chi$ 
will still admit a reasonable probabilistic interpretation (e.g. 
in terms of wave packets) --- and hence fits into a 
{\it true} quantum mechanical scheme, the 
case of exponential increase in some unbounded domain
will lead to a catastrophe, from the point of view of the local 
quantum structure. We will denote such a case by the terminus
{\it false local quantum mechanics}. 
Again this is not a precise formulation, but in practice it should 
suffice to recognize the particular type of behaviour we just 
described. 
The prototypical 
situation is --- in a one-dimensional analogue --- 
a particle in a potential like $V(x)= V_0\, \Theta(x)$ and
a wave function with energy $0<E<V_0$ of the type $\exp(i k x)$ 
for $x<0$ that behaves exponentially increasing for $x>0$.
Does it describe a particle ''tunnelling into nothing''?
It is certainly beyond the scope of reasonable probabilistic
prediction. 
\medskip

To summarize, the {\it false} case appears when --- despite 
of the existence of the effective local quantum mechanical 
structure provided by $H^{\rm eff}$ and $\langle\,|\,\rangle$
--- no reasonable interpretation of a given state $\psi$ in 
terms of ordinary quantum mechanics is possible.
We will encounter an explicit example of this type in
Section 11, related to the question what happens when the 
universe reaches its maximum size in the Hawking model.
\medskip

Let us add the speculation that the non-appearance of {\it false} 
local quantum mechanics related with a wave function $\psi$
might turn out to be related with simple mathematical properties,
such as $|\psi|$ being {\it bounded} (at least in certain
domains). However, since the interplay
between the exact scalar product $Q$ --- to be defined in
Section 6 --- and the local scalar product
in domains with $U<0$ is non-trivial, this is not entirely clear, 
and we will leave this point open. 
\medskip

\section{Bohm-interpretation}
\setcounter{equation}{0}

So far we have dealt with the WKB-approximation.
For completeness we note that the second possibility for the 
use of (\ref{probab1}) in extracting predictions from a
wave function is to assume that the flow lines 
$y_{\rm flow}(\tau)$ themselves provide a congruence of 
''corrected'' (and ''real'') trajectories, no longer 
''classical'' in the 
sense of the original classical equations of motion, but 
''classical'' in the sense of providing a family of
deterministic alternatives on which (\ref{probab1}) defines
a probability measure. This idea is essentially the application of
Bohm's (and de Broglie's) interpretation of quantum mechanics
\cite{Bohm}--\cite{BohmHileyKaloyerou} 
to quantum cosmology, and several attempts to formulate 
it in detail have been made
\cite{Kowalski-GlikmanVink}--\cite{BlautKowalski-Glikman}. 
We should add that 
the WKB- or Bohm-type approaches and the Hawking-Page approach
are {\it not} equivalent even in the WKB-approximation, but 
there is a relation between
them. In a WKB-domain, the probabilities 
(\ref{phawkingandpage}) differ from (\ref{probab1}) 
by the interval of proper time ($N=1$) 
a trajectory spends within a certain region of minisuperspace
\cite{HawkingPage}. 
\medskip

Both interpretations relying on (\ref{probab1}) are limited by
the fact that the current $j$ associated with a
particular solution to the Wheeler-DeWitt equation
may vanish identically (as e.g. for any real $\psi$).
Given that in some region a wave function
$\psi= A \exp(i S)$ (with slowly varying and approximately real $A$) 
represents (in the WKB-sense) a family of expanding universes. 
Then its complex conjugate $\psi^*\approx A \exp(-i S)$ 
describes an
anologous family of contracting universes, and one might 
expect that a superposition of the form
$\psi'=\psi+\psi^*\approx 2 A \cos S$
still contains these two alternatives in a way reminiscent 
of ordinary quantum mechanics. But the Bohm-inspired 
interpretation would just predict a static universe 
\cite{Shtanov}\cite{BlautKowalski-Glikman}. 
It is reasonable to expect that the flow lines of 
each WKB-branch are 
still present in $\psi'$ as a mathematical structure and
as a physical information, but the expression (\ref{probab1}) 
is useless for uncovering it in this case. 
The situation is even worse if a wave function is defined as a
superposition between two WKB-branches that are only
approximately complex conjugates of each other. Then
$j$ could be almost anything, and nothing in it would
give us a hint towards the WKB-branches the wave function has 
been  constructed from.
We will keep in mind that the current (\ref{current1}) 
provides an exact mathematical structure related
with the Wheeler-DeWitt equation, and that it coincides 
approximately with the tangent to the classical paths
in a WKB-situation, but we will {\it not} adopt the point of
view that it is the basic tool for predictions. 
The WKB-interpretation, on the other hand, when confronted
with $\psi'$, would 
have to reconstruct the original WKB-components $\psi$ and 
$\psi^*$, and apply (\ref{probab1}) as approximate assignment 
for alternative classical paths inside each branch, as described 
above. We will try to imbed this latter method into a
''minimal'' but well-defined structure. 
\medskip

\section{The fundamental structure $(\H,Q)$}
\setcounter{equation}{0}

As we have seen in Sections 3 and 4, the interpretation 
of a wave function $\psi$ in terms of physical observables
--- as identified by WKB-techniques --- is 
possible only within local regions of ${\cal M}$ 
at an approximate level. 
Assuming that these approximate identifications
are {\it the only ones} at hand, it follows that 
concepts like observables, time, probability, the 
unitarity of quantum evolution (cf. Refs. 
\cite{Vilenkininter}\cite{HartleIandII})
and the Hilbert space structure (cf. Ref. 
\cite{Louko}) 
are approximate and local in character. One of the goals of 
quantum cosmology is then to extract the conceptual structures
of conventional physics (and also the accuracy to which
they apply) from some underlying mathematical framework. 
\medskip

It is sometimes argued that the mathematical structure
of a fundamental quantum theory of the universe is
still to be found
\cite{BarbourSmolin}. 
Although this may be so, it is not
clear to what extent a general principle would tell one
how conventional physics should be recovered.
It is conceivable that the underlying theory is quite
simple in structure, and that the extraction of
information we need for all sorts of predictions
operates at an approximate level.
We would like to advocate a framework that has a 
small amount of pre-supposed mathematics, and which
makes contact to the realm of observations by {\it structurally
identifying} conventional classical and quantum physics,
using WKB-techniques. 
\medskip

There is a scalar product related with the current 
(\ref{current1})
that seems to be still underexploited in this field.
Due to the definition of $j$, one has to insert the same wave 
function twice. However, if $\psi_1$ and $\psi_2$ are two 
(complex valued) solutions of
the Wheeler-DeWitt equation (\ref{wdw1}), the current
\begin{equation}
j_\alpha(\psi_1,\psi_2) = -\,\frac{i}{2} \Big(
\psi_1^* \partial_\alpha \psi_2 -
(\partial_\alpha \psi_1^*) \psi_2\Big)
\equiv -\,\frac{i}{2}\,\psi_1^*
\stackrel{\leftrightarrow}{\partial_\alpha} \psi_2
\label{current3}
\end{equation}  
is still conserved,
\begin{equation}
\nabla_\alpha\, j^\alpha(\psi_1,\psi_2) = 0,
\label{current4}
\end{equation}  
and admits a congruence of flow-lines analogous to (\ref{flow1}).
Hence, there is also a conserved quantity like (\ref{probab1})
for any ''pencil''. 
In order to remove the reference to such a
pencil, we recall our assumption that $\cal M$ admits a 
foliation by ''spacelike'' hypersurfaces. 
Supposing that at least one of the two wave functions
$\psi_{1,2}$ is sufficiently concentrated on such a
hypersurface $\Gamma$, the expression 
\begin{equation}
Q(\psi_1,\psi_2) = 
\int_{\Gamma} d s^\alpha \,j_\alpha(\psi_1,\psi_2)
\label{Q}
\end{equation}  
will in fact be independent of $\Gamma$. (This is analogous 
to computing the integral 
of $ - \frac{i}{2}\psi_1^* \partial_t \psi_2 + c.c.$
over $\R^3$ at $t=const=$ $arbitrary$
for the Klein-Gordon equation in Minkowski-space.
For the full superspace version of the scalar product 
in the case of pure gravity, see Ref.
\cite{DeWitt}; 
for its use in analyzing the Klein-Gordon equation on a 
curved space-time background, see Ref. 
\cite{BirrellDavies}). 
The scalar product $Q$ is indefinite (but non-degenerate).
In the Hawking-Model, on can choose $\Gamma$ as any hypersurface
of constant $a$, and
\begin{equation}
Q(\psi_1,\psi_2) = -\,\frac{i}{2} \int_{-\infty}^\infty
d\phi\, a \left(
\psi_1^*
\stackrel{\leftrightarrow}{\partial_a} \psi_2
\right)
\label{Qhm}
\end{equation}  
exists and is independent of $a$, as long as at least one of
the two wave functions $\psi_{1,2}$ is sufficiently well-behaved
as $|\phi|\rightarrow\infty$. If $\psi_2$ is understood as
''the'' wave function of the universe and $\psi_1$ as a
''test'' (or ''reference'') state (e.g. a wave packet or an
approximate matter energy eigenstate in the post-inflationary
era), this does not even imply a severe normalization condition 
for $\psi_2$. 
\medskip

Our proposal is that the space $\H$ of complex valued solutions 
to the Wheeler-DeWitt equation (\ref{wdw1}) and the scalar product
$Q$ as defined by (\ref{Q}) provide the {\it only} exact 
mathematical structure at the fundamental level.
We will not try to give a precise definition of $\H$ here
but merely assume that it is reasonably large. Also,
the mutual scalar product of two proper elements of $\H$ shall
exist. In analogy to, say, the momentum eigenstates 
$\exp(i \vec{k} \vec{x})$ in ordinary quantum mechanics, 
it will also be necessary to consider wave functions $\psi$ 
of distributional character, such that $Q(\Xi,\psi)$ exists
for all $\Xi\in\H$. Whereas the candidates for ''the'' wave 
function of the universe will be of this latter type, we think 
of $\H$ as providing an environment for interpretation (hence 
the notation ''reference states'' that we have used already above). 
\medskip 

A key point in our argumentation is now that, in a typical 
WKB-situation, $Q$ reproduces the scalar product 
of the local quantum structure. This is valid at least for
domains in which $U>0$ (which we have assumed to be the relevant 
ones for almost all types of predictions). 
To see this, we consider two 
wave functions $\psi_1$ and $\psi_2$, both of the type 
(\ref{WKB2}) --- inside the same WKB-branch ---
with effective Schr{\"o}dinger type states $|\chi_{1,t}\rangle$
and $|\chi_{2,t}\rangle$. When computing $Q(\psi_1,\psi_2)$,
one encounters an expression proportional to
$\partial_\alpha S$ --- analogous to (\ref{currentWKB}) ---
as the dominant contribution. The corresponding scalar
hypersurface element $ds^\alpha \partial_\alpha S$ on 
the (''spacelike'') hypersurface $\Gamma$ may be
positive or negative definite, according to the orientation
of the flow. Since $U>0$, $\Gamma$ intersects each
classical trajectory only once, inside a domain of ${\cal M}$
in which these form a congruence. By using the local Hilbert 
space structure associated with this WKB-branch one finds
\begin{equation}
Q(\psi_1,\psi_2) \approx 
\int_{\Gamma} d s^\alpha \,(\partial_\alpha S) D^2
\chi_1^* \chi_2 
\equiv \pm \langle
\chi_{1,t}|\chi_{2,t}\rangle\, . 
\label{QQ}
\end{equation}  
This is in accordance with the fact that the right hand side of
(\ref{QQ}) is independent of $t$ on account of the unitarity of 
the evolution (\ref{eff}) ---  
see Section 12, and in particular
equation (\ref{QQQ1}), for details). In Ref.
\cite{DeWitt} 
an equation similar to (\ref{QQ}) was derived in the context of
wave packet-like states. 
The physical role of the two possible signs will concern us later.
Just note here that complex conjugation of the whole
WKB-branch (essentially $S\rightarrow -S$) interchanges them. 
\medskip

Thus, there is some WKB-evidence for using $Q$ as basic
structure. 
In analogy with ordinary quantum mechanics, we expect that
--- forgetting about the indefinite nature of
the scalar product for the moment --- predictions should be
formulated in terms of expressions $Q(\Xi,\psi)$, where $\psi$
is ''the'' wave function of the universe and $\Xi$ represents
an observational device or a ''question'' or
a ''physical property'' of the universe. However, in contrast
to ordinary quantum mechanics, we have no external ''time''
parameter at hand, and thus we assume
that $\Xi$ {\it satisfies the wave equation along with} $\psi$
in order to make $Q(\Xi,\psi)$ a unique number.
One may imagine that the restriction to 
scalar products between solutions of the Wheeler-DeWitt 
equation corresponds to the fact that in a closed system all 
observational devices are subject to the according evolution 
laws. Also, this implies that the notion of an ''observable''
as a well-defined concept (e.g. as an operator on a
Hilbert space) is given 
up at the fundamental level, and is replaced by a wave
function (or a collection of wave functions) that, in a sloppy
way of putting it, encodes certain ''physical properties''
approximately. 
The concept of observables in the usual sense should appear 
only in a WKB-sense, when the structure of
conventional quantum physics is found to be hidden within
the bulk of data provided by $(\H,Q)$. Thus, 
it is the overall structure of all (or many) 
$Q(\Xi_1,\Xi_2)$ that is envisaged to provide an 
{\it approximate} interpretational environment. 
\medskip

Now we can rephrase the different attempts to interpret the 
wave function. The approach of Hawking and Page 
\cite{HawkingPage} 
assumes that all
predictions can be formulated in terms of the numbers
(\ref{phawkingandpage}). The Bohm-type approach 
as described above
\cite{Kowalski-GlikmanVink}--\cite{BlautKowalski-Glikman}
assumes that all predictions can be formulated in terms of
the numbers (\ref{probab1}).
The WKB-interpretation 
\cite{Halliwell3}\cite{Vilenkininter}\cite{Hawking2}\cite{HawkingPage} 
uses numbers of the type (\ref{probab1})
as well, possibly in addition to a 
reconstruction of proper WKB-branches.  
In contrast to these attempts, we will advocate the point of 
view that all resonable predictions should be formulated in 
terms of the numbers (\ref{Q}), i.e. within the mathematical 
framework provided by $(\H,Q)$. The concepts of conventional
physics as well as any limitation of observations
(e.g. the impossibility of any question that implicitly
refers to an {\it external} observer) should emerge from this 
structure, together with appropriate WKB-techniques.
Both objects $\H$ and $Q$ are well-known, and the
underlying idea is not at all new. 
Here, we just try to pursue it as radical as possible. 
Thus, our approach closely communicates with
the WKB-type (semiclassical) interpretation, but the theoretical 
status of the according approximations is shifted and 
those features which depend on particular domains of
minisuperspace are logically separated from the basic structure
$(\H,Q)$ (which, at the fundamental level, contains no
reference to ''local'' things). 
\medskip

In a sense our strategy is to envisage a ''minimal'' 
interpretation (related in spirit to Vilenkin's scheme 
\cite{Vilenkininter}, 
but based on $(\H,Q)$), and to see how far one comes. From 
this point of view, the final solution of the problem to
extract all possible predictions from a quantum theory 
of the universe will, to some extent, depend on the viability of
the particular models at hand rather than on the search for more 
fundamental structures. 
In the following, 
we will not try to be too rigorous, but put emphasis
on the main ideas, the realization of which will certainly
encounter various technical problems when applied to
realistic models. Any model can be viewed 
as providing its own way how (and to which accuracy)
the structures of conventional physics may be uncovered.
We will merely sketch what we believe are the
general features of such a program. 
\medskip

\section{Problems of observables, decomposition and
         almost-classicality}
\setcounter{equation}{0}

Before further considering the issue of interpretation,
we will proceed rather formal for a moment and
take into account the indefinite nature of the
scalar product $Q$ (which can be traced back to the fact
that the Wheeler-DeWitt equation is of second order 
(Klein-Gordon) rather than of Schr{\"o}dinger type).
Since the complex conjugate $\psi^*$ of any wave function
satisfies the Wheeler-DeWitt equation along with $\psi$,
and since $Q$ has the algebraic properties 
\begin{eqnarray}
Q(\psi_1,\psi_2)^* &=& Q(\psi_2,\psi_1)
    \label{algprop1}\\
Q(\psi_1^*,\psi_2) &=& - \, Q(\psi_2^*,\psi_1),
    \label{algprop2}
\end{eqnarray} 
the space $\H$ may be decomposed into two subspaces $\H^\pm$ 
that are orthogonal to each other with respect to $Q$, and on
which $Q$ is positive/negative definite, respectively.
(This is in some analogy to the scalar product associated
with the Klein-Gordon equation in curved space-time 
\cite{BirrellDavies}). 
In other words, one can choose a basis
$\{\Xi^+_r,\Xi^-_r\}$ of $\H$ that is normalized according
to
\begin{eqnarray}
Q(\Xi^\pm_r,\Xi^\pm_s) &=& \pm \delta_{r s}
\label{basis1}\\
Q(\Xi^+_r,\Xi^-_s) &=& 0\,.
\label{basis2}
\end{eqnarray}
For convenience, one may require in addition 
\begin{equation}
(\Xi^+_r)^* = \Xi^-_r \, .
\label{cc}
\end{equation}  
Denoting the spaces spanned by the subsets $\{\Xi^+_r\}$ and
$\{\Xi^-_r\}$ as $\H^+$ and $\H^-$, respectively, we obtain 
the decomposition 
$\H = \H^+ \oplus \H^-$. Complex conjugation maps $\H^\pm$
into $\H^\mp$ in a bijective way. 
This construction is somehow analogous to the
decomposition of solutions to the flat Klein-Gordon equation
into negative and positive frequency states (as e.g.
$\psi = \exp( \pm i m c^2 \,t/\hbar)$ for zero three-momentum). 
Expanding a wave function $\psi$ with respect to such
a basis,
\begin{equation}
\psi = \sum_r ( c^+_r\, \Xi^+_r  + c^-_r\, \Xi^-_r )
\equiv \psi^+ + \psi^-\, , 
\label{expansion2}
\end{equation}  
the coefficients may be found simply by
\begin{equation}
c_r^\pm = \pm Q(\Xi^\pm_r,\psi)\, . 
\label{expansioncoeff}
\end{equation}  
Thus, the expansion of a given wave function $\psi$ in terms of 
a basis of other solutions is traced back to the computation
of numbers of the type $Q(\Xi,\psi)$, as long as the 
normalization condition (\ref{basis1})--(\ref{basis2}) 
is adopted. This supports the use of $(\H,Q)$ as the fundamental
framework in our program  towards a ''minimal'' 
interpretational scheme.
\medskip

We can formally simplify the notation of a decomposition by
introducing the operator $K : \H\mapsto\H$ which acts
as $\pm 1$ on $\H^\pm$. It has the properties
\begin{eqnarray}
K^2 &=& 1 \label{K1}\\
K(\psi^*) &=& -\, (K\psi)^* \,. \label{K2} 
\end{eqnarray}
By
\begin{equation}
Q_{\!{}_K} (\psi_1,\psi_2) = Q(\psi_1, K \psi_2) 
\label{QK}
\end{equation}  
it defines a positive definite scalar product 
$Q_{\!{}_K}$ on $\H$ (that coincides
with $\pm Q$ on $\H^\pm$ and renders $\{\Xi^+_r,\Xi^-_r\}$
an orthonormal basis). Moreover, K is a hermitean operator
both with respect to $Q$ and $Q_{\!{}_K}$, and the subspaces 
$\H^\pm$ are orthogonal to each other with respect to 
$Q_{\!{}_K}$ as well.  
Conversely, given an operator $K : \H\mapsto\H$ which satisfies
(\ref{K1}) and for which the scalar product 
$Q_{\!{}_K}$
defined by (\ref{QK}) is positive definite, then
a decomposition of the above type is induced
by defining $\H^\pm$ as the eigenspaces of $K$ with
respect to the eigenvalues $\pm 1$
(and (\ref{K2}) follows as a consequence). 
Hence, the notation of such a decomposition 
$\H=\H^+\oplus\H^-$ 
is identical to the notation of an operator $K$ with the 
properties listed above. We will call $K$ a
decomposition (operator).  
\medskip

This construction is not unique. Any decomposition operator 
$K$ gives rise to a positive definite scalar product, and 
for any $K$ one may formally apply the
standard quantum mechanical assignment of probabilities
and give $\H$ (possibly redefined so as to contain
all states with $Q_{\!{}_K}(\Xi,\Xi)<\infty$) 
the role of a quantum mechanical Hilbert space. In 
this sense, any triple $(\H,Q,K)$ corresponds
to a scheme upon which the interpretation of wave functions 
can be based. 
The identity (\ref{QQ}) guarantees that $Q_{\!{}_K}$ translates 
into the local quantum mechanical scalar product within a 
WKB-branch, as long as the whole branch (\ref{WKB2}) and its
complex conjugate are associated with $\H^+$ and $\H^-$, 
respectively. 
\medskip

The degree of non-uniqueness of $K$ is 
a Bogoljubov-type freedom
\cite{BirrellDavies}: 
Let $K$ and $\widetilde{K}$ be two decomposition operators
and $\H=$ $\H^+\oplus\H^-$ 
$= {\widetilde{\H}}^+\oplus{\widetilde{\H}}^-$ the
corresponding decompositions. Then 
$\widetilde{K}={\cal S}K{\cal S}^{-1}$,
where the linear operator ${\cal S}: \H\mapsto \H$
acts isometrically, $Q({\cal S}\psi_1,{\cal S}\psi_2)=$
$Q(\psi_1,\psi_2)$, and commutes with complex conjugation,
${\cal S}(\psi^*)=({\cal S}\psi)^*$. If ${\cal S}$ leaves 
$\H^\pm$ invariant,
it is just a unitary transformation on these subspaces,
leaving the decomposition invariant, 
$\widetilde{K}=K$. Otherwise we have
${\cal S}:\H^\pm\mapsto{\widetilde{\H}}^\pm$ and a true change,
$\widetilde{K}\neq K$. 
It may as well be expressed as a change of a basis
of the type (\ref{basis1})--(\ref{cc}) to an other
one by ${\cal S}\, \Xi^\pm_r\equiv$ $\widetilde{\Xi}^\pm_r=$
$\sum_s (a_{rs}\Xi^\pm_s + $ $b_{rs}\Xi^\mp_s)$
for (at least some) non-zero $b_{rs}\,$. (In general, the
collection of all $(a_{rs},b_{rs})$ must obey a 
Bogoljubov-type normalization condition). 
\medskip

This non-uniqueness of $K$ provides 
a major obstacle in formulating a consistent quantum 
cosmological framework (see e.g. Ref. 
\cite{Kuchar}).
This is in contrast to the Klein-Gordon
equation in Minkoswki space, where the existence of a timelike
Killing vector field ensures a {\it unique} Lorentz-invariant
definition of negative/positive frequency wave functions,
and thus a unique decomposition. 
One can imagine several possible ways out of the dilemma
(which we call ''decomposition problem'').
\medskip

To begin with, one may suspect that some fundamental principle
should uniquely specify one particular decomposition $K_0$. 
It is hard to see how this can be done in general without
introducing a lot of additional structure, but we
cannot exclude it, and this matter is presumably
model-dependent. There seems to be a naturally preferred
decomposition in the Hawking model (see Section 11), 
relying on the analycity structure of wave functions in the
asymptotic region $a\rightarrow\infty$, 
but it is not clear to what extent this feature carries
over to more sophisticated models. 
Another attempt to define a unique decomposition (based
on a ''positive frequency'' condition in the limit of zero
spatial volume) was presented in Refs. 
\cite{Wald}\cite{HiguchiWald}. 
(As a further possible point of view, one may of course expect 
the fundamental problems of interpretation not to be solved 
{\it in general} but by {\it the particular} structure of the
underlying classical theory: general relativity plus
matter. Thus the (non)existence of a preferred $K_0$ 
might be envisaged as a criterion to discriminate between
different models for matter). 
In case it is reasonably possible to single out a unique 
decomposition, 
the triple $(\H,Q,K_0)$ should provide a consistent
basis for deriving all sorts of predictions. 
Note that even in this case the identification of wave functions 
with observational data is not pre-supposed, and local
WKB-techniques will certainly still play some role. 
Also, the appearance of {\it false} local quantum mechanical
situations in domains with $U<0$ may occur. 
\medskip

Another possible line of thought 
is to recall that in the minisuperspace
approximation we have truncated the major bulk
of degrees of freedom contained in the full theory. Although 
classically these degrees of freedom may appear to be of 
''small scale'' type and are irrelevant for cosmology, 
this need not be the case in a quantum theory. The mere fact 
that such degrees of freedom are not included into observation 
may in principle --- by means of a reduced
density matrix formulation and a ''decoherence'' mechanism, 
e.g. in a path integral or a consistent histories approach
\cite{JoosZeh}--\cite{Kiefer3} 
--- give rise to a superselection rule 
\cite{Kiefer6}\cite{GiuliniKieferZeh} 
at the minisuperspace level 
that simply forbids certain superpositions of wave functions to 
be observed, and hence to be used as a basis of reference
states.
(Recall that ''reference states'' --- at least in the
strong sense of the word --- should be related with possible
observations). This in turn could show up as an additional 
effective structure, e.g. as a condition for preferred 
decompositions, possibly depending on the particular WKB-domain 
in ${\cal M}$ to which an observation refers.
\medskip

We would prefer not to incorporate explicitly the bulk of 
original degrees of freedom, but merely work with
solutions of the minisuperspace Wheeler-DeWitt equation, 
and to implement the ''trace over unobservables'' locally. 
Let us try how far one comes if the ''minimality'' 
of our approach is retained, and the decomposition 
problem is treated {\it at an equal footing} with 
two other important questions: The interpretation of the
reference states in $\H$, and the problem of the experience
of an almost-classical (quasi-classical) world.
\medskip

Although we reject any fundamental (mathematically
exact) principle telling us what the ''properties'' represented 
by a particular state $\Xi\in\H$ are, we may structurally 
extract the physics of observations by
WKB-techniques from certain types of states. This may
appear to be a ''logical circle'' in that the wave
function of the universe $\psi$ is interpreted in terms
of reference states whose interpretation is not
pre-supposed either! However, taking the closedness of the 
universe serious, such a ''logical circle'' is 
quite plausible (if not unavoidable). In the ''minimal''
interpretation the answer is that ''properties''
and concepts like observables, 
time and unitarity may be identified with limited accuracy
just because they {\it are no} accurate concepts. 
\medskip

Even if for some wave function a WKB-interpretation has
been performed successfully in some local domain, there is
still a fundamental problem. To any WKB-type wave function
there is a large collection of classical trajectories
contributing, and the spread of $\chi$ over these may
in principle be interpreted as ''quantum fluctuations''.
Moreover, as we already know, {\it given} a WKB-type wave function 
$\psi$, there will be a range of WKB-branches $(D,S)$ that only 
differ slightly from each other and may all be associated
with $\psi$ to any reasonable accuracy. 
However, trajectories that are distinguishable at large
(macroscopic) scales are never experienced according to 
probabilistic laws, but rather as elements of a classical world, 
even if several repeated measurements are performed in order
to find out ''which trajectory is realized''. It is only for
very close trajectories that the spread of $\chi$ and the
uncertaintly in the WKB-branch is experienced as a genuine 
quantum feature. In terms of the standard WKB-split into 
classical and quantum variables, one would say that the universe 
primarily evolves according to {\it one} of the evolutions 
$y_{\rm cl}(t)$ described by the action $S_0(y_{\rm cl})$, and 
that the rest of the variables $y_{\rm q}$ are described by a 
quantum theory {\it on} this background. In terms of our modified 
WKB-approach, one can express classicality by considering
collections (''tubes'') of various trajectories $y(t)$ that are 
close to each other so as to represent the same evolution with 
respect to macroscopic observations. Genuine quantum fluctuations 
are experienced only for alternatives within such tubes, i.e.
with respect to small scale (microscopic) measurements. No matter
how 
this fact is formulated in terms of variables, it just means that 
macroscopic (quasiclassical) alternatives ''decohere''
(see Refs. \cite{JoosZeh}--\cite{Kiefer3}). 
This provides the possibility of ''time'' as
an ordering structure, together with at least some
almost-classical variables that altogether guarantee the
experience of a world retaining its ''identity'' as an 
alternative with respect to ''other worlds''. Without such a 
structure we cannot formulate predictions in the sense of 
statements about observations that can be {\it prescribed} and 
{\it performed}. The reason behind
the decoherence of macroscopic alternatives is commonly 
regarded as the fact that one does not observe all variables.
In ''tracing out'' most of them, an according reduced density
matrix becomes an almost-classical statistical mixture with 
respect to the quasiclassical variables. 
Partially, these questions should not provide conceptual
problems for quantum cosmology. Inside each WKB-branch,
associated with a local quantum structure, the non-observation 
of variables should be performed by tracing these out in the
sense of a Hilbert space. 
However, there must one thing be guaranteed in order for these
methods to work: The existence of a viable local quantum
structure enabling one not only to perform traces formally but 
also to obtain results that are sufficiently well-behaved 
so as to allow for interpretation. 
This condition is far from being trivial (due to the
approximate and local nature of the WKB-methods), and far from
being of minor importance only. 
\medskip

Thus we have formulated three conceptual problems: 
{\it (i)} How is the bulk of states in $\H$ to be identified 
with physical quantites? {\it (ii)} How is the non-uniqueness of the 
decomposition $K$ to be dealt with? {\it (iii)} What can be inferred 
about almost-classicality on macroscopic scales which is necessary to 
formulate predictions in terms of local quantum physics?
\medskip

We will try to treat these problems of an equal footing. Due
to the ''minimality'' of the philosophy pursued here, the
answer to all of them is given it terms of the identification
of conventional physics by WKB-techniques. 
The way to attack {\it (i)} has been announced already: 
By identifying local quantum structures to the accuracy
admitted by the WKB-techniques, the ''properties'' described
by states $\Xi\in\H$ which display WKB-behaviour in some 
domain of $\cal M$, and the observables associated with
them (within this domain) are uncovered. The principal 
limitation in the accuracy of this identification is provided 
by the nature of the oscillations in $\Xi$ (e.g. by the size
of the ''wave lengths'', as compared to the size of the domain).
{\it Predictions are always stated with respect to a domain},
hence locally in $\cal M$. There is no
ultra-local (pointwise) formulation of physics, because
any WKB-domain has a minimal size (below which the oscillations
in $\Xi$ cannot be resolved). Thus, for a very deep reason, 
conventional physics is approximate in practically all its 
concepts.
\medskip

To illustrate this point, two possible questions in the Hawking
model are {\it ''What is the probability to observe the matter energy
$E$ when the universe is in its post-inflationary 
--- $|\phi|\,\lsim\, 1$ --- era?''} and 
{\it ''What is the probability to find the universe expanding when 
it is in the region $|\phi|\,\lsim\,1$ and has 
$E\approx E_{\rm given}$ ?''}. 
It might happen for some wave function that the answer to the
second question is different from the answer to
{\it ''What is the probability to find the universe expanding when 
it is in the region  $|\phi|\gg 1$ ?''} where the wave function
may even make it necessary to specify the region in more 
detail (as e.g. prescribing $\phi\approx $ $\phi_{\rm given}\gg 1$, 
$a\approx $ $a_{\rm given} \gg1/(m \phi)$). 
As a further example, a question like
{\it ''What is the probability for the universe to
be expanding?''} is in general not well-posed 
since it must be assigned to a WKB-domain associated with the
concrete observation whose outcome is to be
predicted (although it might be admissible for certain wave
functions). Questions of this type refer to
the macroscopic (large scale) state and provide the
typical motivations a theory of the universe is 
usually concerned with (see e.g. Ref.
\cite{Vilenkin3}). 
Another class of admissible questions may refer to 
particular things happening ''in our universe'', and this is 
just the realm of local quantum physics, where the word 
''our universe'' refers to a particular domain as well. 
\medskip

The answer to {\it (ii)}, the decomposition problem, may be 
given at a similar level. Suppose that in some domain $U>0$. 
In this case,
we have already encountered a relation between the fundamental
structure $Q$ and local quantum physics: equation (\ref{QQ}) and 
its double sign gives us a hint. 
Obviously, given a congruence of classical paths that make up a 
WKB-branch by their action $S$, there are {\it two} possible
signs in the relation between the locally
assigned Hilbert space structure and $Q$. Moreover,
the $Q$-scalar product between two wave functions corresponding
to two WKB-branches that are complex conjugate to each
other is likely to vanish approximately if the oscillations 
are rapid enough. 
(By direct computation one may infer that the scalar product
between complex conjugate branches lacks the ''large'' 
contribution from $\partial_\alpha S$ in the integrand, as compared to
products within one branch).
Also, the proper (single-branch) WKB-states $\psi\sim $ $ A \exp(i S)$ 
are the {\it only ones} around which an effective local quantum 
mechanics 
(\ref{eff0}) or (\ref{eff}) is defined (as opposed to their 
superpositions $\psi\sim$ $\sum A_j \exp(i S_j)$). 
This is so because any local quantum structure needs an
evolution parameter $t$, whereas several macroscopically different
WKB-branches would provide several such parameters.
Hence, requiring that all those wave functions around (or
''near'') which a local quantum structure is possible are in 
one of the two local subspaces $\H^\pm$, we have almost 
solved the  problem: The only reasonable candidates for elements of 
$\H^\pm$ at the level of accuracy limited by the WBK-techniques
are the (single-branch) states $\psi\sim A \exp(\pm i S)$. 
Since we have assumed $U>0$, the congruences of classical
paths clearly fall into two classes: ''past'' (incoming) and
''future'' (outgoing) directed ones, with respect to the 
DeWitt metric. It is locally clear what the physical difference 
between these two possibilities is. In many models one may 
associate these with contracting and expanding 
universes. The details of such assignments depend on the
structure of $\cal M$ and its geometry, but they have always
a clear physical meaning. This provides a list of
candidate elements of two locally and approximately
defined subspaces $\H^\pm$ of $\H$. The superpositions thereof
will be oscillating as well, but not of the proper 
(single-branch) WKB-type. There will also be elements of 
$\H$ containing only exponential rather than oscillating 
contributions within our domain. 
These are representing tunnelling phenomena and do not
correspond to observable features (at least not 
in the domain under consideration). In terms of the $Q$-product, 
we expect all these states to be approximately orthogonal to 
the WKB-contributions, and for convenience we decompose the 
subspace of $\H$ corresponding to these tunnelling states as 
well (in fact it does not matter how, but there may be 
reasonable choices, e.g. dictated by smoothness with respect to 
adjacent domains).
We end up with an {\it approximate decomposition}
$\H\approx\H^+\oplus\H^-$, associated with the {\it local
domain} in which the analysis was performed. This provides
the interpretational environment for the wave function of the
universe with respect to the domain. 
In practice one even needs less: Since the $Q$-product between 
largely distinct WKB-branches is expected to be negligible, the 
choice of $K$ may be attached to the particular observation 
one is interested in, and restricted to the subspace of 
$\H$ spanned by the WKB-branches of interest and their
complex conjugates.
Note that, although a classically unique notion
of outgoing/incoming paths exists, this does not enable us to 
assign outgoing/incoming properties for wave functions at the
{\it exact} level, so the essential difference to the flat
Klein-Gordon equation remains intact. 
\medskip

At the technical level, we may write down equations defining
the approximate decomposition. Consider a domain $\cal G$ of
$\cal M$ in which $U>0$, and let $\Gamma$ be a ''spacelike''
hypersurface in $\cal G$. Then there is a uniquely defined
family of (outgoing) classical trajectories starting from 
$\Gamma$ orthogonally. The corresponding action may be computed by
solving the Hamilton-Jacobi equation (\ref{HamJac}) with
initial condition $S|_\Gamma=0$ in a vicinity of $\Gamma$.
The sign of $S$ is fixed by choosing the time coordinate $t=-S$ 
along the paths, such that it increases in the outgoing direction. 
The orthogonal derivative along the unit normal 
of $\Gamma$ may expressed in this time gauge as 
$\partial_\perp=$ $U^{1/2}\,\partial_t$ (but has of course a
geometric significance independent of any gauge). Now consider 
wave functions of the WKB-form (\ref{WKB2}). The prefactor $D$ may 
be computed in a vicitity of $\Gamma$ (it is unique only up to a function
that is constant along each path, but this freedom will cancel out
in what follows --- it can always be chosen as
$D|_\Gamma = 1$). Furthermore let us assume that $\chi$ approximately 
satisfies the effective Schr{\"o}dinger equation (\ref{eff}) and
dies off near the boundary of $\Gamma$, so that it
lives only on a domain in which $U>0$. 
All wave functions of this type make up
a space $\H^-(\Gamma)$, which is thus defined approximately. 
An according space $\H^+(\Gamma)$ is provided by the complex
conjugate wave functions, i.e. by $\psi=$ $\chi D \exp(-iS)$,
subject to the conjugate effective Schr{\"o}dinger equation
$i\partial_t\chi=$ $-H^{\rm eff}\chi$. Here we use the same
time coordinate $t$ as before, hence produce an additional 
minus sign. It is a straightforward application of the
formalism described in Section 12 to rewrite the equations
$i\partial_t \chi= \mp H^{\rm eff}\chi$ defining $\H^\pm(\Gamma)$
in terms of $\psi$. The result is
\begin{equation}
\partial_\perp\psi = \left( \pm\,\, i\, U^{1/2} 
-\,\frac{\partial_\perp U }{2 U} + \,\frac{1}{2} \,K 
\pm \,i\, U^{-1/2} H^{\rm mod} \right) \psi \, , 
\label{apprdec}
\end{equation}
where $K$ is the trace of the extrinsic curvature 
on $\Gamma$, and $H^{\rm mod}$ is the operator
$N^{-1} D H^{\rm eff} D^{-1}$ which --- in the present gauge ---
reads $- \frac{1}{2} U^{1/2}{\cal D}^a U^{-1/2}{\cal D}_a$
(cf. equation (\ref{Heff}) in Section 12). ${\cal D}_a$
is the covariant derivative with respect to the induced
metric on $\Gamma$. 
The identities necessary to establish this result are
$\partial_\perp S = $ $-U^{1/2}$ and
\begin{equation}
\frac{\partial_\perp D}{D} =
- \frac{\partial_\perp U}{2 U} + \frac{1}{2} K\, . 
\label{id}
\end{equation}
Each of the two equations (\ref{apprdec}) is just a restriction 
among the initial data $(\psi,\partial_\perp \psi)$ on $\Gamma$
for the Wheeler-DeWitt equation (\ref{wdw1}). 
The condition for $\chi$ to satisfy the effective Schr{\"o}dinger
equation to a good accuracy in a vicinity of $\Gamma$ is that 
$H^{\rm mod}\psi$ is in a sense much ''smaller'' than $U\psi$. 
(Let us note in parentheses that --- if the behaviour of $\chi$
at the boundary of $\Gamma$ is ignored --- the simplest
possibility to specify a wave function is to set
$\psi|_\Gamma=\psi_0=const$. At $\Gamma$, we thus have
$H^{\rm mod}\psi=0$, and (\ref{apprdec}) defines initial
conditions for an exact solution near $\Gamma$. In the case of 
the flat Klein-Gordon equation and $\Gamma$ a spacelike plane
in Minkowski-space, $\psi$ turns out to be a negative/positive
frequency plane wave). 
This procedure is performed for all ''spacelike'' hypersurfaces
$\Gamma$ in $\cal G$. If $\cal G$ is not too large, we expect 
--- to some reasonable accuracy --- the
scalar product $Q$ to be positive/negative definite on the linear 
span of the union $\cup \H^\pm(\Gamma)$ over all $\Gamma$. 
This defines an approximate decomposition 
for oscillating modes in a local domain in which $U>0$.  
\medskip

If the potential is negative, $U<0$, in some domain, any 
WKB-branch defines a local 
quantum structure as well. Although in this case the trajectories
do not fall into two clearly separated categories (because they are
''spacelike'' now), the double structure of mutually complex 
conjugate WKB-branches still exists, but the
relation between $Q$ and the local scalar product is lost due
to the non-spacelike character of a hypersurface on which
$\langle\,|\,\rangle$ is defined. Completing such a local hypersurface
to a global one, $\Gamma$, which is admissible in the definition
(\ref{Q}) of $Q$, a wave packet constructed by
appropriate initial conditions for (\ref{eff}) might leave
the $U<0$ domain, but intersect $\Gamma$ somewhere else 
and produce a further contribution to $Q$ about which we
have no control. However, we will use the $U<0$ domains
only for a particular condition that refers to the notation
of {\it true} and {\it false} local quantum mechanics 
(see below), and there is no need for a decomposition
of $\H$ there. 
\medskip

Summarizing, the decomposition problem is solved at the level
of accuracy that enables one {\it (a)} to distinguish between
time-reversed congruences of classical paths and {\it (b)}
to implement these into local quantum structures.
For the use of extracting predictions with reference to some 
domain, one defines the triple $(\H,Q,K)$, where 
$\H$ and $Q$
are the fundamental structure of the theory and $K$ is the
local approximate decomposition. Although in practice a
decomposition of the whole space $\H$ is not necessary,
we will retain this notation for convenience. 
Due to our ''minimal'' point of view, there is no 
possibility to transcend the approximate character of this 
construction, just as there is no possibility to identify 
physical properties in an oscillating state better than 
allowed by the WKB-methods. 
\medskip

So far our scheme allows for local approximate probabilistic
interpretation, but rather in the sense of conventional
quantum mechanical expectation than quantum cosmology: With 
regards to large scales (semiclassical 
variables), a wave function contains macroscopic
alternatives that never get mixed in observations. 
Let us consider now a state and several adjacent WKB-domains in 
$\cal M$. Inside each, one has a local quantum structure, and 
these structures are
expected to ''match'' smoothly (e.g. by the existence of
intermediate domains, covering part of two adjacent ones,
such that there is a common local quantum structure). 
Mathematically, one could define adjacent domains by
a local foliation such that any relevant classical trajectory
intersects any foliation hypersurface only once. 
The particular domains are provided by appropriate 
neighbourhoods of the hypersurfaces (which may often, but
not always, be chosen ''spacelike''). 
One is tempted to thinking about short pieces of semiclassical
evolution. Alternatively, one may consider 
families of macroscopically equivalent trajectories
in $\cal M$, followed over a short period of time.
These local structures should ''match'' so as to provide ''the
quasiclassical history'' of the universe as it can be described 
in retrospect.
Our point is now that the possibility of 
predictions concerning the large-scale almost-classical
variables implies the existence of a chain of adjacent 
local quantum structures {\it without interruption}.
We admit that our
formulation is not very precise, but the idea is that
local quantum structures are considered 
''along'' any piece of a local WKB-branch (e.g.
by following the hypersurfaces of a foliation), 
or ''along'' any piece of ''quasiclassical history''
(expressed in terms of the classical variables $y_{\rm cl}$). 
The crucial requirement may now be formulated as follows:
All these adjacent local structures must be of the
type of {\it true local quantum mechanics}, because
otherwise we cannot talk about predictions ''on the
way'' of the experienced macroscopic history of the 
universe. In principle, we should have formulated 
this statement already when giving the answer to 
{\it (i)}. In order to emphasize the {\it local} character
of the identification of quantum structures and 
''properties'' of states, we have swept it under the
carpet. In the present context, however, it seems
unavoidable to impose a feature that is a bit more
{\it global} in nature and ensures (or limits) the realm 
of predictions ''in'' a universe. 
\medskip

As already remarked at the end of Section 4,
it is conceivable that this condition may be re-expressed
in a mathematically more rigorous way, such as the requirement
that the wave function $\psi$ be {\it bounded} ($|\psi|<k<\infty$),
at least in certain domains. 
Such a condition is added to the tunneling proposal
\cite{Vilenkinout}\cite{DelCampoVilenkin}
as well, the motivation probably being a quite 
similar one than here. 
However, we do not consider it as an ultimate part of
the definition of states. Any wave function that does {\it not} 
meet this criterion is admissible but 
--- as soon as the {\it false} situation occurs ''along'' 
a quasiclassical history --- provides an ''interruption'' 
that would not occur in a purely classical world. 
\medskip

The principle of relating {\it predictability} with
{\it true local quantum mechanics} is important only 
in situations where the existence of $U<0$ domains 
may not be ignored. 
Since the classical paths are ''timelike'' if $U>0$,
the only possibility for an outgoing trajectory to {\it return}
and become an ingoing one (or {\it vice versa}) is to enter a 
region with $U<0$ and leave it with reversed orientation.
The typical situation for this is the case when the
(classical) universe has reached its maximum size and starts
to recontract. Also, these returning points are just
those where the scale factor becomes needless as a time 
parameter, and where the intrinsic nature of time in
quantum cosmology must show up
\cite{Zeh}. 
Our conclusion is thus that the fate of the universe
near its maximum expansion depends on whether the
wave function of the universe $\psi$ gives rise to
{\it true} or {\it false} local quantum mechanics
there. In Section 11, we will treat this problem as it 
appears in the Hawking model. In more sophisticated
models, there may appear more general situations of this
type. Let us just summarize that a history of the
universe that may be experienced macroscopically
(and which is experienced ''continuous'' in evolution)
requires the existence of a chain of reasonably matching 
adjacent {\it true} local quantum mechanics structures. A 
given wave function may describe {\it several} such ''worlds'', 
that must all be regarded as alternatives to each other. 
In general, there will contribute
many WKB-branches to a wave function in a local domain,
and many distinct large-scale histories are possible
as alternatives. However, it may happen that within
{\it one single} WKB-branch, i.e. even along the classical 
paths in this branch, a logical separation into alternatives 
occurs! In such a case, the associated local quantum mechanics 
will typically be {\it false}. 
The appearance of {\it false} local quantum mechanics in 
a wave function is by no means inconsistent or forbidden, 
but is just an indication for a logical separation of 
macroscopic histories from each other that is not always
recognized easily. (If {\it false} situations occur
well-separated from the domains in which the universe
is quasi-classical, e.g. in a domain describing the 
nucleation of the universe, they will be comparably harmless,
because there is nothing to be ''interrupted'' there).
This possibility is missing in many 
discussions about the interpretation of quantum cosmology,
and our criterion implies that the common interpretation
of the no-boundary wave function is likely to be wrong.
By applying our criterion to the Hawking model in Section 11, 
we will find hints that the no-boundary wave function allows 
for a cosmic catastrophe. 
\medskip

Let us mention that the {\it true chain requirement}
is not a complete substitute for a decoherence mechanism. 
We have already mentioned that macroscopically equivalent
trajectories (''tubes'' of paths in $\cal M$) display observable
quantum fluctuations, whereas macroscopically distinct
trajectories decohere. In order to work out the actual
scales at which these phenomena take place, one should
invoke the techniques of tracing out unobserved 
quantities within the WKB-domains. This issue
--- although of major importance --- is not in particular
our concern, once it is brought down to familiar
Hilbert space structures.
\medskip

Thus we have presented our key idea how the three problems 
{\it (i) -- (iii)} might be solved in particular models, 
in an interpretational scheme that is as ''minimal'' as possible. 
\medskip

Let us at the end of this section insert a comment
on possible relations between the selection of
''the actual'' wave function and the decomposition problem. 
In the discussion about Vilenkin's outgoing mode (tunnelling) 
proposal
\cite{Lindeout}--\cite{DelCampoVilenkin} 
it is sometimes regretted that we do not know how to define
precisely what an ''outgoing mode'' is --- in particular
in domains of minisuperspace which may not be relevant for
observations. When interpreted with respect to our scheme
this is clear, because even if there is a classically
well-defined definition of ''outgoing'' paths (which we
have assumed) 
this provides only an approximate concept for wave functions. 
Hence the proposal can at best determine a wave function 
approximately. It is certainly preferable to have a
proposal which singles out a unique wave function at the
exact level, and logically independent of the approximate 
reconstruction of conventional physics.
This would in any case require the introduction of an
additional structure, and we leave it open how 
it can be done and what consequences it would
have for the decomposition non-uniqueness problem.
Also, incuding path integral arguments --- as is
necessary if the no-boundary wave function
\cite{HartleHawking} 
is considered as the ''true'' state --- could enrich
the mathematical possibilities for an alternative treatment of 
the problem
(see also Refs.
\cite{HalliwellLouko}--\cite{Halliwell4}). 
\medskip

\section{''Minimal'' interpretation}
\setcounter{equation}{0}

We can now formulate how relative and approximate
probabilities are assigned.
Let $K$ be a decomposition associated with some local
domain of $\cal M$. We proceed according
to the formal rules of ordinary quantum mechanics, using
the scalar product $Q_{\!{}_K}$. 
Therefore we assume some wave function $\psi$ to 
represent the actual quantum state of the
universe. It need not be in $\H$
(neither $Q(\psi,\psi)$ nor $Q_{\!{}_K}(\psi,\psi)$ 
are required to be finite), but it shall have
well-defined scalar product with any $\Xi\in\H$.
In terms of a 
basis $\{\Xi^+_r,\Xi^-_r\}$, normalized according to
(\ref{basis1})--(\ref{cc}), 
the assignment of approximate relative probabilities associated 
with each $\Xi^\pm_r$ is given by
\begin{equation}
P^\pm_r = |c^\pm_r|^2 \equiv |Q(\Xi^\pm_r,\psi)|^2\, . 
\label{probr}
\end{equation}  
In principle, the relative probabilities of finding the universe 
in the $\pm$ modes are 
\begin{equation}
P^\pm = \sum_r |c^\pm_r|^2 \equiv \pm Q(\psi^\pm,\psi^\pm)
\equiv Q_{\!{}_K}(\psi^\pm,\psi^\pm),
\label{pmprob}
\end{equation}  
although these two numbers may be infinite. 
All expressions above emerge when one formally expands 
\begin{equation}
Q(\psi,\psi) = Q(\psi^+,\psi^+)+Q(\psi^-,\psi^-)
\label{eeexp}
\end{equation}  
in terms of the components of $\psi$ and changes the sign
of the $\psi^-$-term.
Thus, one never has to deal with the original expression 
$Q(\psi,\psi)$, and therefore one will not encounter any 
problem if $j=0$, as opposed to the Bohm-type interpretation.
In general, one may assign a relative probability with
respect to {\it any} normalized state $\Xi\in\H$, 
with arbitrary non-zero $+$ and $-$
components, just by $P(\Xi) =$ $| Q_{\!{}_K}(\Xi,\psi)|^2$. 
\medskip

The type of predictions that may be stated in our
formalism are to a large extent just those emerging from
the standard WKB-treatment of a given wave function.
For these, our scheme provides a clearer way of writing
things down. 
In this sense our program can be regarded as an attempt
to give the local WKB-techniques some common and global
structure.
However, as soon as different WKB-branches and different
WKB-domains in $\cal M$ are involved, the standard
WKB-formalism becomes less well-posed, and one is
obliged to introduce more or less heuristic {\it ad hoc}
considerations which will in effect just uncover
pieces of the structure defined by $\H$, $Q$ and $K\,$.
As an example, Vilenkin, when defining his interpretational 
scheme
\cite{Vilenkininter} 
treats the (macroscopically) different WKB-branches 
separately, which implies that the orientation of the
hypersurface element $ds^\alpha$ in (\ref{probab1}) 
depends on the particular branch one is integrating. 
Hence, his argument that the cross-terms (involving
different branches) in (\ref{probab1}) are small due to
rapid oscillations of the integrand is not really 
well-posed, since $ds^\alpha$ is not specified 
in this case. As a consequence, it is conceptually unclear 
how to estimate the accuracy of these cancellations. 
(In contrast, his cross-terms appear in our formalism in the
form $Q(\psi_1,\psi_2)$, which is at least in principle
well-posed and may serve as a check of orthogonality of
the different branches). 
The technically relevant point in Vilenkin's scheme
is an adjustment of signs in order to set up
a computational framework. In practice this adjustment 
does the same job as our decomposition $K$ of $(\H,Q)$,
but it does not operate {\it within} a fundamental and
(hopefully) appealing structure. 
\medskip

Once having identified elements of $\H$ with physical 
properties of the universe as observable in the
respective domain, our interpretational
scheme may reproduce the probability
and Hilbert space structure as assocated with the
local WKB-branches and the effective Schr{\"o}dinger
equations (\ref{eff0}) or (\ref{eff}). In this sense, the 
concept of probabilities in conventional quantum physics is
recovered, and with it the standard WKB-interpretation
if $\psi$ is locally of the WKB-type with a single
branch, $\psi=$ $A \exp(i S)$. In domains in which 
$\psi$ turns out to be a superposition 
\begin{equation}
\psi\approx \sum_j A_j e^{i S_j}\, , 
\label{WKBsuper}
\end{equation} 
our scheme automatically implies the correct
book-keeping of relative normalizations in the different
branches. This is due to the positivity of the
scalar product $Q_{\!{}_K}$. 
Moreover, our interpretation tells us in a simple
way whether two
observations associated with two states $\Xi_1$ and
$\Xi_2$ are in fact {\it alternatives} to each other.
This is the case if $Q_{\!{}_K}(\Xi_1,\Xi_2)=0$
(a non-trivial statement if the two states are elements
of slightly different WKB-branches: it is only for actions 
$S_1$ and $S_2$ fairly different from each other that one can 
expect the scalar product between the whole branches to be zero 
on account of destructive interference in the integral (\ref{Q})), 
and it shows us that the notation of alternatives in a 
measurement is a local one, depending on $K$. Again, this
provides the possibility of estimating the accuracy at
which standard physics is valid. 
Clearly, our scheme admits $\psi$ to be any superposition 
of elements of $\H$, out of both (local) subspaces $\H^\pm$.
However, it is able to do even more, in particular
when some wave function $\psi$ is considered in different
local WKB-domains. The treatment of $(\H,Q)$ as a {\it fixed}
fundamental structure and $K$ as a {\it local} one is crucial 
in identifying how {\it local physics changes} when
the evolution times are long. The effects showing up here
transcend to some extent the reach of the local 
WKB-interpretation. 
The identification of ''logical
interruptions'' along the large-scale history by means
of the {\it true}/{\it false} quantum mechanics criterion
is more than what is usually done, although it is achieved
by no more than the application of WKB-techniques.
\medskip

There is another type of effects related with long 
evolution times, and it is the final theoretical part in our 
scheme. So far, we have mainly talked about the reconstruction 
of local conventional quantum physics. 
The identification of the local scalar product and the 
properties of states works at the Schr{\"o}dinger approximation 
(\ref{eff0}) or (\ref{eff}). Only with respect to it we can talk 
about probabilites. However, at the next higher order in the 
(standard or modified) WKB-expansion we encounter a unitarity
violating (as well as a hermitean) contribution to the 
effective Hamiltonian. How do we deal with this situation? 
Our answer is that we just have to apply the local WKB-techniques
(in addition to a standard decoherence mechanism, i.e.
taking traces in local Hilbert spaces) in order to account for
unitarity-violating and quantum gravitational corrections! 
Suppose that a {\it given} wave function $\Xi$ is analyzed in 
adjacent WKB-domains ${\cal G}_j$, each one providing its local 
decomposition $K_j$  and its ({\it true}) local quantum structure. 
In each domain one will find several relevant WKB-branches. 
The probabilities 
associated with these branches may (even if the branches
exist in a domain larger than the individual ${\cal G}_j$'s) 
differ as one runs from one domain to another. This happens for
example if $\Xi$ represents a congruence of expanding universes
in some domain, and a superposition of expanding and contracting
universes in another one. Also, if only one congruence of
classical paths suffices to describe $\Xi$, the evolution of
probabilities may be such that it does not fit into a
unitary wave equation when followed over many local
WKB-domains. In other
words, even if the unitarity-violating contribution to the
Hamiltonian is small, it may give rise to observably large 
inconsistencies if (\ref{eff0}) or (\ref{eff})
are retained over long times. Technically speaking, the 
evolution of the wave
function according to the Schr{\"o}dinger approximation
may simply be in conflict with the fact that it solves the
Wheeler-DeWitt equation. Note however that these effects 
--- as well as the hermitean correction at the higher 
orders --- are too small to show up at short (proper) time 
scales. The corrected Schr{\"o}dinger equation 
--- equations (\ref{wdweps1}) or
(\ref{wdweps2}) in Section 12 --- implies a 
quantum gravity correction to any energy scale encountered
in local physics, but these corrections can only
be verified experimentally during a long period of time. 
In Section 12, a rough estimate of the relevant time scales 
is given. 
\medskip

It is precisely features like this that are accounted for by our 
local decomposition construction. 
The crucial point with such situations is that {\it one
never has to propagate a wave function with respect to the
local quantum structure}. We suppose that $\Xi$ is
prescribed, i.e. we know from the outset that it satisfies
the Wheeler-DeWitt equation. 
The WKB-domains necessary for our interpretation must only
be large enough to allow for the identification of
WKB-branches (i.e. of $\chi$) and the definition of local 
physical concepts. In our universe, this corresponds to extremely 
small time scales, so that $\cal M$ can be thought of as being 
covered by a huge number of ${\cal G}_j$'s. 
The interpretation can be performed 
in any of these domains independently.
As a consequence, in any domain a probability interpretation
is at hand. The long-term change of local physics is
incorporated automatically. The only problem that remains
is how these structures {\it match} in order to describe a
''continuity'' that can be experienced as
the quasiclassical ''history'' of the universe. 
However, applying a local decoherence mechanism, provides
a local notation of quasiclassical history, for each
domain ${\cal G}_j$ independently. In terms of the
classical variables $y_{\rm cl}$, as used in the standard
WKB-expansion (supposing for simplicity that they remain
intact over the total region considered), 
one would find a family of small pieces of
background evolutions $y_{\rm cl}(t)$, the ''length''
of each piece never exceeding the size of the respective
domain in these coordinates. Any piece is given a (relative) 
probability. These pieces match to a family of
background histories $y_{\rm cl}(t)$ that can be experienced. 
Unitarity-violation 
as well as quantum gravitational corrections 
should show up as a slight departure from the 
local unitary evolution (\ref{eff}) when 
one passes through a sufficiently large number of adjacent 
domains ${\cal G}_j$. 
Working out these features in detail
is certainly a difficult issue in a realistic model.
Concerning their conceptual status, we expect that all
uncertainties in realizing this program (with regards to 
the words ''approximate'', ''local'' and ''smooth'') arise 
just at the same level of accuracy which is limited by the 
WKB-philosophy altogether. There are exceptions such as the
case of a flat DeWitt metric and positive constant potential $U$
(this is in fact the flat Klein-Gordon equation).
There, the WKB-expansion can be performed
to all orders without encountering any unitarity-violating
term. As already remarked, this is due to the fact that a 
timelike Killing vector ensures the existence of a unique 
decomposition $K$ into negative/positive frequency states. 
Thus, the approximate nature of interpretation in
quantum cosmology can be traced back to the very nature
of any realistic Wheeler-DeWitt equation, as opposed to the
flat Klein-Gordon case. 
\medskip

We thus do not allow for ''jumps'' between different
(quasiclassical) backgrounds
(cf. Ref. 
\cite{Zeh1}).
Unitarity-violation is
usually considered as an indication of instability 
\cite{Kiefersemi}, 
and the relevant time scales are computed
(see e.g. Ref.
\cite{Kiefer7}).   
In our view only smooth transitions wlll occur, where the
word ''smooth'' means that adjacent local quantum structures 
match, as described above. No abrupt jump between macroscopically
different WKB-branches is possible. 
To some accuracy the slight change of local physics
(the long-term departure of the quasiclassical history from
classical solutions)
may be described by the version of semiclassical
gravity that includes the ''back-reaction'' according to
the modified Einstein-equations 
\cite{Kiefersemi}  
$G_{\mu\nu}=\langle T_{\mu\nu}\rangle$ or some minisuperspace
version thereof. 
This ''principle of continuity'' matches perfectly
our interpretation of {\it false} quantum mechanical
situations as logically separating pieces of trajectories from 
each other (as opposed to predicting a jump).
This does not mean that the universe behaves
peacefully: Drastic events are possible, if clearly
predicted by a wave function (see Section 11 for an
example). 
Also, sometimes very drastic tunnelling phenomena 
--- such as the tunnelling of the universe into a
de Sitter phase --- are supposed to be possible in some
models (see e.g. Ref.
\cite{HartleHawking}), 
although with strongly suppressed probabilities. 
Most of them are admitted by our scheme as well, in particular
if they arise from a (local) standard quantum mechanical
interpretation (e.g. under-barrier tunnelling). 
Thus, our scheme does not at all
truncate the variety of things that might happen. 
\medskip

To sum up, our approach to long-term phenomena
is simply to treat the higher orders in a 
WKB-expansion --- a soon as unitarity-violation appears --- 
as irrelevant {\it inside} a local WKB-domain. 
\medskip

It is conceivable that any further attempt to formulate
predictions more accurate than allowed by
WKB-techniques would mean to 
leave the realm of conventional physics and all its concepts. 
We emphasize that this is a consequence of our conceptual
(''minimal'') point of view, and that other positions 
are of course possible. Also, we must say 
that there is certainly some form of ''experience'' possible
in regions where all WKB-concepts begin to break down heavily 
(shortly ''before'' the big crunch, say). But it is hard
to say whether this has anything to do with ''observations''
in the sense of physics. At least we feel that it would
be a realm of nature in which questions formulated in terms of
the language of conventional physics cannot be posed. 
This sort of possible breakdown of predictability
has no counterpart in ordinary quantum mechanics (where,
e.g., a momentum or energy eigenstate is an unambigously
well-defined concept). Formally, it arises because
most relevant fundamental observables of interest (e.g. 
the momentum of the matter field $p_\phi = - i \partial_\phi$)
are represented by operators that 
do not commute with the constraint operator ${\cal H}$ defining
the Wheeler-DeWitt equation (\ref{wdw1}). They are thus not
operators acting on the space of solutions $\H$, and their
eigenfunctions are not allowed as wave functions. In terms of
the general theory of quantization of parametrized systems,
they do not even qualify as ''observables''
\cite{Isham}\cite{Kuchar}, 
although at the 
WKB-level there is certainly some information about their
distribution in a wave function. Measuring ''the value'' of
such an observable to a degree more accurate than allowed
by its identification in a reasonable WKB-approximation is simply 
impossible, because at such an accuracy there is no
unique notion of what it should mean in terms of observations.
(If, on the other hand, an operator like $p_\phi$ happens to
commute with ${\cal H}$ in a simple model, we have the exceptional
case of a completely well-defined observable
whose eigenstates satisfy the Wheeler-DeWitt equation and need
no approximate WKB-arguments for their identification).
Maybe a final answer to the question what kind of experience 
one could make in almost genuine quantum gravitational 
situations
amounts to feed the theory with information about all the
particular objects present there, in particular
the human body, including the brain and the like.
\medskip

On the conceptual level (leaving cosmology for the moment), 
one might object that at least the problem of the final state
of black holes should provide an observational
''window'' towards a full quantum gravity
\cite{Kieferpriv}. 
This is certainly true, but the experimental device by which
physical observations of quantum black holes are performed will, 
for example, be at some distance from these objects 
(or ''separated'' from them by the appearance of largely different 
energy scales) and thus in a WKB-type environment.
Nevertheless, the wave function satisfies the Wheeler-DeWitt equation 
exactly. In this sense, the final state of black holes is
''described'' by full quantum gravity, and the non-WKB features
of the wave function in certain domains of (mini)superspace will
of course be essential. 
Although the mathematical details are far from being clear,
this seems to be a beautiful example how an exact underlying
mathematical framework might interplay with the approximate
nature of extracting physical information. 
Posing questions that refer to a situation too ''close'' to 
such a quantum gravity process would run into the problem that the 
precise description of some ''measurement'' (that {\it should} be 
performed in order to test the theory) becomes impossible 
{\it on account of the nature of the process itself}. 
\medskip 

Summarizing, a ''minimal'' interpretation of
quantum cosmology that relies on $(\H,Q)$ as the
only fundamental mathematical structure makes a quite
radical point of view possible: 
Quantum cosmology (quantum gravity) virtually destroys the
language of physics and limits the realm of nature
in which we can reasonably talk about observations 
--- independent of who performs them  --- and
thus about physics. A true ''Planck scale physics'' 
would exist primarily as a mathematical theory, and
its relation to ''experience'' is unclear and tends
to transcend what is usually called the ''physical world''. 
We have presented our arguments only in the minisuperspace 
approximation, but in principle one can try to implement
analogous ideas --- in particular the ''minimality'' of the
scheme --- to a full quantum cosmological framework.
The main goal such a framework could possibly achieve 
is to show {\it how} the concepts of usual physics are hidden 
therein, and to extract predictions for all observations that 
can be formulated in the conventional physics' language {\it and}
that actually {\it can be performed}. 
We admit that our approach does not tell us {\it why} just
these identifications between approximate mathematical
structures and observations have to be made.
We leave it open whether a ''completion'' is possible 
and which philosophical status it would have.
\medskip

\section{Comments}
\setcounter{equation}{0}

Despite the extensive use of WKB-techniques, the
underlying structure of our program consists of inserting 
solutions
of the Wheeler-DeWitt equation into the scalar product $Q$.
This was our starting point, and it
raises the question how a formalism based on 
numbers $Q(\psi_1,\psi_2)$
relates to a Bohm-type interpretation using 
(\ref{probab1}) as a measure on the set of flow-lines
$y_{\rm flow}(\tau)$ of $j$. In terms of this
interpretation, 
it is possible to associate with each pencil of flow
lines a relative probability, irrespective of how narrow the 
pencil is. However, the expression (\ref{probab1}) 
{\it cannot} simply be written down in terms
of numbers of the type $Q(\Xi,\psi)$, 
with $\Xi$ being a solution of the Wheeler-DeWitt
equation. It is rather of the type $Q(\psi',\psi')$, where
$\psi'(y)$ is a function that is defined to be identical to
$\psi(y)$ for all points $y$ through which a trajectory of the
pencil ${\cal B}$ passes, and zero otherwise. Hence, $\psi'$
does not in general satisfy the Wheeler-DeWitt equation.
Among all functions that arise from $\psi$ by cutting away
its values outside a tube ${\cal T}$, it is just the 
functions of
the type $\psi'$ (i.e. with ${\cal T}={\cal B}$) for which
the integral (\ref{probab1})
is independent of the hypersurface $\Gamma\,$! 
Hence $\psi'$
mimicks a solution without actually being one, but this
is enough to define a conserved current. 
One may in fact use this situation as a {\it definition} of 
the flow lines of $j$. One encounters
precisely the same structure in ordinary quantum mechanics,
where the trajectories considered in the Bohm-interpretation
\cite{Bohm}--\cite{BohmHileyKaloyerou}
just serve for the same purpose as $y_{\rm flow}(\tau)$. 
Interpreting these flow lines as describing an actual motion (in
particular when the wave function is not approximately
of the WKB-type)
is not necessary in quantum mechanics, nor is it in quantum
cosmology. When this situation is examined within our scheme,
it turns out that we cannot assign relative probabilities
to pencils of flow-lines, because an arbitrarily
cut-out pencil does not correspond to an exact solution
of the Wheeler-DeWitt equation. In contrast, we can compute
relative probabilities of the type $|Q(\Xi^\pm,\psi)|^2$,
where $\Xi^\pm$ is a solution that represents an
incoming or outgoing wave packet, 
concentrated near $\cal T$. When this packet is considered 
to be very narrow on a hypersurface $\Gamma$ as used in 
(\ref{probab1}), it will spread out very fast 
in a neighbourhood of $\Gamma$. It is {\it this} 
property which is associated with the reference state 
$\Xi$, and it has nothing to do with a pencil-like
structur, nor with a conserved current.
The current $j$ may be viewed as a mathematical object which 
is of minor importance, just as this point of view
is possible in ordinary quantum mechanics. In this sense,
the interpretations based on (\ref{probab1}) are not
equivalent to ours. 
However, they have a reasonable place within
our framework, and the relative probabilities
(\ref{probab1}) may be regarded as expressing a Bohm-type
interpretation of the ordinary quantum mechanics structure
(\ref{eff}) inside the local WKB-branches.
This may be read as an argument in favour of the 
Bohm-interpretation, but at the level of quantum mechanics it 
is just a matter of talking about the formalism, and may
(but need not) be adopted. 
However, all arguments we gave in this discussion rely on
the WKB-approximation. There is a lower bound in the
width of packets and initial conditions, below which
the approximation gets spoiled. This provides a further
conflict between the Bohm-interpretation and ours.
In Section 10, we will discuss this point when having
a simple example at hand. 
\medskip

We would like to add a remark concerning the formalism of
ordinary quantum mechanics and the way how it
emerges. Adopt the phrase ''everything shall be expressed
in terms of $Q(\psi_1,\psi_2)$'' and retain it even
in the local WKB-branches and their associated
quantum structure. To be specific, let $\psi_{1,2}$ be two 
approximate WKB-type wave functions of the type (\ref{WKB2}). 
As we have seen in
equation (\ref{QQ}), the scalar product $Q$ boils 
down locally to the scalar product of ordinary
quantum mechanics, but with two solutions of the
{\it time-dependent} Schr{\"o}dinger equation inserted.
Taking our proposal serious that probabilities should
be expressed in this way, the question arises whether
one can reconstruct the standard formulae for amplitudes
which are rather of the type
$\langle a|\chi_t\rangle$, where $|a\rangle$ is a 
{\it time-independently} 
defined eigenstate of some operator $A$ to an eigenvalue $a$.
The answer is quite simple, and it makes use
of the particular role time plays in quantum mechanics: 
Let $|\eta_t\rangle$
be a solution of the time-dependent Schr{\"o}dinger equation 
that coincides
with $|a\rangle$ at $t=t_0$. To the approximation at which
quantum mechanics is valid, this state is unique.
Then the scalar
product $\langle \eta_t|\chi_t\rangle$ is just
$\langle a|\chi_{t_0}\rangle$, and it is of the required form.
Thus, the conventional formalism is in accordance with
our proposal. The practical use of wave functions like
$|\eta_t\rangle$ for predictions relies on our ability
to ''prepare'' states {\it strictly at} $t=t_0$. This is possible 
within the 
realm of conventional quantum mechanics, where time plays the 
role of an external parameter and, moreover, the
underlying wave equation is of first order in $\partial_t\,$.
\medskip

As a further remark --- that shall prevent
misunderstandings --- we emphasize that the status
of quantum cosmology in the approach using the
Wheeler-DeWitt equation is that of a second-quantized field
theory in the ''position representation''. Although in the
minisuperspace approximation most of
the degrees of freedom have been truncated, the wave
function $\psi(y)$ is just a simplified version of
$\psi[g_{ij}(x),\phi(x)]$. Thus, our particular interpretational
scheme to use only scalar products of solutions to the field 
equation does {\it not} carry over to
the Klein-Gordon equation, which may be regarded as the
{\it first}-quantized model for a scalar field, or as a 
classical model whose quantization is {\it second}-quantized.
In the wave functional $\psi$, the field
$\phi(x)$ just serves as a variable
and is never subjected to a Klein-Gordon equation. 
Whenever we talked about this equation so far, we had in mind
the mathematical similarities and differences to the
Wheeler-DeWitt equation, not
an alternative interpretation of quantum field theory.
\medskip

It is sometimes proposed that a proper theory of quantum
cosmology should be {\it third}-quantized (see e.g. Refs.
\cite{HosoyaMorikawa}\cite{McGuigan}). 
In such a case,
our interpretation as it was formulated will clearly need
some modification.
Nevertheless, the idea of using reference wave functions 
whose physical contents is known as far as observations are
concerned is in perfect accordance with the needs of
third quantization. In an expansion of the type
(\ref{expansion2}), the coefficients $c^\pm_r$
would become operators associated with the ''modes''
$\Xi^\pm_r$. A fundamental theory 
could be defined by a Fock space structure with
respect to any decomposition $K$ (the superscripts $\pm$ 
on the operators $c^\pm_r$ denoting creation and
annihilation, respectively). An appropriate set of 
commutation relations (for any
given $K$) is provided by 
\begin{equation}
[c^-_r,c^+_s] = \delta_{r s}\qquad\qquad 
[c^\pm_r,c^\pm_s]=0\, . 
\label{comm}
\end{equation}  
A change of decomposition would have to be represented
as Bogoljubov-transformation, just as what is usually
done in the formalism of quantum field theory in 
curved space-time
\cite{BirrellDavies}. 
The notation of states, i.e. superpositions of objects 
of the type 
\begin{equation}
|\Phi\rangle =
c^+_{r_1} c^+_{r_2}\dots c^+_{r_N}
|0\rangle_{\!{}_K}\, ,  
\label{fock}
\end{equation}  
should thus exist at the exact fundamental level.
Writing down a given state $|\Phi\rangle$
in this way as element of a Fock space, a decomposition
$K$ must be selected. Here, the approximate nature
of conventional physics comes in exactly as it
does in the {\it second}-quantized version:
Taking into account what kind of observations we have
in mind, we choose the approximate local decomposition
as constructed in Section 7. However,
a {\it third}-quantized formalism as sketched here is 
usually regarded as a {\it free} theory, and 
the major new possibility one has is to introduce
interactions (''between'' different universes, such
as joining and splitting). Such a construction 
could possibly accomodate for topology fluctuations 
and the creation and annihilation of baby-universes,
thereby using as little pre-supposed mathematical structure
as is reasonably possible,
in accordance with the spirit of the rest of this work. 
\medskip

\section{Wave packets in the Hawking model}
\setcounter{equation}{0}

So far, we have used the Hawking model just to illustrate some
expressions in a simple context. Let us look at it now in a 
bit more detail. In this section we will analyze approximate
wave functions in the inflationary domain --- which is specified
by the conditions $|\phi|\gg 1$ and $m^2 a^2\phi^2\gg 1$ ---
and work out the associated local quantum structure as an example. 
There is a family of classical paths that is particularly 
important, and we will confine our consideration to these.
A typical trajectory of this family starts at some
$\phi=\phi_0$, along one of the two curves (hypersurfaces) 
at which $U=0$,
i.e. $m^2a^2\phi^2=1$. In a time gauge $N\neq 0$, its
initial condition is $\dot{a}=\dot{\phi}=0$, and from
its initial point $(a_0,\phi_0)$ it runs into the region $U>0$.
If $|\phi_0|\gg 1$, it represents the inflationary era of a 
classically expanding universe, until eventually
$|\phi|$ becomes of order unity (and inflation comes to 
end, followed by oscillations of $\phi$ with decreasing
amplitude --- see Section 11). 
These paths appear in the context of the no-boundary proposal 
as those dominating the Euclidean path integral in the 
inflationary domain 
\cite{HartleHawking}\cite{Hawking2} 
(by which they may be called
no-boundary trajectories), 
as well in the context of the outgoing mode proposal 
\cite{Vilenkinout} 
and chaotic inflation
\cite{LindeInfl}. 
\medskip

Without loss of generality, we restrict ourselves to positive 
$\phi$. In addition to the WKB-approximation (which is perfectly
applicable within the inflationary region), 
we assume $\phi\gg 1$ and $U\approx m^2 a^3 \phi^2\gg a$.
The inflationary nature of the expansion tends to swallow 
information along the trajectories, and the mathematical aspect 
of this feature is that retaining a reasonable degree of 
accuracy is quite cumbersome. So we will write down
only the very leading order in most expressions 
(and use the $\approx$ sign only in case of very crude 
approximations). The trajectories defined above 
have been studied in very detail by Page
\cite{Page}. 
The corresponding action is given by
\begin{equation}
S=  -\,\frac{1}{3 m^2\phi^2} \left( m^2 a^2 \phi^2-1\right)^{3/2}
+\, \frac{\pi}{4} 
\approx -\, \frac{1}{3} m \phi a^3\, . 
\label{action}
\end{equation}
The trajectories (which, according to our notation, are outgoing 
ones, describing expanding universes), are given by the curves
\begin{equation}
a = \frac{1}{2 m \phi_0^{2/3}\phi^{1/3}}
\, \exp\left(\frac{3}{2} ( \phi_0^2 - \phi^2 )\right)
\approx 
\exp\left(\frac{3}{2} ( \phi_0^2 - \phi^2 )\right).
\label{trajhm1}
\end{equation}
This expression is valid as long as 
$1\ll\phi\ll\phi_0 - 1/(3 \phi_0)$.
In the proper time gauge we have $N=1$, the evolution parameter 
along each trajectory being $t_{\rm proper} = 3(\phi_0-\phi)/m$. 
We will now write down the local quantum structure associated
with this congruence, and simply apply the general formalism as 
described in Section 12. The most convenient coordinates
for the quantum case are $\phi_0$ (corresponding
to the $\xi^a$ of Section 12) and $t\equiv -S$. The
transformation from $(a,\phi)$ to $(t,\phi_0)$ is provided
by (\ref{action}) and (\ref{trajhm1}).
To leading order, the DeWitt metric (\ref{dwhm}) reduces to
\begin{equation}
ds^2_{\rm DW} = - \,\frac{1}{a^3 m^2 \phi^2}\, dt^2 + 
\frac{a^3 \phi_0^2}{\phi^2}\, d\phi_0^2\, .
\label{wdhminlf}
\end{equation}
The ''lapse'' of this metric is thus given by 
${\cal N}=$ $U^{-1/2}$, the ''shift'' (corresponding to 
${\cal N}_a$ of Section 12) vanishes identically. The 
(physical) lapse defining the negative of $S$ as time parameter 
is given by 
$N = U^{-1}$, and the prefactor from (\ref{WKB2})
--- see also equation (\ref{Dsolution}) --- is given by 
$D^2=$ $m^{-1} a^{-3} \phi_0^{-1} f(\phi_0)$, where $f(\phi_0)$ 
is an arbitrary function defining the particular representation. 
Choosing $f(\phi_0)\equiv 1$, the wave function can be 
written as
\begin{equation}
\Xi(t,\phi_0) =  m^{-1/2} a^{-3/2} \phi_0^{-1/2} \,
\chi(t,\phi_0)\, \exp(-i t)\, .
\label{psiwkb}
\end{equation}
The prefactor coincides in this approximation with the one 
given in Ref. 
\cite{Page}. 
The local scalar product takes the form
\begin{equation}
Q(\Xi_1,\Xi_2) \approx -\int d\phi_0 \,\chi_1^*(t,\phi_0)
\chi_2(t,\phi_0) \equiv 
- \langle\chi_{1,t}|\chi_{2,t}\rangle
\label{Qhmwkb}
\end{equation}
where the minus sign comes from the fact that we are
considering the $\H^-$ subspace of expanding universes. 
(The approximate local decomposition $K$ in terms of 
incoming/outgoing modes is quite clear here, and
all wave functions constructed around this
family of trajectories are in $\H^-$).
Hence, the probability density associated with a state
in the WKB-interpretation is just 
\begin{equation}
dP(t,\phi_0) = |\chi(t,\phi_0)|^2\, .
\label{probabhm}
\end{equation}
This quantity is essentially the integrand of
(\ref{probab1}) and is usually computed by using curves of 
(small) constant $a$ or $U$ to play the role of $\Gamma$. 
It plays a major role in the
discussion about the likeliness of sufficient inflation
\cite{Halliwell3}\cite{GrishchukRozhansky},
\cite{GrishchukSidorov}--\cite{Lukas}
and the comparison of the no-boundary versus the 
tunnelling proposal. 
Inserting (\ref{psiwkb}) into the WheelerDeWitt equation
(\ref{wdwhm}) --- or likewise computing the
effective Hamiltonian as given by (\ref{Heff}) 
straightforwardly ---  we find as the dominant contribution
\begin{equation}
i \,
\frac{\partial}{\partial t}\, \chi(t,\phi_0) =
-\,\frac{1}{18\, t^2}
\left(\frac{\phi}{\phi_0}\right)^{1/2}
\frac{\partial}{\partial \phi_0}\,
\frac{\phi}{\phi_0}\,
\frac{\partial}{\partial \phi_0}
\left(\frac{\phi}{\phi_0}\right)^{1/2}
\, \chi(t,\phi_0)\, .
\label{schrhm}
\end{equation}
The Hamiltonian is hermitean with respect to (\ref{Qhmwkb}).
Also, recall that $\phi$ is understood as a function of
$t$ and $\phi_0$ (one finds $\phi^2(t,\phi_0) \approx$
$\phi_0^2 - 2 \ln(t)/9 $).
This is an example for a local quantum structure of
the type (\ref{eff}), i.e. an approximate unitary 
evolution, whose Hamiltonian is not at all of the standard form. 
Since $t$ grows rapidly as the evolution goes on, the
action of the effective Hamiltonian on wave functions will
be suppressed, and $\chi$ becomes essentially a function
of $\phi_0$ alone. The probability density (\ref{probabhm})
also becomes $dP(\phi_0)$, and this is the approximation
that is usually invoked in the discussion of the
inflationary domain. The fact that (\ref{probabhm})
contains a slight dependence on $t$ --- and thus does not
exactly provide a probability density on the set of
classical trajectories --- just reflects the difference
between classical paths and the flow lines of the
current (\ref{current1}). 
\medskip

Let us discuss whether the WKB- or Bohm-inspired
interpretations which consider $dP(\phi_0)$ as an
approximate (relative) probability density on the set of 
classical paths is equivalent to ours. In Section 9,
we have already considered an aspect of this question,
and we would like to add another one here. 
We may use (\ref{schrhm}) to construct wave packets.
Note that this evolution equation implies a very tiny amout
of spread if $t$ is large. Given a wave function $\psi$, 
the typical (relative) probabilities as used in our scheme are
$P=|Q(\Xi,\psi)|^2$, where $\Xi$ is an outgoing (hence $\H^-$) 
wave packet-like state. One might expect that a notation of 
probability of this type, based on the use of various wave packets
$\Xi$, is equivalent to stating that $dP(\phi_0)$ is a probability 
density. This would be perfectly true in conventional
quantum mechanics. 
\medskip

However, in our case the WKB-approximation was invoked in order
to obtain (\ref{schrhm}). In this approximation,
$\chi(t,\phi_0)$ may be prescribed at $t=t_0$, and the
Schr{\"o}dinger time evolution fixes $\chi$ uniquely. 
This enables one to prescribe $\chi$ to be non-zero only 
in an arbitrarily narrow interval of $\phi_0$ a $t=t_0$.
(Compare the discussion of ordinary quantum mechanics in
Section 9). The expression $P=|Q(\Xi,\psi)|^2$ is just
the standard quantum mechanical probability for finding
$\phi_0$ in the given interval, when measured at time $t_0$.
The drawback is that --- as we have already mentioned --- 
if the width of the packet becomes too narrow
(i.e. the change of $\chi$ being too large), the
WKB-approximation based on $S$ from (\ref{action}) will break 
down. Prescribing a wave function $\Xi$ at the
''initial'' curve $t=t_0$ to be a narrow peak in some interval of
$\phi_0$, and using the full Wheeler-deWitt equation to
describe the dynamics, some information about the
time-derivative of $\Xi$ is missing. Without adding
this information to the initial conditions, the state
$\Xi$ is simply unknown! 
In a sense, this situation corresponds to the uncontrolled 
production of particles in quantum field theory 
if one likes to look at some process too accurate. 
In the WKB-approximation, our interpretation tells
us that the time-derivative of $\Xi$ at $t_0$ must be chosen
such that it is compatible with (\ref{schrhm}).
This makes $dP(\phi_0)$ effectively a probability density
for sufficiently large intervals of $\phi_0$.
Hence, the use of very narrow initial conditions at $t_0$ 
is limited by the requirement of assigning a clear 
{\it operational} significance to the situation. 
As opposed, in the WKB-approximation such a significance is 
provided by our ability to identify the ''instant of time'' $t_0$.
Thus, no predictions are possible for 
intervals of $\phi_0$ whose size are below a certain bound. 
This is a conceptually important (though not very practical) 
difference to the standard WKB-scheme based on the notation of
a probability density. The limitation of observations --- caused 
by the breakdown of local quantum mechanics --- is reflected
automatically in our interpretation: No prediction is
possible that is not expressed in terms of the
$Q$-product between solutions of the Wheeler-DeWitt
equation. In this sense the breakdown of the WKB-approximation
for very narrow wave packets is not just a shortcoming of a 
convenient formalism, but rather a crucial piece of our 
interpretation. 
\medskip

\section{Approximate matter energy eigenstates in the Hawking model}
\setcounter{equation}{0}

The probabilistic interpretation of wave functions in the
Hawking model is usually based on the variable $\phi_0$,
labelling the no-boundary trajectories, as used in the 
foregoing Section. From the knowledge of a distribution
in $\phi_0$, sometimes conclusions are drawn to other 
quantities, such as the density parameter
\cite{HawkingPage}. 
It would however be convenient to have some basis of
$\H$ at hand which is related more directly to 
features we can observe in the post-inflationary domain. 
Such a family is provided by the approximate eigenstates
of the matter energy after inflation. Therefore, we
consider trajectories that enter the domain $|\phi|\lsim 1$
after an inflationary era (these trajectories may --- but need 
not --- be in the no-boundary family whose behaviour in the
inflationary domain was analyzed in the foregoing
Section). Any such trajectory begins to oscillate with 
decreasing amplitude. This forces the scale factor $a(t)$ to 
decelerate (matter dominated phase). Eventually the
universe reaches a maximum size $a_{\rm max}$ (this happens when 
the amplitude of $\phi$ falls into the $U<0$ domain, i.e. inside 
the pair of curves $\phi=$ $\pm 1/(a m)$), and recontracts. 
The matter energy $E$ (defined to be measured in proper
time, i.e. in the gauge $N=1$) is approximately conserved during 
this era, and its value is related to the maximum size of the
universe by $a_{\rm max}=2 E$. Quantum mechanically, it is
given by the operator
\begin{equation}
E = -\,\frac{1}{2 a^3} \,\partial_{\phi\phi} +
\frac{1}{2}\, a^3 m^2 \phi^2 
\label{Equ}
\end{equation}
which appears (multiplied by $2a$) in the Wheeler-DeWitt equation
(\ref{wdwhm}). 
Since $E$ does not commute with the Hamiltonian constraint,
there will be no exact energy eigenstates in $\H$.
However, one may find approximate eigenstates, and this is
what we sketch now. Since the details about this issue will
be published elsewhere 
\cite{FE9} 
we will be brief, and mainly write down the results. 
The mass $m$ of the scalar field shall be
of the order of magnitude $10^{-6}$ 
(in the dimensionless units of Ref.
\cite{Hawking2}), 
in order to account for
the necessary amount of density fluctuations
\cite{HalliwellHawking}\cite{LindeInfl}. 
\medskip

Using the fact that (\ref{Equ}) is of harmonic oscillator
type in $\phi$, with frequency $m$ and mass
$a^3$, we infer that the eigenstates of $E$
have eigenvalues $E_n = (n+\frac{1}{2}) m$. The classical universes
with energy $E_n$ enter the post-inflationary domain
at a size $a_{\rm min}\approx$ $(n/m)^{1/3}$ and reach a maximum
size $a_{\rm max}\approx$ $2 m n$. This immediately
implies $n \gg m^{-2}\approx 10^{12}$ 
as condition for $E_n$ to describe the energy of a real
universe. Any such eigenvalue is associated with
two approximate solutions of the Wheeler-DeWitt equation,
given by
\begin{equation}
\Xi^\pm_n(a,\phi)\approx
\left(\frac{m}{2 \,E_n}\right)^{1/4}
\exp\left(\pm
\frac{2\, i}{3}\, (2E_n)^{1/2}\, a^{3/2}\right)
\Psi_n(m^{1/2} a^{3/2} \phi)\, ,
\label{Xihm}
\end{equation}
where $\Psi_n$ are the unit harmonic oscillator
eigenfunctions (e.g. $\Psi_0(x)=$ 
$\pi^{-1/4}\times$ $\exp(-x^2/2)$). The above form is valid for
$a_{\rm min}\,\lsim\, a \ll a_{\max}$ and extends in $\phi$
over the range in which the $\Psi_n$ are oscillating.
For even larger $|\phi|$, the $\Xi_n^\pm$ become strongly
suppressed. 
These wave functions have been used by Kiefer
\cite{Kiefer1} 
to construct wave packets, and a particular
combination of $\Xi_0^\pm$ (called $\Omega_0^+$,
see equation (\ref{Omegan}) below) is mentioned by Page
\cite{Page} 
as the dominating part of the no-bounary wave function
in the Euclidean domain $|\phi|<1/(m a)$. 
Moreover, Hawking and Page have encountered wave functions
related to these 
\cite{HawkingPage2} 
in a wormhole context. 
The $\Xi_n^\pm$ are normalized according to
(\ref{basis1})--(\ref{cc}). By appropriate continuation
to $a>a_{\max}\,$, one may infer the asymptotic 
as well as the analyticity structure
of these wave functions. In Ref. 
\cite{FE9} 
we provide evidence that $\Xi_n^\pm$ may be 
defined {\it exactly}. 
Also, small values of $n$ (integer and non-negative)
may be included, but these wave functions 
are mere tunnelling states. One thus seems to end up with
an exact basis of $\H$ into which any given wave function
may be expanded.
\medskip

The physical interpretation of the $\Xi_n^\pm$ is clear:
they represent contracting and expanding universes,
respectively, whose
approximate value of the matter energy in the domain 
$|\phi|\lsim 1$ is $E_n$. They also define a
decomposition $K$ which corresponds to finding the universe
contracting ($\Xi_n^+$) or expanding ($\Xi_n^-$) when the 
observation is done in
the domain $|\phi|\lsim 1$ (i.e. in the ''post''- or
''pre''-inflationary era, according to the orientation of
the classical paths). Thus, we have a perfect set
of reference states at hand (just note that for
$n\,\lsim\, m^{-2}$ a relation to an observation is
not possible).
\medskip

If it is actually possible to define a basis $\Xi_n^\pm$ {\it exactly}
by means of analyticity arguments, one may think of considering 
the resulting decomposition $K$ as a {\it preferred} one
upon which all interpretational issues are based. 
(We have pointed out this possibility in Section 7, the preferred 
decomposition was called $K_0$ there). 
It is however not clear
whether $K$ coincides with the approximate local decompositions
as constructed in Section 7 in other domains (i.e. for
small $a$ and large $|\phi|$). Also, we do not know to what extent
the appropriateness of the asymptotic analycity structure
for defining an exact decomposition carries over to other models. 
It will not be necessary here to specify the philosophy
so as to consider $K$ as an exact or an approximate object, but
this choice provides an interesting issue for future research. 
\medskip

As an example of an expansion of a state into 
$\Xi_n^\pm$, we consider the no-boundary wave function
. In the inflationary domain it can be written as 
\cite{Page} 
\begin{equation}
\psi_{\rm NB}(a,\phi) \approx 
\frac{1}
{\pi^{1/2} a \, \sqrt[4]{a^2 m^2\phi_0^2-1}} 
\left(
   \sqrt{6} + 2 \,\Big( \exp(\frac{1}{3m^2\phi_0^2})-1 \Big) 
\right) 
\cos(S)
\label{nobound}
\end{equation}
with $S$ from (\ref{action}), and $\phi_0$ now interpreted
as a function of $a$ and $\phi$. 
Expanding $\psi_{\rm NB}$ in terms of the basis
(see equations (\ref{expansion2})--(\ref{expansioncoeff}))
is a bit tricky because not much details about the wave
function in the domain $|\phi|\lsim 1$ are known.
Without any computation we just state that a
rough estimate 
\cite{FE9} 
yields for $n\gg m^{-2}$
\begin{equation}
c_n^+ = (c_n^-)^* = 
\frac{1}{\sqrt{6\pi n}}\,
\frac{1}{\sqrt[4]{2\ln(n m^2)}}
\left(
   \sqrt{6} + 2 \,\Big( \exp (
\frac{3}{2m^2\ln(n m^2)} )-1 \Big) 
\right) k_n 
\label{psinbcn}
\end{equation}
for even $n$, where the $|k_n|$ are of order unity, 
and $c_n^\pm=0$ for odd $n$. 
The first equality is exact, it stems from the fact that
$\psi_{\rm NB}$ is real. The result shows that very large
$n$ are suppressed (note that the associated probabilities
are $P_n^\pm = |c_n^\pm|^2$), at least 
as long as universes nucleating with densities above the 
Planck scale ($n\gsim m^{-2} \exp(\frac{9}{2m^2})$) 
are excluded. 
This is a well-known problem of
the no-boundary wave function to produce sufficient inflation
and a universe of realistic size 
\cite{Halliwell3}\cite{GrishchukRozhansky},
\cite{GrishchukSidorov}--\cite{Lukas}
(On the other hand, in the limit $n\rightarrow\infty$ the energy 
spectrum becomes $|c_n^\pm|\sim n^{-1}$, which makes
arbitrarily large $n$ to dominate as long as no upper cutoff 
for $n$ is introduced. Also cf. Ref. 
\cite{HawkingPage}). 
\medskip

There are two particularly interesting real superpositions 
$\Omega^\pm_n$ of the $\Xi_n^\pm$, defined by
\begin{equation}
\Xi_n^\pm(a,\phi) = \exp\left(\pm i\, (\frac{\pi}{2}\,E_n^2+
\frac{\pi}{4}) \right)\, \left(
\Omega_n^+(a,\phi) \mp i \Omega_n^-(a,\phi)\right)\, .
\label{Omegan}
\end{equation}
For $a\gg a_{\rm max}$ they have the asymptotic behaviour
$\Omega_n^\pm\sim \exp(\pm a^2/2\mp a E_n)$. Expanding 
$\psi_{\rm NB}$ in terms of these, one finds --- up to some 
uncertainty in the approximations involved --- that the expansion 
coefficients with respect to both types of states 
$\Omega^\pm_n$ are non-zero. This matches
the well-known fact that $\psi_{\rm NB}$ grows exponentially
with $a$ when $|\phi|$ is sufficiently small. The
dominating part for large $a$ and $\phi=0$ is $\Omega_0^+$
(a pure tunnelling state), and it 
is just this function that has been displayed by Page
\cite{Page} 
as the exponentially growing piece of $\psi_{\rm NB}$.  
\medskip

Let us insert here a remark about the tunnelling wave
function 
\cite{Vilenkinout}. 
It is usually required to be bounded,
$|\psi_{\rm T}(a,\phi)|<\infty$. The outgoing mode proposal
operates at the ''singular'' boundary of $\cal M$, i.e.
it is concerned with the contribution of trajectories
describing a singular behaviour $|\phi|\rightarrow\infty$
as $a\rightarrow 0$. This makes it difficult to compute
$\psi_{\rm T}(a,\phi)$ in the post-inflationary domain.
However, if all necessary conditions may consistently 
be implemented (which does not seem evident to us;
at least Hawking and Page 
\cite{HawkingPage2} 
provide a hint for the existence of wave functions that are 
everywhere bounded), 
the tunnelling wave function is appearently 
a superposition of the $\Omega_n^-$ alone (although
with complex coefficients). 
\medskip 

We will now use the Hawking model to discuss the issue of 
{\it true} and {\it false} local quantum mechanics, as 
introduced in Section 4. The major part of the 
post-inflationary phase provides an example of a local 
quantum structure for which reasonable wave functions
lead to {\it true} local quantum mechanics. 
Consider a family of universes
with (post-inflationary) matter energy $E$, in the domain
$a_{\rm min}\approx (E/m^2)^{1/3}\ll $ $a \ll$
$a_{\rm max}= 2E$. This automatically guarantees 
$|\phi|\ll 1$. The curvature part of the potential
can completely be ignored, as well as the fact that
the hypersurfaces of constant $a$ have a small
part in which they are not ''spacelike''.  
The approximate evolution (in the proper
time gauge $N=1$) is given by
\begin{eqnarray}
a(t) &=& \left( \frac{3}{2} \sqrt{2E}\, t \right)^{2/3}
\label{acl}\\
\phi(t) &=& \pm \,F(a(t)) \sin( m t + m \xi)\, , 
\label{phicl}
\end{eqnarray}
where the (slowly changing) amplitude of the oscillations is 
\begin{equation}
F(a)=\frac{\sqrt{2E}}{m\, a^{3/2}}\, , 
\label{ampl}
\end{equation}
thus $F(a(t))=2/(3 m t)$. The scale factor evolution
is of the matter dominated type. (If $a$ becomes of the
order of magnitude of $a_{\rm max}$, the expansion settles
down. The equation describing this behaviour is
$\dot{a} = $ $\sqrt{(2E/a)-1}$, but we will stick to the domain 
in which (\ref{acl}) is valid). The parameter $\xi$ labels
the trajectories, and $E$ is regarded as fixed (the
condition $E\gg m^{-1}$ ensures $a_{\rm min}\ll a_{\rm max}$).
Adjacent trajectories intersect each other at the caustic 
$\phi=F(a)$, which allows us to follow a single path only during 
half a period in which the sign of $\dot{\phi}$ is non-zero. 
The action associated with these trajectories is given by
\begin{equation}
S_\pm = -\,\frac{2}{3}\sqrt{2E}\, a^{3/2} \pm
\frac{E}{m} \left(
\arcsin\left(\frac{\phi}{F(a)}\right) +
\frac{\phi}{F(a)} 
\sqrt{ 1 -\,\left(\frac{\phi}{F(a)}\right)^2} 
\,\, \right)\, , 
\label{actosc}
\end{equation}
where the double sign denotes the sign of $\dot{\phi}$. 
The sign of the first term is chosen such that
$\dot{a}>0$. This first part of the action --- let us
denote it by $S_0(a)$ --- suggests itself as describing a 
(semiclassical) expansion around the matter dominated
background evolution $a(t)$. It also appears 
in the exponential of (\ref{Xihm}). 
The second part $S_{\rm q}(a,\phi)$ would then belong to 
the (quantum) fluctuations of $\phi$. 
Upon exponentiating
$i S_{\rm q}$ and taking the real part,
one just recovers the $\Psi_n$-part in (\ref{Xihm}) with
$n=E/m$ and small $\phi$. Hence, the $\Xi_n^-$ are associated 
with the two WKB-branches given by $S_\pm$. One may now
straightforwardly apply the formalism of Section 12 to
reveal the according local quantum structure. 
We use the coordinates $(t,\xi)$ from (\ref{acl})--(\ref{phicl}),
and restrict ourselves to a small domain of $\xi$ such
that $\cos(m t + m \xi)>0$. 
The DeWitt metric becomes (approximately) decomposed as
\begin{equation}
ds^2_{\rm DW} = -2 E\, dt^2 + 2 E
\cos^2(m t + m \xi) \, (d\xi + dt)^2
\label{dwosc}
\end{equation}
(cf. equation (\ref{admdecom})), 
and the effective Schr{\"o}dinger equation reads 
(for both $S_\pm$ branches) 
\begin{equation}
i\, \frac{\partial}{\partial t} \,\chi(t,\xi) = 
-\,\frac{1}{4E} \left( f(\xi) \cos(X) \right)^{-1/2}
\frac{\partial}{\partial\xi}\, 
\frac{\sin^2(X)}{\cos(X)}\, 
\frac{\partial}{\partial\xi}
\left(\frac{f(\xi)}{\cos(X)}\right)^{1/2}
\chi(t,\xi)\, , 
\label{schrhmosc}
\end{equation}
where $X\equiv m (t+\xi)$, and $f(\xi)$ is the arbitrary function 
related to the prefactor 
$D^2=$ $f(\xi)/(2E\cos(X))$ from (\ref{Dsolution}) 
specifying the representation. This is a particular
example for a local quantum structure (\ref{eff}). In order to 
display it 
in terms of more convenient coordinates, we 
rewrite this equation in terms of $(t,\phi)$. 
Setting $\chi(t,\xi)\equiv\widehat{\chi}(t,\phi)$ 
and $f\equiv 1$, the result is 
\begin{eqnarray}
i\, \frac{\partial}{\partial t} \, \widehat{\chi}(t,\phi) = 
\Bigg(
&-& \frac{m^2}{4E}\, \Big( F(t)^2-\phi^2 \Big)^{1/2}\, 
\frac{\partial}{\partial \phi}\, 
\phi^2 \, 
\frac{\partial}{\partial \phi}\, 
\Big( F(t)^2-\phi^2 \Big)^{-1/2} \nonumber\\
&+& i\,\Bigg( \mp m\, 
\Big( F(t)^2-\phi^2 \Big)^{1/2} 
-\frac{\phi}{t} \Bigg) 
\,\frac{\partial}{\partial \phi} \, 
\Bigg) \, \widehat{\chi}(t,\phi) 
\label{schrhmaphi}
\end{eqnarray}
for the branch $S_\pm$, where $F(t)$ is sloppy for 
$F(a(t))=$ $2/(3 m t)$. 
This is a special case of equation (\ref{eff2}). As compared
to (\ref{schrhmosc}) it has the advantage that its range of 
applicability is not constrained to small tubes around
a trajectory. It may be regarded as the Schr{\"o}dinger
equation for approximate eigenstates of $E$ in an
expanding universe around trajectories with prescribed
sign of $\dot{\phi}$. At $\phi=$ $F(t)$ (the location 
of the caustic) the WKB-approximation breaks down. Classically,
it is just here that the sign of $\dot{\phi}$ changes. 
For $\phi\ll F(t)$ and large $t$, the first term in the
effective Hamiltonian becomes suppressed, and we get 
$(3t \partial_t\pm$ $ 2\partial_\phi) \widehat{\chi}=$ $0$,
admitting a constant $\widehat{\chi}$. 
The physical wave function is a superposition
of an $S_{+}$ and an $S_{-}$ solution 
that behaves reasonable for $\phi>F(t)$. For a
{\it given} wave function (such as $\Xi_n^-$) this is clear,
since it is well-behaved as $|\phi|\rightarrow \infty$. 
This setup may be interpreted as a {\it true} quantum
mechanical structure. 
\medskip

If one wishes to further analyze each of the two equations 
(\ref{schrhmaphi}) separately, some simplification is possible.
The first part of the Hamiltonian is $H^{\rm eff}$ from
(\ref{schrhmosc}), rewritten in terms of the coordinates
$(t,\phi)$. The total Hamiltonian may --- by means of a simple
integration --- be rewritten as
$e^{-i G_\pm}(H^{\rm eff}+ V)e^{i G_\pm}$ with
$G_\pm\equiv$ $G_\pm(t,\phi)$ and $V\equiv$ $V(t,\phi)$ being
pure functions. $V$ has a small imaginary part stemming from 
the particular operator ordering that may presumably be 
neglected. Redefining the wave function as 
$\overline{\chi}=$ $e^{i G_\pm}\widehat{\chi}$, and
setting $W=$ $V - $ $\partial_t G_\pm$, the equations
(\ref{schrhmaphi}) attain the more familiar form 
$i \partial_t \overline{\chi}(t,\phi) =$ 
$(H^{\rm eff} + W)\overline{\chi}(t,\phi)$. 
\medskip

In order to pass over to the
semiclassial picture, one would either identify 
$S_0=$ $-\frac{2}{3}\sqrt{2E} a^{3/2}=$
$-2 E t$ as the classical
''background'' action or perform a picture changing
transformation on the Hamiltonian from (\ref{schrhmaphi}). 
The first method amounts to impose the semiclassical WKB-ansatz
(\ref{WKB0}), i.e. (ignoring the prefactor $D_0$)
$\psi=$ $\widetilde{\chi}\exp(i S_0)$ (although $S_0$ has 
not been constructed by an $\ell_P\rightarrow 0$ or some
related limit). This results to
leading order into (using (\ref{acl}) for the conversion
between $t$ and $a$ as variables, hence
$\widetilde{\chi}\equiv \widetilde{\chi}(t,\phi)$ or
$\widetilde{\chi}\equiv \widetilde{\chi}(a,\phi)$, as
one prefers)
\begin{equation}
i\,\partial_t \widetilde{\chi}
\equiv i\, \sqrt{\frac{2E}{a}}\,\partial_a \widetilde{\chi}
= (E^{\rm op} - E ) \widetilde{\chi}\, , 
\label{semiclschrhm}
\end{equation}
where $E^{\rm op}$ is the operator (\ref{Equ}) for the matter
energy, as opposed to $E$ as the classical parameter characterizing
the background action $S_0$ and hence the quasiclassical
evolution (\ref{acl}). In a sense, this equation is a special
case of a local quantum structure (\ref{eff0}) as emerging in the
conventional WKB-approach. 
Taking into account the harmonic
oscillator type of $E^{\rm op}$ and imposing an adiabatic 
approximation, we obtain solutions of the form
\begin{equation}
\widetilde{\chi}(t,\phi) \approx \exp\big(-i(E_n - E)t\big)
\Psi_n\left(\frac{3}{2}\, t \phi\,\sqrt{2 E m} \right)
\label{fjgscd}
\end{equation}
for any non-negative integer $n$ and 
$E_n=(n+\frac{1}{2})m\approx n m$. The WKB-appoximation is
good if the oscillations induced by the $t$-dependence
of $\widetilde{\chi}$ are milder than those of the
prefactor $\exp(i S_0)\equiv$
$\exp(-2 i E t)$ by which $\widetilde{\chi}$
is multiplied to give $\psi$, i.e. if $m n \approx E$. 
This in turn gives back wave functions of the type
$\Xi_n^-$ (cf. equation (\ref{Xihm})). Also, the physical
significance of the proper time $t$ (defined by (\ref{acl}))
as appearing 
in the semiclassical context is clarified: it is the time
variable associated with the difference 
$\widetilde{H}^{\rm eff}=E^{\rm op}-E$. 
\medskip

The second method invokes a picture changing
transformation on the Hamiltonian in order to
put back a factor $\exp(i(S-S_0))$ into $\chi$. 
Naively one would expect to arrive at 
$\widetilde{\chi}$ from (\ref{semiclschrhm}). 
However, this is only possible if the two branches $S_\pm$ are
macroscopically equivalent. A straighforward implementation
of a picture changing transformation gives identical results 
for the two branches 
only to the very crude approximation of small $\phi/F(t)$ and 
large $t$. In order to recover the semiclassical picture
at the same accuracy as above, one must perform in addition 
a superposition of (i.e. an average between) the 
{\it two} branches $S_\pm$, thereby unifying the two underlying 
modes (positive and negative $\dot{\phi}$) 
in terms of one single quasiclassical quantum structure. 
Technically, this should restore a peaceful behaviour
for large $|\phi|$, as in
(\ref{Xihm}) and (\ref{fjgscd}). 
However, the details are far from being clear, and
the issue deserves further study. 
(Also note that the ''wave lengths'' induced by the two
pieces of the action (\ref{actosc}) are of comparable order 
of magnitude
when expressed in terms of the DeWitt metric, although
$S_0$ varies in ''timelike'' and $S_{\rm q}$ varies mainly in
''spacelike'' direction. This might indicate that the 
above interpretation of the sum $S_0 + S_{\rm q}$
leads us close to the breakdown of the semiclassical 
WKB-approximation, and trying another decomposition of the action 
into two parts might be reasonable. 
Moreover, it is not totally clear to what extent the use of the
proper time gauge is appropriate). 
\medskip

Let us now consider another question and ask for the 
significance of the exponentially increasing part of 
$\psi_{\rm NB}$ for small $|\phi|$ and large $a$. This
part shows up as the contribution with respect to the
basis elements $\Omega^+_n$. The simplest philosophy is
that an exponential contribution in a wave function does
not refer to observations, and so may essentially be
ignored. 
This would imply (and has often been stated for the 
no-boundary wave function) that any classical trajectory 
provides a possible quasiclassical (macroscopic) ''history'' 
of the universe,
including the feature that it ''returns'' after attaining
maximal size, and recontracts. 
The according probabilities may be computed by WKB-methods
in the post-inflationary or even in the inflationary domain.
On the other hand, it has been argued by Zeh and Kiefer
\cite{Zeh}\cite{Kiefer1}\cite{KieferZeh}
that a viable wave function $\psi$ should asymptotically
vanish for $a\gg a_{\rm max}$, and they talk about
a ''final condition''.
Only in this case, $\psi$ has a chance to be decomposed 
into wave packet-like states that are peaked around classical 
trajectories describing returning universes 
(i.e. paths in which $a(t)$ attains its maximum value and
thereafter decreases again). These wave packets should 
provide a true intrinsic notation of ''time'' in a universe. 
However, the issue of expanding $\psi$ near the turning 
region into proper wave packets provides problems in some 
models (in particular in the Hawking model, as has been 
shown by Kiefer 
\cite{Kiefer1}). 
How does our interpretational scheme judge on this 
situation? 
\medskip

Restricting oneself to one particular energy value
$n\gg m^{-2}$, the asymptotics of the wave functions
$\Omega^\pm_n \sim$ $\exp(\pm a^2/2)$ for $a$ exceeding
the maximum radius associated with $E_n$ clearly
shows two different sorts of behaviour. Without
going into the details, it is likely that $\Omega^\pm_n$,
when WKB-analyzed with respect to the local quantum 
structure for the domain $a\,\gsim \,a_{\rm max}$ and 
$|\phi|\,\lsim\, 1/(m a)$, provides effective wave functions with 
the asymptotics $\chi^\pm \sim$ $\exp(\pm a^2/2)$. Note
that $a$ is now a coordinate labelling the classical
paths, whereas $\phi$ can be chosen as ''time''
coordinate $t$. Thus, in case the argument holds,
$\Omega^-_n$ gives rise to a {\it true}, and $\Omega^+_n$ 
belongs to a {\it false} local quantum mechanical 
situation. In other words, $\Omega^-_n$ 
admits expansion, maximum size and recontraction
to be experienced as the macroscopic 
history of the universe.
The fact that wave packets cannot be 
formed by superpositions of $\Omega_n^-$ for various $n$
does not necessarily spoil this argument, but just adresses the
details to a decoherence mechanism, as also remarked by
Kiefer in his study 
\cite{Kiefer1} 
(although this issue seems to be still somewhat open: 
loosely speaking, it is not yet clear whether decoherence 
effects near the turning point are ''strong enough'' to 
ensure almost-classicality there 
\cite{Kiefer4}; cf. also Ref. \cite{KieferZeh}; 
also cf. Ref. 
\cite{Keski-VakkuriMathur} 
for a recent discussion of the validity of the semiclassical 
approximation near turning points). 
On the other hand, $\Omega^+_n$ interrupts the
chain of reasonably {\it true} local quantum structures
just in the region of return. Hence, we conclude that
any wave function 
\begin{equation}
\psi(a,\phi) =\sum_{n=1}^\infty
(C^+_n\, \Omega^+_n(a,\phi) +
C^-_n \, \Omega^-_n(a,\phi) )
\label{genwf}
\end{equation}
contains several possibilities for the experience of a
macroscopic history of the universe.
If all $C_n^+ =0$, the universe will be experienced as
expanding and recontracting. If $C_n^-=0$, all (or most) 
possible macroscopic histories are likely to consist of ''broken''
almost-classical evolutions (the details depending on a 
proper decoherence analysis). 
The broken histories have the following properties:
Either, the universe is 
created (i.e. becomes real) in the inflationary domain,
near the curve $U=0$, expands and reaches a maximum size,
thereby undergoing a transition into a pure tunnelling 
state (i.e. ''tunnelling into nothing''). 
As already mentioned in Section 8 we cannot tell 
much about what actually will be experienced 
when conventional physics is just about to break down.
Or --- as a logical alternative --- 
the universe is created at its maximum size, contracts,
and disappears into ''nothing'' at the edge of the
inflationary domain. According to our interpretation,
no relation between these alternatives can be
experienced. Even the question whether the second
possibility occurs ''after'' the first one has no
meaning. The radicality in this statement relies on 
the type of scheme we apply, and we admit that a different 
interpretation of quantum cosmology may match just those
macroscopic histories that we have separated. 
\medskip

As an example, 
the no-boundary wave function $\psi_{\rm NB}$ has both 
non-zero $C^+_n$ and $C^-_n$, and thus contains all 
three variants (one type of ''unbroken'' and two 
types of ''broken'' histories)
with according probabilities (which we suspect to be 
roughly of equal order). In contrast, any bounded
wave function (such as the tunnelling wave function,
if it exists) 
has $C_n^+ =0$ and should thus contain only
''unbroken'' histories. 
Due to our interpretation, the {\it false} contributions
in a wave function imply a ''crack of doom'' ---
the breakdown of physical laws in the universe
in a way unexpected by its inhabitants (cf. Ref.
\cite{MatznerMezzacappa} 
for the use of this notion in Kaluza-Klein theory, where
a singularity in the compactified dimension leads
to a totally unexpected cosmic catastrophe). 
However, this does by no means provide a logical 
inconsistency. With regards to the question whether
''our universe'' will survive its era of maximum 
expansion, one just can hope that the wave function 
is only a superposition of the $\Omega_n^-$.
\medskip

\section{Modified WKB-expansion}
\setcounter{equation}{0}

This final Secton serves as an appendix. 
We outline the treatment of rapidly oscillating
solutions of the Wheeler-DeWitt equation (\ref{wdw1}), and the
local reconstruction of an approximate unitary evolution of the
standard quantum mechanical type.
Our procedure does {\it not} assume an {\it \`a priory}
separation of the coordinates $y^\alpha$ into classical (heavy)
and quantum (light) ones, nor does it assume a particular structure
of the Hamiltonian with respect to small parameters such as
$\ell_P$ or $\hbar$. Moreover, it is formulated in terms
of geometric quantities on $\cal M$ and on the equal-time
hypersurfaces of the classical trajectories, and may
in this respect be considered more appealing than the standard
(Born-Oppenheimer type) approach. 
\medskip

Let $\psi$ be written as in $(\ref{WKB2})$ and 
suppose that $S$ is a classical action function,
i.e. it satisfies the {\it exact} Hamilton-Jacobi
equation (\ref{HamJac}) in some domain $\cal G$ of $\cal M$ 
and therefore defines a congruence of classical paths.
Inside $\cal G$ the trajectories do not intersect each other, 
and no caustics are present. The sign of the potential $U$ may
be of either type, or even changing, as long as the 
trajecories still form a congruence, i.e. as long as
they altogether cross the zero-potential surface without
mutual intersection or turning points. 
However, for simplicity, we will only deal with the
cases where $U>0$ or $U<0$ in the whole of $\cal G$, and lay
emphasis on the former.
\medskip

By choosing the lapse function $N(y)$,
an evolution parameter $t$ is defined for each path via equation
(\ref{classequ}). The derivative of a scalar function $F(y)$ 
on $\cal G$ along the trajectories with respect to 
$t$ is then provided by 
$\dot{F}=$ $N (\nabla^\alpha S)\nabla_\alpha F$. Moreover, 
suppose that $D$ is real and satisfies the conservation law 
\begin{equation}
\nabla_\alpha( D^2 \nabla^\alpha S) = 0\, . 
\label{cons}
\end{equation}
This fixes $D(y)$ only up to $D(y)\rightarrow $ $k(y) D(y)$ 
where $k$ is constant along the classical 
paths (which corresponds to a pure ''picture changing''
transformation on the effective wave function $\chi$). The
local WKB-branch is defined by $S$, and the representation of
the hidden quantum mechanical structure will be fixed 
by a choice of $D$.
The prefactor $\chi$ represents the particular state. 
\medskip

Inserting (\ref{WKB2}) into the Wheeler-DeWitt equation
(\ref{wdw1}), we easily obtain an {\it exact} equivalent of the
latter 
\begin{equation}
i \, \dot{\chi} = {\cal K} \chi\, , 
\label{wdw2}
\end{equation}
where
\begin{equation}
{\cal K} = -\,\frac{1}{2}\, N\, D^{-1}\,
\nabla_\alpha \nabla^\alpha\, D \equiv
-\,\frac{1}{2}\, N\, D^{-1} \Big(
D \nabla_\alpha \nabla^\alpha +
2 (\nabla_\alpha D) \nabla^\alpha +
(\nabla_\alpha \nabla^\alpha D) \Big)\, . 
\label{wdwham}
\end{equation}
Note that we distinguish operators like
$\nabla^\alpha D:$ $F\rightarrow \nabla^\alpha(D F)$ from
$(\nabla^\alpha D):$ $F\rightarrow (\nabla^\alpha D) F$
by appropriate brackets.
The origin of the operator $\cal K$ may be illustrated by 
condensing the manipulations performed above in the form
\begin{equation} 
D^{-1} e^{-i S} (\nabla_\alpha\nabla^\alpha - U)
D e^{i S} = 
\frac{2}{N} \Big( 
i N (\nabla^\alpha S)\nabla_\alpha - {\cal K} \Big)
\label{condens}
\end{equation}
which is valid when acting on a scalar function on $\cal M$. 
It represents a canonical transformation. 
\medskip

In order to make contact with ordinary quantum mechanics we
define $\Gamma_t$ to be the hypersurfaces of constant $t$, 
thus defining a foliation of $\cal G$.
The scalar product $Q$ leads us the way to the hidden Hilbert 
space structure. 
Let $\psi_1$ and $\psi_2$ be two solutions of the Wheeler-DeWitt
equation of the type (\ref{WKB2}) for the same $S$ and $D$, but 
with two functions $\chi_1$ and $\chi_2$, respectively. Then
(\ref{Q}) is a sum 
\begin{equation}
Q(\psi_1,\psi_2) = \langle\chi_1|\chi_2\rangle_\Gamma +
(\chi_1|\chi_2)_\Gamma
\label{separ}
\end{equation}
where
\begin{eqnarray}
\langle\chi_1|\chi_2\rangle_\Gamma &=&
\int_{\Gamma} d s^\alpha \,(\partial_\alpha S) D^2
\chi_1^* \chi_2 \, ,
\label{QQQ1}\\
(\chi_1|\chi_2)_\Gamma &=&
-\,\frac{i}{2} \int_\Gamma 
d s^\alpha\, D^2 (
\chi_1^*
\stackrel{\leftrightarrow}{\partial_\alpha} \chi_2). 
\label{QQQ2}
\end{eqnarray}
For the moment, we leave the orientation of the
hypersurface element $d s^\alpha$ unspecified. 
Obviously, the sum (\ref{separ}) is independent of $\Gamma$,
whereas the individual terms are in general not. By
setting $\Gamma=\Gamma_t$ inside $\cal G$, 
and by assuming that the contribution to the integral from 
the part of $\Gamma$ that lies outside $\cal G$ is
negligible, the scalar product (\ref{QQQ1}) becomes a 
function of $t$. As we will see,
this already defines the Hilbert space structure associated with
the WKB-branch $(S,D)$: The identification is based on the fact 
that for WKB-type wave functions
($\,\exp(i S)$ oscillating much faster than $D$ and the $\chi$'s) 
the scalar product $\langle\chi_1|\chi_2\rangle_{\Gamma_t}$ 
is {\it almost independent} of $t$, and 
$(\chi_1|\chi_2)_{\Gamma_t}$ from (\ref{QQQ2}) is negligible. 
\medskip

In order to reveal the relation between the definition
(\ref{QQQ1}) and the conservation law (\ref{cons}), we note that
\begin{equation}
\frac{d}{dt}\, \langle\chi_1|\chi_2\rangle_{\Gamma_t} =
\langle\dot{\chi}_1|\chi_2\rangle_{\Gamma_t} +
\langle\chi_1|\dot{\chi}_2\rangle_{\Gamma_t}\, . 
\label{deriv}
\end{equation}
There are many ways to see this easily. Maybe the most
explicit one is to introduce coordinates $\xi^a$ 
($a=1,2,\dots n-1$) that label the trajectories
($\dot{\xi}^a=0$). Choosing $(y^\alpha)\equiv(t,\xi^a)$ as
coordinate system in $\cal G$, makes 
$\dot{F}\equiv$ $\partial_t F\equiv$ $\partial_0 F$ just 
the partial derivative and
$\xi^a$ a coordinate system on each of the hypersurfaces
$\Gamma_t$. 
(Analogous coordinates have, in the standard semiclassical 
formulation, been used by Cosandey for the sector of 
''classical variables'' 
\cite{Cosandey}). 
The identity (\ref{condens}) then becomes
\begin{equation} 
D^{-1} e^{-i S} (\nabla_\alpha\nabla^\alpha - U)
D e^{i S} = 
\frac{2}{N} \left( i\,\partial_t - {\cal K} \right)\, .
\label{condens2}
\end{equation}
{}From now on we have to distinguish two cases.
\medskip

If $U(y)>0$ inside $\cal G$, the trajectories are 
''timelike'' and the $\Gamma_t$ should reasonably be chosen as
''spacelike'' with respect to the DeWitt metric
$ds^2_{\rm DW} =$ $\gamma_{\alpha\beta}dy^\alpha dy^\beta$. 
(Note that, even if $N$ has been fixed, there is still the
freedom to reset the evolution parameter as
$t\rightarrow t-t_0$ in each trajectory individually). 
As a consequence, the metric induced on $\Gamma_t$ will
be positive definite. 
If $U(y)<0$ inside $\cal G$, the trajectories are
''spacelike'', and one will reasonably choose the
$\Gamma_t$ so as to contain one ''timelike'' and
$n-2$ ''spacelike'' directions. In this case the induced
metric will be of Lorentzian-type signature
$(-,+,+,\dots)$. Although these two cases may appear 
quite similar from the local point of view (refering to $\cal G$ 
only), they are related differently
to the global structure: In the $U>0$ case, any $\Gamma_t$ may
be considered as just the part of a globally ''spacelike''
hypersurface $\Gamma$ --- as used in (\ref{separ})
--- that happens to lie in $\cal G$. Also, one may construct
wave functions with negligible values on the part of $\Gamma$
outside $\cal G$ (just by prescribing appropriate {\it initial
conditions} for the Wheeler-DeWitt equation on $\Gamma$). 
This establishes the approximate relation between $Q$ and 
$\langle\,|\,\rangle_{\Gamma_t}\,$. Also, when analyzing a 
{\it given} WKB-state with respect to this structure, one does 
not expect to encounter catastrophic (e.g. exponential)
non-normalizibility effects.
In the case $U<0$ however, a similar construction is not 
possible because $\Gamma_t$ is no longer ''spacelike''. One may 
of course extend $\Gamma_t$ to a hypersurface $\Gamma$ that
qualifies for being used in the computation of (\ref{Q}). But
it will in general pass through other domains (outside $\cal G$), 
giving rise to non-zero contributions, and the relation between
$Q$ and $\langle\,|\,\rangle_{\Gamma_t}\,$ becomes less
transparent. The latter may still be defined by (\ref{QQQ1}),
and the reconstruction of local quantum structures is to
some extent possible as well, although a {\it given}
WKB-state, when analyzed with respect to this structure,
may display unacceptable ({\it false}) --- see Section 4 ---
quantum behaviour with respect to normalization issues, and 
the interpretation in terms of observations may be 
impossible. In Section 7, we have related this 
possibility to the issue of interpretation.
\medskip

For the (local) purpose of this Section, the two cases
are not very different. From now on, no reference to global 
issues like $\Gamma$ and $Q$ will be important. 
We first discuss the $U>0$ case, and afterwards just comment 
on the changes necessary to carry the results over to $U<0$. 
The case of $\cal G$ containing a hypersurface on 
which $U=0$ is a bit trickier (and so we will not
explicitly go into its details), but in principle
there is no objection against it either.
\medskip

As announced, we specify $\Gamma_t$ to be ''spacelike''. 
With respect to the coordinates
$(y^\alpha)\equiv (t,\xi^a)$, we can write down the 
DeWitt metric in an $(n-1)+1$-ADM-type decomposition
\begin{equation}
ds^2_{\rm DW} = - {\cal N}^2 dt^2 + h_{ab}
(d\xi^a +{\cal N}^a dt) (d\xi^b +{\cal N}^b dt)
\label{admdecom}
\end{equation}
where ${\cal N}$ and ${\cal N}_a \equiv
h_{ab}{\cal N}^b$ play the role 
of lapse and shift (all these quantities depending
on $(t,\xi)$, and the former not to be confused with the
lapse $N$ of the space-time metric). 
The (positive definite) induced metric 
(first fundamental form) on
$\Gamma_t$ is $h_{ab}(t,\xi)$, its determinant being denoted
by $h$, the inverse by $h^{ab}$. Hence 
$\gamma = - {\cal N}^2 h$. 
The volume element on $\Gamma_t$ is $d^{n-1}\xi\,\sqrt{h}$.
The hypersurface element $d s^\alpha$ is given by
$d^{n-1}\xi\,\sqrt{h} n^\alpha$, where $n^\alpha$ is the
unit ($n^\alpha n_\alpha=-1$) vector orthogonal to
$\Gamma_t\,$.
We choose its orientation to be $n_0>0$, hence $n^0<0$.
This makes (\ref{QQQ1}) a {\it positive definite} scalar 
product. Note that with this choice the hypersurface 
element is correlated with the orientation of the classical
trajectories, as defined by increasing $t$. Under
complex conjugation ($S\rightarrow -S$, hence 
$t\rightarrow -t$), it becomes reversed (although $n_0>0$
remains intact when expressed with respect to the
new coordinates $(-t,\xi)$). If one likes to work
with a {\it prescribed} orientation of $\Gamma$ in (\ref{Q}),
this has to be taken into account by changing the orientation
for all outgoing WKB-branches, say. The result is just the
$\pm$ sign in (\ref{QQ}), which does not show up
here. 
Moreover, we denote the covariant derivative with respect
to $h_{ab}$ by ${\cal D}_a$, and set of course 
$h^{ab}{\cal D}_b = {\cal D}^a$. 
The trace of the extrinsic curvature (second fundamental
form) on $\Gamma_t$ is given by 
\begin{equation}
K = \nabla_\alpha n^\alpha = 
\frac{1}{{\cal N}}\,{\cal D}_a {\cal N}^a 
- \frac{\dot{h}}{2 {\cal N} h} \, , 
\label{trace}
\end{equation}  
and 
\begin{equation}
\partial_\perp = - n^\alpha\nabla_\alpha =
\frac{1}{{\cal N}}\,\partial_t - \frac{{\cal N}^a}{{\cal N}}\,
\partial_a \equiv - {\cal N} \nabla^0
\label{derivorth}
\end{equation}  
is the derivative orthogonal to $\Gamma_t$.
\medskip

The definition of the coordinate system immediately implies
\begin{eqnarray}
{\cal N}_a &=& N \partial_a S
\label{Na}\\
{\cal N}^2 &=& N^2 U + {\cal N}_a {\cal N}^a =
N^2( U + ({\cal D}_a S) {\cal D}^a S) = 
N^2 (\partial_\perp S)^2\, , 
\label{NN}
\end{eqnarray}
the second chain of equalities just expressing the 
Hamilton-Jacobi equation (\ref{HamJac}) in several forms. Also 
we find $\dot{S}= - N U$ and $\partial_\perp S = - {\cal N}/N$.
The components of the DeWitt metric are 
$\gamma_{00}=$ $ - N^2 U$, $\gamma_{0a}=$ ${\cal N}_a$ 
and $\gamma_{ab}=$ $h_{ab}$. The inverse metric is
$\gamma^{00}=$ $-{\cal N}^{-2}$,
$\gamma^{0a}=$ ${\cal N}^{-2} {\cal N}^a$, and 
$\gamma^{ab}$ is given by equation (\ref{gammaab}) below. 
Note that if ${\cal N}_a\neq 0$ the vector field $\partial_t$
is not orthogonal to $\Gamma_t$. Hence, the surfaces
of constant $S$ need not coincide with the surfaces of constant
$t$. 
\medskip

The equation (\ref{cons}) for $D$ may now easily be written
down in these coordinates, the general solution being
\begin{equation}
D(y)^2 = \frac{N(y) f(\xi)}{{\cal N}(y) \sqrt{h(y)}}
\label{Dsolution}
\end{equation}
with $f$ an arbitrary (positive) function of constant
value along each of the classical paths. 
We are now in position to display some relations that
clarify the structure emerging here. Remember that 
the total volume element on $\cal M$ with respect to the
DeWitt metric is given by 
$d\mu =$ $ d^n y\,\sqrt{-\gamma}$.
Using the previons results, we find
\begin{equation}
ds^\alpha\, \partial_\alpha S =
d^{n-1}\xi\,\sqrt{h}\, n^\alpha \partial_\alpha S =
d^{n-1}\xi\,\sqrt{h}\,\frac{{\cal N}}{N}
\label{vol1}
\end{equation}  
and
\begin{equation}
d\mu = dt\, {\cal N}\,d^{n-1}\xi\,\sqrt{h} = 
dt \,N\, ds^\alpha\,(\partial_\alpha S)\, . 
\label{vol2}
\end{equation}  
Hence, any surface integral with measure as in (\ref{QQQ1})
is given by ($F\equiv F(y)\equiv F(t,\xi)$)
\begin{equation}
\int_{\Gamma_t} ds^\alpha\, (\partial_\alpha S)
D^2\, F = 
\int d^{n-1} \xi\, f(\xi) F(t,\xi) 
\label{surfaceint}
\end{equation}  
where the second integral is over the whole parameter space
labelled by the $\xi^a$ (in fact, we require all integrands to
be well-concentrated on $\Gamma_t$, so that neither the local
nature of the construction nor the possible existence of
boundaries of $\Gamma_t$ give rise to any problem). 
Differentiating with respect to $t$, it becomes clear that 
$\partial_t$ acts on $F$ alone 
\begin{equation}
\frac{d}{dt} \int_{\Gamma_t} ds^\alpha\,(\partial_\alpha S)
D^2 \, F = \int d^{n-1}\xi\, f(\xi)\dot{F}(t,\xi) = 
\int_{\Gamma_t} ds^\alpha\,(\partial_\alpha S)
D^2 \, \dot{F} \, ,
\label{surfaceint2}
\end{equation}  
{}from which (\ref{deriv}) follows immediately. We thus 
obtain an alternative form of the scalar product 
(\ref{QQQ1})
\begin{equation}
\langle\chi_1|\chi_2\rangle_{\Gamma_t} =
\int d^{n-1}\xi\, f(\xi) \chi_1^*(t,\xi)\chi_2(t,\xi)
\equiv \int d^{n-1}\xi\, \sqrt{h}\,
\frac{{\cal N} D^2}{N}\chi_1^*\chi_2 \, , 
\label{surff}
\end{equation}
where the second expression reflects the covariant character
of the integral. 
One may achieve a result analogous to (\ref{surfaceint2})
by computing the difference between two integrals of the
type (\ref{surfaceint}) for two hypersurfaces $\Gamma_{t_1}$
and $\Gamma_{t_2}$. Denoting by $G$ the region between these,
and using Gauss' theorem, one encounters the volume
integral
\begin{equation}
\int_G d\mu \nabla^\alpha\Big(
D^2(\partial_\alpha S) \, F \Big) .
\label{surfaceint3}
\end{equation}
Using (\ref{cons}), one again sees that 
$N^{-1}\partial_t\equiv$ $(\nabla^\alpha S)\partial_\alpha$ just
acts on $F$. Using (\ref{vol1}) and (\ref{vol2}), one
recovers (\ref{surfaceint2}) in various forms. 
It also becomes clear how the function $f(\xi)$ specifies 
the representation of operators and the local scalar 
product. From (\ref{surff}) we see that the choice $f(\xi)\equiv 1$ 
makes $\langle\,|\,\rangle$ the standard scalar product on
(a domain of) $\R^{n-1}$. Also, once a choice of $f$ has been
fixed, one may redefine the $\xi^a$ among themselves so as 
to achive $f\equiv 1$ in the new variables. 
\medskip

The Wheeler-DeWitt equation (\ref{wdw2}) looks like a
Schr{\"o}dinger equation, but the operator ${\cal K}$
contains derivatives off $\Gamma_t$. The general
structure we encounter here is that 
any hermitean operator acting on the space of
square integrable functions $\chi(\xi)$ 
(with respect to the scalar product(\ref{surff}))
defines a unitary evolution. 
Let us emphasize that such a unitary evolution is accociated 
with only {\it one} Schr{\"o}dinger equation for $\chi(t,\xi)$. 
This provides a difference 
to the standard semiclassical formalism: There, one is faced 
with {\it several} Schr{\"o}dinger equations, belonging to several
classical backgrounds and thus several definitions of the
classical time flow. Thus, the effective Schr{\"o}dinger 
equation (\ref{eff0}) of the standard approach is essentially 
unique only if the classical backgrounds contributing are 
strongly peaked. This may be implemented by a peak in the 
prefactor $D_0$ from (\ref{WKB0}); cf. Ref. \cite{Cosandey}. 
No such peak in $D$ is needed here, when  
a local quantum structure is formally extracted from the 
Wheeler-DeWitt equation. 
This confirms the nature of our formalism as an intermediate
structure between the full Wheeler-DeWitt equation and 
observational quantum mechanics (which operates in a 
quasiclassical background), as has been mentioned in 
Section 3. 
\medskip

We thus raise the question 
whether --- to some approximation --- $\cal K$ may be replaced 
by an operator $H^{\rm eff}$ such that 
{\it (i)}  $H^{\rm eff}$ acts purely tangential to
$\Gamma_t$ (i.e. for any given $t$ it acts in the space
of functions on $\Gamma_t$ --- in practice, it is a 
combination of multiplication operators and ''spatial'' 
derivatives $\partial_a\,$.), and
{\it (ii)} for any $t$ it is hermitean with respect to 
the scalar product (\ref{QQQ1}) or, equivalently,
(\ref{surff}).
The second condition just means that $f H^{\rm eff}$ is
hermitean with respect to the standard scalar product in
$\R^{n-1}$.  
It may be replaced by a ''covariant'' version, namely by 
requiring that for any $t$ the combination
$N^{-1} {\cal N} D^2 H^{\rm eff}\equiv$ 
$f h^{-1/2} H^{\rm eff}$ is hermitean with respect
to the scalar product
\begin{equation}
\int_{\Gamma_t} d^{n-1}\xi\,\sqrt{h}\,\chi_1^*\chi_2 
\label{covariantscalarproduct}
\end{equation}  
defined by the proper volume on $\Gamma_t$. 
If this can be achieved, the Wheeler-DeWitt
equation (\ref{wdw2}) approximately becomes (\ref{eff}),
and the unitarity of the evolution defined by
$H^{\rm eff }$ is evident.
\medskip

In order to find $H^{\rm eff}$, we begin writing down
the $(n-1)+1$-decomposition of the full Laplacian (when
acting on scalars functions on $\cal M$) 
\begin{equation}
\nabla_\alpha \nabla^\alpha = 
-\partial_\perp \partial_\perp + K \partial_\perp
+ \frac{1}{{\cal N}}\, {\cal D}_a {\cal N} {\cal D}^a \, .
\label{laplace}
\end{equation}  
Here, ${\cal D}_a$ acts on everything to the right of it.
This may be inserted into $\cal K$. By explicitly shuffling 
the derivatives $\partial_t$ to the right, one 
achieves the decomposition
\begin{equation}
{\cal K} ={\cal U}\, \partial_{t t}
+{\cal V}\, \partial_t+ {\cal W}
\label{HH}
\end{equation}  
with $\cal U$, $\cal V$ and $\cal W$ operators that act purely
tangential to $\Gamma_t\,$. Before evaluating them, it is
appropriate to discuss the kind of expansion we will perform.
\medskip

Rapid oscillations in a WKB-type wave function $\psi$ are 
induced by a rapidly varying action $S$. In the conventional
WKB-expansion approach this is accomodated for by assuming that
the action $S_0$ of the classical variables is
large. One usually sets 
$S_0(y_{\rm cl}) = \sigma_0(y_{\rm cl})/\epsilon$
with $\epsilon$ a small expansion parameter (e.g. $\ell_P$ or
$\hbar$). We will not introduce such a quasiclassical
''background'' action,
but work with a solution $S(y)$ of the {\it exact} Hamilton-Jacobi
equation (\ref{HamJac}). The latter is expressed in the
$(n-1)+1$-split by equation (\ref{NN}). We may 
assume that $S$ is large (even if it contains contributions
in addition to the ''true'' large part $S_0$). Setting
$S(y)=\sigma(y)/\epsilon$ amounts to formally assume 
$U(y) = u(y)/\epsilon^2$
in a small $\epsilon$ expansion in order to retain the form
of (\ref{HamJac}). By viture of (\ref{Na}) and (\ref{NN}) this
makes $N^{-1} {\cal N}\sim$
$\epsilon^{-1}\sim$ $N^{-1} {\cal N}_a$ large quantities. 
Dividing the Wheeler-DeWitt equation (\ref{wdw2}) by $N$, 
the time derivative at the
left hand side appears in the combination $i N^{-1}\partial_t$. 
At the right hand side, we find all time derivatives
to occur in the combination 
${\cal N}^{-1}\partial_t$ which is 
$({\cal N}^{-1} N)N^{-1}\partial_t$ and thus scales as
$\epsilon N^{-1}\partial_t$ in a small $\epsilon$
expansion. This is the mechanism that suppresses the time
derivatives at the right hand side (as opposed to the
one at the left hand side). We will use $\epsilon$ as a formal 
book-keeping parameter indicating small contributions. 
$D$ appears in the Wheeler-DeWitt equation homogeneous of degree 
zero, and so will not contribute any $\epsilon$. 
The ratio ${\cal N}_a/{\cal N}$ will be kept in all expressions,
although it may be small. 
The net effect is thus to give any time-derivative at the
right hand side a factor $\epsilon$. 
This is achieved by expressing the Wheeler-DeWitt
equation in terms of $h_{ab}$, $N$, $D$, ${\cal N}$,
${\cal N}_a$ and the derivatives 
$\partial_t$ and $\partial_a$, and by {\it formally} substituting
\begin{equation}
{\cal N}\rightarrow \frac{1}{\epsilon}\, {\cal N} \qquad
{\cal N}_a \rightarrow \frac{1}{\epsilon}\, {\cal N}_a
\qquad 
{\cal N}^a \rightarrow \frac{1}{\epsilon}\, {\cal N}^a\, , 
\label{replace2}
\end{equation}
leaving all other quantities unchanged. 
This does not necessarily mean that ${\cal N}$ and
${\cal N}_a$ are large, but simply accounts for the
insertion of appropriate $\epsilon$'s in the
operator $\cal K$. Thus, we have to expand $\cal K$ with 
respect to $\epsilon$. The only relevant pieces 
getting $\epsilon$'s 
are the contributions $\partial_\perp$ and $K$ in 
(\ref{laplace}). Both terms decompose into an
$O(\epsilon)$ and an $O(\epsilon^0)$ contribution.
Note that the $\epsilon^{-1}$ from ${\cal N}$ cancel in the
last expression of (\ref{laplace}). 
In terms of (\ref{HH}) one finds 
\begin{eqnarray}
{\cal U} &=& \frac{\epsilon^2 N}{2\, {\cal N}^2} 
\label{U1}\\
{\cal V} &=& \frac{\epsilon^2 N}{2\,{\cal N}^2}
\left(
\frac{2\, \dot{D}}{D} +
\frac{\dot{h}}{2h}
- \frac{\dot{{\cal N}}}{{\cal N}}
\right)
-\,\frac{\epsilon N}{2\,{\cal N}}\,
D^{-1} {\cal A} D
\label{V1}\\
{\cal W} &=& \frac{\epsilon^2 N}{2\,{\cal N}^2}
\left(\frac{\ddot{D}}{D} +
\Big(\frac{\dot{h}}{2h} -
\frac{\dot{{\cal N}}}{{\cal N}} \Big)
\frac{\dot{D}}{D}   \right)
\nonumber\\ 
& & - \,\frac{\epsilon N}{2\,{\cal N}} \left(
D^{-1} {\cal A} \dot{D} + D^{-1} A^a\,{\cal D}_a\, D \right)
+ H^{\rm eff}
\label{W1}
\end{eqnarray}
where
\begin{equation}
{\cal A} =
{\cal D}_a \,\frac{{\cal N}^a}{{\cal N}} +
\frac{{\cal N}^a}{{\cal N}}\, {\cal D}_a
\label{anticomm}
\end{equation}
is an anticommutator (both ${\cal D}_a$'s acting on every 
function to the right of them), and
\begin{equation}
A^a = \frac{1}{\sqrt{h}}\,\partial_t
\left(\sqrt{h}\,\frac{{\cal N}^a}{{\cal N}}\right) 
\label{Aa}
\end{equation}
is a vector field, with $\partial_t$ acting only on the
expression in the bracket. 
Note that the rule of assigning a factor $\epsilon$ 
to any time-derivative in $\cal K$ includes the
dot-terms in the above expressions. 
When time-derivatives of $D$ are involved, one may use
the identity
\begin{equation}
\frac{\dot{D}}{D} =
\frac{\dot{N}}{2N}-\frac{\dot{h}}{4h}- 
\frac{\dot{{\cal N}}}{2\,{\cal N}}
\label{Ddot}
\end{equation}
which follows from (\ref{Dsolution}). 
The last operator in (\ref{W1}) is given by
\begin{equation}
H^{\rm eff} = -\,\frac{N}{2\,{\cal N}D}\,
{\cal D}_a \,{\cal N} \gamma^{ab}{\cal D}_b\, D
\equiv 
-\,\frac{N}{2\,{\cal N}D\sqrt{h}}\,
\partial_a \,\sqrt{h} {\cal N} \gamma^{ab}
\partial_b\, D\, ,
\label{Heff}
\end{equation}
where the ''spatial'' contravariant components of the
DeWitt metric, when expressed as a tensor object on
$\Gamma_t\,$, read
\begin{equation}
\gamma^{ab} = h^{ab} - 
\frac{{\cal N}^a {\cal N}^b}{{\cal N}^2}\, . 
\label{gammaab}
\end{equation}
We have thus expanded $\cal K$ with respect
to $\epsilon$ and explicitly displayed the coefficient
operators. Setting 
${\cal U}=\epsilon^2\, {\cal U}_2$,
${\cal V}=$ $\epsilon^2\,{\cal V}_2\,+$ $\epsilon\,{\cal V}_1$ 
and ${\cal W}=$ $\epsilon^2\,{\cal W}_2\,+$ 
$\epsilon\, {\cal W}_1 + H^{\rm eff}$, the
WheelerDeWitt equation --- with the replacement (\ref{replace2})
performed --- reads
\begin{equation}
i\,\dot{\chi}=
\epsilon^2 \,{\cal U}_2\, \ddot{\chi} +
(\epsilon^2\,{\cal V}_2+\epsilon\,{\cal V}_1)\dot{\chi}
+(\epsilon^2\,{\cal W}_2 +\epsilon\,{\cal W}_1 +
H^{\rm eff} )\chi \, .
\label{wdw3}
\end{equation}
Setting $\epsilon=1$ gives back its {\it exact} form
(\ref{wdw2}). 
Also, after performing a
small $\epsilon$ expansion to arbitrary order, in the end
one should set $\epsilon=1$ as well, because we
have introduced $\epsilon$ in a formal way as a
book-keeping parameter. 
\medskip

There is still the question open how to choose the coordinate
$t$, i.e. the lapse $N$ and the hypersurfaces $\Gamma_t$. 
In general, one should choose $t$ as a coordinate 
that conveniently serves as ''time'' in the local description of 
the evolution of the universe and the laws of nature. In some
situations a ''good'' choice of time is quite evident, in others
one may have a lot of freedom. In fact, there is a wide range
of admissible choices, and only too drastic transformations 
thereof would spoil the structure of the small $\epsilon$ expansion. 
The $\Gamma_t$ should approximately coincide with the surfaces 
of constant $S$, so as to shift the dominant oscillations from 
$\chi$ to the WKB-phase $\exp(i S)$. 
The according local quantum structures are essentially 
equivalent (just as the scalar product 
$\langle\,|\,\rangle$ is largely independent of the
hypersurface at which it is computed). The time-derivative 
$\partial_t$ is along the trajectories, and it transforms 
under a change of time gauge and a change of the equal-time 
hypersurfaces simply as 
$\partial_t =$ $(N/\overline{N})\partial_{\,\overline{t}}$.
The operators acting purely tangential to $\Gamma_t$
may be regarded as time-dependent operators acting on a space
of functions which are defined on the set of trajectories.
Sufficiently small transformations just interpolate between 
different representations (or ''pictures'') of this situation. 
If the equal-action hypersurfaces differ too much from the
equal-time hypersurfaces, the pattern defined by the
small $\epsilon$ expansion may break down. 
Once the foliation provided by the $\Gamma_t$ has been fixed, 
a further change of time coordinate leaving this foliation
invariant (i.e. $t\leftrightarrow \overline{t}$, hence
$N(y)/{\overline{N}(y)}$ being a function of $t$ only) 
yields an identical structure with respect to $\epsilon$ 
(apart from a mixture between $\cal U$ and $\cal V$ 
which corresponds to the transformation of
$\partial_{tt}$ into a 
$\partial_{\,\overline{t}\overline{t}}$ and a 
$\partial_{\,\overline{t}}$ term). 
In this case one finds $N^{-1} H^{\rm eff}=$ 
${\overline{N}}^{-1} {\overline{H}}^{\rm eff}$. 
In order to understand the emergence of quantum mechanics
better, it is certainly necessary to study the issue of the
appropriate time gauge and the accuracy of the
approximations involved in more detail. 
\medskip

In the time gauge provided by the condition
${\cal N}_a =0$ our formalism becomes
somewhat simplified. On account of (\ref{Na}), this
is equivalent to $S\equiv S(t)$, i.e. the evolution 
coordinate $t$ 
along each trajectory is only a function of $S$. As a
consequence we find ${\cal A}=A^a=0$, and (\ref{NN}) reduces
to ${\cal N}^2 = N^2 U$. The relation between $t$ and $S$
is provided by $\dot{S}(t)=-N U$, which implies that the
combinations $N U$ and $N^{-1} {\cal N}^2$ depend only on
$t$ as well. The operators (\ref{U1})--(\ref{W1}) simplify,
and we obtain ${\cal V}_1 =  {\cal W}_1 = 0$, hence all
contributions to $O(\epsilon)$ vanishing.
\medskip

One may specify such a time gauge by setting $S(t)=-t$,
i.e. by using the (negative of the) classical action 
as time parameter along the paths. In this case, the
$(n-1)+1$-split as well as the space-time lapse $N$ become
uniquely determined, and one finds (in addition to
${\cal N}_a =$ ${\cal A}=$ $A^a=0$)
the further simplifications $N=$ ${\cal N}^2=$ $U^{-1}$ and 
$D^2=$ $U^{-1/2} h^{-1/2} f(\xi)$. The operators
(\ref{U1})--(\ref{W1}) become
\begin{equation}
{\cal U}=\frac{\epsilon^2}{2}\, ,\qquad
{\cal V}=0,\qquad
{\cal W}=H^{\rm eff} +\frac{\epsilon^2}{2}
\left(\frac{\ddot{D}}{D}-2\,\frac{{\dot{D}}^2}{D^2}\right),
\label{UVW}
\end{equation}
The $O(\epsilon^2)$-term in ${\cal W}$ may also be written
as $ -\,\epsilon^2 D (\partial_{tt} D^{-1})/2$. This 
is just a real function of the coordinates $(t,\xi)$, and thus 
represents a hermitean operator.
\medskip

Ordinary quantum mechanics is now recovered easily. To
lowest order ($\epsilon\rightarrow 0$), the operator $\cal K$
is just $H^{\rm eff}$ as given in (\ref{Heff}), and 
the Wheeler-DeWitt equation (\ref{wdw3})
reduces to the effective Schr{\"o}dinger equation (\ref{eff}). 
$H^{\rm eff}$ acts purely tangential to
$\Gamma_t$ which is just condition {\it (i)}, as defined above.
Moreover, the combination $N^{-1} {\cal N} D^2 H^{\rm eff}$ 
is clearly hermitean with respect to the scalar product 
(\ref{covariantscalarproduct}) and thus obeys
condition {\it (ii)}. As a consequence, (\ref{eff}) defines
a unitary evolution of the conventional quantum mechanical
type. As already mentioned, no severe normalization problems
are expected with the effective Schr{\"o}dinger wave
functions $\chi$ in the case $U>0$ that we have dealt with so far.
{\it Constructing} a sufficiently large collection of wave 
packets that nicely fit into the approximate Hilbert space 
structure, a {\it given} wave function can be expanded (and
hence interpreted) in terms of these. This is what we have
denoted {\it true local quantum mechanics}.
\medskip

Let us now comment on the case $U<0$. If the $\Gamma_t$ 
are chosen so as to make the induced metric Lorentzian, 
the $(n-1)+1$-split (\ref{admdecom}) has to be replaced by
\begin{equation}
ds^2_{\rm DW} =  {\cal N}^2 dt^2 + h_{ab}
(d\xi^a +{\cal N}^a dt) (d\xi^b +{\cal N}^b dt)\, . 
\label{admdecomU}
\end{equation}
The following analysis applies almost identically, and
most of the formulae are correctly transformed by 
replacing ${\cal N}^2\rightarrow$ $-{\cal N}^2$ and 
${\cal N}\sqrt{h}\rightarrow$ ${\cal N}\sqrt{-h}$.
In particular, (\ref{Na}) remains valid, (\ref{NN}) 
becomes
\begin{equation}
{\cal N}^2 = - N^2 U - {\cal N}_a {\cal N}^a =
- N^2( U + ({\cal D}_a S) {\cal D}^a S) 
\label{NNU}
\end{equation}
and the contravariant components of the DeWitt metric
tangential to $\Gamma_t$ (equation (\ref{gammaab}))
carry over to
\begin{equation}
\gamma^{ab} = h^{ab} + 
\frac{{\cal N}^a {\cal N}^b}{{\cal N}^2}\, . 
\label{gammaabU}
\end{equation}
With this definition, the expression for
the effective Hamiltonian is still given by (\ref{Heff}),
and the various forms of the Wheeler-DeWitt equation as
well as the general discussion still apply. 
The unit normal is of course changed into
$n^\alpha n_\alpha=1$, and the transformation of all
formulae containing it (and $\partial_\perp$)
explicitly is a simple exercise. Its two possible 
orientations imply a double sign like in 
(\ref{QQ}) if a {\it fixed} hypersurface $\Gamma$ is used, and
the $Q$-products are modified so as to integrate over
$\Gamma$ only locally. 
As in the $U>0$ case, one may {\it construct} a
collection of local wave packets that fit into the
approximate Hilbert space structure. However, a 
{\it given} state, when expanded into such reference
wave packets, may spoil any reasonable local 
normalizability. Such a feature could possibly prevent any
probabilistic interpretation. Hence, as outlined
in Section 4, the $U<0$ case may give rise to {\it true} 
as well as to {\it false local quantum mechanics}, 
depending on the wave function under consideration. 
For the rest of this Section we will again restrict
ourselves to the $U>0$ case. 
\medskip 

When the issue of operator ordering is ignored, the structure 
of the effective Hamiltonian (\ref{Heff}) may
be understood heuristically in terms of the classical system. 
Let us for simplicity restrict ourselves to the case $S=-t$.
Choose a solution $S$ of the Hamilton-Jacobi equation 
(\ref{HamJac}), parametrize the according paths by $t=-S$
and select variables $\xi^a$ to label them. Then
$(y^\alpha)\equiv$ $(t,\xi^a)$ is a coordinate system
on $\cal M$. If the classical system is rewritten in terms
of these coordinates, an evolution (the parameter
being denoted by $\tau$) is given by $(t(\tau),\xi^a(\tau))$. 
The special cases $t(\tau)=\tau$, $\xi^a(\tau)=$ $const$ 
give back the original family of paths, but these are of course
not the only possible ones. Rescaling the lapse as 
$N=\nu\,U^{-1}$, the Lagrangian of the classical system reads 
(the dot denoting $d/d\tau$) 
\begin{equation}
{\cal L} = \frac{1}{2}\left(
-\nu^{-1}\, \dot{t}^2 - \nu + \nu^{-1} U h_{ab} \dot{\xi}^a \dot{\xi}^b
\right)\, . 
\label{classlagr}
\end{equation}
Variation with respect to $\nu$ yields the constraint equation, 
which, in the Hamiltonian form, reads
\begin{equation}
-p_t^2 + 1 + U^{-1} h^{ab} p_a p_b = 0\, . 
\label{classconstraint}
\end{equation}
In the gauge $\nu=1$ (which we fix once the
constraint is established), we find $p_t=-\dot{t}$. 
Upon identifying $p_a$ with $-i\partial_a$ and ignoring 
operator ordering, the last term in 
(\ref{classconstraint}) just has the structure of
$2 H^{\rm eff}$ (recall $U^{-1}=N$ for $\nu=1$)
and thus shall be denoted by the same
name. 
Redefining the variables as $t(\tau) =$ $\tau + $ $T(\tau)$ 
and the momenta as $p_t=$ $-1+p_T$, hence $p_T=$ $-\dot{T}$, 
we obtain the constraint in the form 
\begin{equation}
-p_T = - \,\frac{1}{2}\, p_T^2 + H^{\rm eff}\, . 
\label{classconstraintT}
\end{equation}
Quantizing by means of the substitution 
$p_T\rightarrow$ $-i\partial_T$, 
$p_a\rightarrow$ $-i\partial_a$ gives 
\begin{equation}
i\,\dot{\chi} = \frac{1}{2}\,\ddot{\chi} + H^{\rm eff}\chi
\label{wdwheur}
\end{equation}
which has the overall structure of (\ref{wdw3}) with $\epsilon=1$
in the gauge $S=-t$ (the additional operator ${\cal W}_2$ appearing 
there may be considered as arising by operator ordering and related
issues). 
The above redefinition of the momenta corresponds
to the transition from $\psi$ to $\chi$ (or likewise to 
the ansatz $\psi=\chi\exp(-it)$). 
The WKB-approximation declares $\ddot{\chi}$ to be negligible,
and we have reproduced the essentials of (\ref{eff}). 
The classical counterpart of the effective Hamiltonian thus governs 
the evolution of those degrees of freedom which describe the
deviation of generic trajectories from the paths generated by $S$.
The classical counterpart of the WKB-approximation is
the condition $p_T^2 \ll |p_T|$, thus 
$|p_T|\ll 1$ (or equivalently $|\dot{T}|\ll 1$). As a 
consequence, we must have $|H^{\rm eff}|\ll 1$ in order to
be close to the original $S$-trajectories. A similar analysis
may be performed using a general gauge for the $S$-trajectories,
the classical effective Hamiltonian becoming 
$H^{\rm eff}=$ $\frac{1}{2}N \gamma^{ab}p_a p_b$
(cf. equation (\ref{Heff})). It 
arises as part of $\gamma^{\alpha\beta}p_\alpha p_\beta$, 
which explains the appearance of $\gamma^{ab}$ instead of
$h^{ab}$. 
\medskip

If one tries to go beyond the Schr{\"o}dinger 
approximation derived so far, one may
iteratively get rid of the time-derivatives of $\chi$ in
(\ref{wdw3}) to any order in $\epsilon$. 
In order to find the $O(\epsilon)$ approximation, one
may just insert (\ref{wdw3}) itself into the term 
$\epsilon\, {\cal V}_1 \dot{\chi}$, and the result is
\begin{equation}
i\,\dot{\chi} = 
\Big(H^{\rm eff} + \epsilon\, ({\cal W}_1 - i\, {\cal V}_1
H^{\rm eff}) \Big) \chi
\label{wdweps1}
\end{equation}
with $\epsilon=1$ inserted. This provides the first
post-quantum mechanics approximation. As is clear
if ${\cal N}_a\neq 0$, the
Hamiltonian appearing here does not meet the
hermiticity condition {\it (ii)}. This follows from the
fact that part of ${\cal W}_1$ is a first order
''spatial'' derivative operator with real coefficients,
and thus provides an anti-hermitean contribution that
cannot be cancelled by any other term. 
\medskip

One may proceed even further. Iteratively differentiating 
(\ref{wdw3}) with respect to $t$ and performing appropriate
substitutions yields a Schr{\"o}dinger type
equation without time-derivative on the right hand side
(but with non-hermitean Hamiltonian)
to arbitrary order in $\epsilon$. If one uses a time
gauge in which ${\cal N}_a=0$ (see above), the
$O(\epsilon)$ corrections in (\ref{wdweps1}) are in fact
absent, because ${\cal V}_1 =$ ${\cal W}_1=0$. In this case
we obtain at $O(\epsilon^2)$ 
\begin{equation}
i\,\dot{\chi} = 
\Big(H^{\rm eff} + \epsilon^2 ( B + i\, C) \Big) \chi
\label{wdweps2}
\end{equation}
with
\begin{eqnarray}
B &=& -\, {\cal U}_2 \, (H^{\rm eff})^2 + {\cal W}_2
\label{B}\\
C &=& - \, {\cal U}_2 \,\dot{H}^{\rm eff} - 
{\cal V}_2 \,H^{\rm eff}
\nonumber\\
&\equiv& - \,\partial_t \left(
\frac{D^2\sqrt{h}}{2\,{\cal N}^2} H^{\rm eff} 
\right)
\equiv - \,\partial_t \left(
\frac{N}{2\,{\cal N}^2} f(\xi) H^{\rm eff} 
\right)
\label{C}
\end{eqnarray}
and $\epsilon=1$ inserted. 
The operators $B$ and $C$ are hermitean with respect to the
local scalar product. (In order to see this, one has to use the 
fact that $N^{-1} {\cal N}^2$ depends only on $t$ in this gauge). 
As a consequence, the term $\,i\, C$ violates unitarity. 
Using in addition the gauge $S=-t$ (see above), 
further simplifications occur, and one finds
${\cal U}_2 = \frac{1}{2}$, ${\cal V}_2=0$ and
${\cal W}_2 = - \frac{1}{2} D (\partial_{tt} D^{-1})$
(see equation (\ref{UVW})). 
In this gauge, the corrections are negligible if
--- in a sloppy formulation --- $(H^{\rm eff})^2$, 
$\dot{H}^{\rm eff}$ and ${\cal W}_2$ are ''smaller'' than 
$H^{\rm eff}$. Note that the first of these statements 
(which seems to be the essential one) directly corresponds
to the classical analogue of the WKB-approximation in the
gauge $\nu=1$, as was discussed above (below equation 
(\ref{wdwheur})). 
\medskip

Thus we see that unitarity violation appears at 
$O(\epsilon)$ or, by relating the time parameter to
the action, at $O(\epsilon^2)$. Although we did not make 
use of higher orders in our interpretational scheme, they may 
be of some help if one likes to demonstrate explicitly how 
accurate the effective Schr{\"o}dinger equation
(\ref{eff}) describes the local time evolution, and 
of what order of magnitude the deviations are (as compared to 
solutions of the full Wheeler-DeWitt equation). 
\medskip

Let us at this point perform a rough estimate of the
order of magnitude of the correction terms for the case
that the physical time parameter $t$ depends mainly on $S$.
The dominant contribution may be inferred from (\ref{wdweps2}). 
Upon performing the expectation value (acting with 
$\langle\chi|$ from the left), setting 
$\varepsilon=\langle H^{\rm eff}\rangle$, assuming widths to 
be small (omitting the brackets) and neglecting ${\cal V}_2$
and ${\cal W}_2$,
this implies a quantum gravitational shift of the 
effective Hamiltonian 
\begin{equation}
\delta\varepsilon \approx - \,\frac{\varepsilon^2}{2N U} 
\label{corr1}
\end{equation}
and an imaginary contribution $\,i\,\varepsilon_I$ with 
\begin{equation}
\varepsilon_I \approx -\, 
\frac{\dot{\varepsilon}}{2N U} 
\label{corr2}
\end{equation}
indicating --- in the usual language --- an instability. 
The approximation $S\approx$ $S(t)$ is in particular
well-applicable during inflation if $t$ is the proper time
(the exponential expansion suppresses the dependence 
of the action on variables other than the scale factor). 
This is the situation 
of a quantum field in de Sitter space. Formally, we can use
as background the framework of the Hawking model as described in
Section 10, in the domain $|\phi|\gg 1$ 
with $\Lambda^{\rm eff}\approx $ $m^2\phi^2\approx$
$H_0^2$, $H_0$ the Hubble constant, 
$U\approx a^3 H_0^2$ and $N=1$. From the inflationary (outgoing) 
trajectories (equation (\ref{trajhm1})) we find
$\delta\varepsilon\approx$ $-a^{-3} H_0^{-2}\varepsilon^2$.
This expression agrees in structure with results achieved 
in the conventional WKB-expansion and by using more accurate 
methods 
\cite{Kiefer7}.
However, one should not forget that our $\varepsilon$
does not coincide with 
$\langle H_{\rm matter}\rangle$ from the semiclassical
approach. 
\medskip

We can now estimate under which condition the hermitean 
correction (\ref{corr1}) is small. The WKB-approximation itself 
is characterized mainly
by the condition $|\dot{\chi}|\ll |\chi \dot{S}|$, which
means in the language used above that $\varepsilon \ll N U$.
The order of magnitude of the correction term
(\ref{corr1}) relative to the uncorrected Hamiltonian
in the Schr{\"o}dinger approximation is
$\delta\varepsilon/\varepsilon\approx$ $\varepsilon/(N U)$
Hence, the hermitean corrections are small if and only if
the WKB-approximation holds, i.e. if the change of $S$ is 
large enough. Denoting by $\delta t_\varepsilon$
$\approx \varepsilon^{-1}$ the
typical time scale set by the effective Schr{\"o}dinger
equation, and by $\delta t_{\rm min}\approx$ $(N U)^{-1}$
the minimal resolution time in the WKB-approximation
(cf. the estimate at the end of Section 3), we find
$\delta\varepsilon/\varepsilon\approx$
$\delta t_{\rm min}/\delta t_\varepsilon$. Thus
$\delta t_\varepsilon \gg \delta t_{\rm min}$, which
states that the time scale relevant for the local quantum
structure is much above the minimum. 
Moreover, the correction itself may be associated with a time
scale $\delta t_{\rm corr}\approx$
$\varepsilon^{-2} N U$. 
At least from a heuristic point of view, 
this is the time that must 
be available in order to measure an effect stemming from
the hermitean correction to the Hamiltonian. 
Thus $\delta t_{\rm corr}/\delta t_{\rm min}\approx$
$(\varepsilon/\delta\varepsilon)^2 \gg 1$, and we expect 
never to run into the problem that the correction plays
a role in sufficiently small local WKB-domains. If this is actually 
true (the details may be model-dependent),
the quantum gravity corrections can be neglected 
in the local effective Schr{\"o}dinger equation, as long as it is 
followed over moderate time intervals. 
\medskip 

The anti-hermitean correction (\ref{corr2}) can be
expected to be small 
when the prefactor $D$ varies slowly as
compared to the WKB-phase, $|\dot{D}|\ll|D \dot{S}|$,
because, roughly (cf. (\ref{Ddot})), 
$\dot{\varepsilon}/\varepsilon\sim \dot{D}/D$. 
This is reasonable, because $D$ was chosen so as to
guarantee the local unitarity. Hence, if the
WKB-approximation is good, both corrections to the
effective Schr{\"o}dinger equation are locally suppressed.
This does {\it not necessarily} mean that quantum gravity 
effects are unobservably small, but the time scales relevant for 
observations may be very large. 
What we suggest is that these effects
are automatically incorporated by following the time evolution
through a large number of adjacent domains, as described in
Section 8. The particular modification of the 
(hermitean) Hamiltonian and the local quantum structure for 
long-range time evolution
(as long as possible) and the time scale above which unitary 
quantum mechanics would not account for viable predictions may 
in principle be determined by this procedure. 
\medskip

There is an alternative way to look at the condition under which
the WKB-approximation is good, and the corrections to the
effective Schr{\"o}dinger equation are small. The essential
inequality is that $(N U)^{-1} (H^{\rm eff})^2 $ is negligible
as compared to $H^{\rm eff}$. This in turn implies that
$H^{\rm eff}$ is ''small'' as compared to $N U$, and we
obtain a scale in the coordinates $\xi^a$ below which 
$\chi$ must not change significantly. Using only the
essential structure of $H^{\rm eff}$, and applying it to
a wave function of the form
$\chi\sim$ $\exp(-A_{ab}\delta\xi^a\delta\xi^b)$
with $\delta\xi^a=$ $\xi^a-{\overline{\xi}}^a$, we find that
the maximum eigenvalue of $h^{ac}A_{cb}$ must be much smaller 
than $U$. As a consequence, the width of wave packets is
bounded from below. In terms of the ''proper length''
defined by the DeWitt metric we have 
$s_{\rm min}\gg U^{-1/2}$, a result that was also mentioned in
Section 3. Once the variations of the wave function $\chi$ in
$\xi^a$ are well below the bound, we expect the correction terms
to be neligible and the WKB-approximation to hold. 
This implicitly assumes that the equal-time hypersurfaces
do not differ too drastly from the equal-action hypersurfaces,
and that the contributions to 
$H^{\rm eff}$ stemming from operator ordering issues
(i.e. $D$, $\cal N$ and $\sqrt{h}$ in equation
(\ref{Heff})) are suppressed as well. 
A more detailed analysis will depend on the particular situation 
and the model one is considering. 
\medskip

Ultimately, the result of iterating the 
elimination of time-derivatives at the right hand side of
(\ref{wdw3}) 
beyond any finite order in $\epsilon$ is an equation of the
form $i\,\dot{\chi}=H^{\rm tot}\chi$, where $H^{\rm tot}$
is an infinite sum of generally non-hermitean
operators acting purely tangential on $\Gamma_t$ and
being of arbitrarily high order in the ''spatial'' derivatives. 
The validity of such an expansion is guaranteed if 
the sum converges in some sense, even after setting 
$\epsilon=1$. (A sililar --- although much simpler --- 
procedure arises when the flat Klein-Gordon equation is 
written down in the Schr{\"o}dinger form by
expanding the square root $\sqrt{m^2-\triangle}=$
$m-$ $\triangle/(2m)-$ $\triangle\triangle/(8 m^3)+\dots$.
In this example, unitarity is not violated, but this
is due to the fact that negative/positive frequencies
may unambigously be identified). We do not know for which
wave functions $\chi$ the expansion converges (we don't 
know either much about convergence in the
standard WKB-approach), but at least formally we have 
obtained a {\it unique} expansion scheme to all orders 
in $\epsilon$. In terms of physics, this is probably 
not of much value because unitarity breaks down already at 
$O(\epsilon)$ or $O(\epsilon^2)$. 
\medskip

Nevertheless, one could think of integrating 
$i\dot{\chi}=H^{\rm tot}\chi$ and its complex conjugate
equation 
for some initial value $\chi(0,\xi)$ in order to obtain
solutions of the Wheeler-DeWitt equation
that may be associated with $\H^-$ and $\H^+$, respectively.
This would in fact be the attempt to exactify 
equation (\ref{apprdec}) of Section 7. In the case of 
the flat Klein-Gordon equation this leads to 
the formal appearance of the operator $\pm\sqrt{m^2 - \triangle}$,
which can be given a well-defined meaning in the Fourier 
representation (i.e. with respect to a particular 
regularization, or with respect to a basis in which $\triangle$ 
is diagonal) and reduces to the
function $\pm(m^2 + \vec{k}^2)^{1/2}$. The result is 
the standard decomposition into 
negative and positive frequencies. In the general case of the 
Wheeler-DeWitt equation, one would not expect to find a 
distinguished regularisation for $H^{\rm tot}$, 
and there seems to be very little
chance to obtain a tool that would distinguish in some
exact sense between outgoing and incoming modes at the 
quantum level. 
However, we cannot exclude that a more
intense study of the expansion leading to $H^{\rm tot}$
sheds new light on the problem. 
\medskip

The local quantum structure, based on the effective
Schr{\"o}dinger equation (\ref{eff}), has so far been 
formulated in terms of coordinates $(t,\xi^a)$ on $\cal M$. 
The $\xi^a$ are {\it conserved} quantities along any 
classcial trajectory belonging to the congruence described by
$S$ (although they are not conserved along other classical
trajectories). This defines a very special type of
formulation of quantum mechanics, which is only in few
cases in the familiar representation. (It may be illustrated 
by describing the harmonic oscillator 
$L =$ $(\dot{x}^2-$ $x^2)/2$ in terms of the variable $\xi$, 
defined by $x=$ $\sin(t+\xi)$, instead of $x$. Then 
$\xi(t)=const$ provides a family of classical solutions, 
but all other solutions in the domain $|x|\le 1$ emerge as well). 
One may however reformulate the effective Schr{\"o}dinger
equation (\ref{eff}) in terms of other coordinates $(t,z^a)$. This 
gives us the possibility to define a local quantum structure in
a quite general coordinate system on $\cal M$, 
provided the small $\epsilon$ expansion holds (i.e.
provided the $\Gamma_t$ do not differ too drastly from the 
hypersurfaces of constant action). The systematic way to achieve
this (after $S$ has been selected) is to choose a coordiate
$t$ (i.e. a scalar function $t(y)$ on $\cal M$) such that
the equal-time hypersurfaces $\Gamma_t$ are spacelike. (Even 
this condition may be somewhat relaxed). After having made 
this choice, the lapse $N(y)$ may be determined. The coordinates
$\xi^a$ labelling the classical trajectories may be introduced
as {\it auxiliary} quantities, and the $(n-1)+1$-split of the
DeWitt metric as well as the determination of $D$ (choice
of $f(\xi)$) and the definition of the scalar product
and the effective Hamiltonian $H^{\rm eff}$ proceed as above. 
One may now select coordinates $z^a$ such that
$(t,z^a)$ make up a coordinate system on $\cal M$. In other
words, the $z^a$ serve as coordinates on each individual 
$\Gamma_t$, and we may write down their definition as
$z^a \equiv z^a(t,\xi)$. From now on, the effective
Schr{\"o}dinger equation may simply be {\it re-written} 
in terms of these new coordinates. Introducing
\begin{equation}
\widehat{\chi}(t,z) \equiv \chi(t,\xi)\, , 
\label{ztransf}
\end{equation}
the result is 
\begin{equation}
i\, \partial_t \widehat{\chi} = (H^{\rm eff} +
i\, {\cal C}^a {\cal D}_a) \widehat{\chi}\, , 
\label{eff2}
\end{equation}   
where ${\cal C}^a {\cal D}_a\equiv {\cal C}^a \partial_a$ is 
the vector field 
$- (\partial z^a/\partial t) \partial/\partial z^a$ 
on $\Gamma_t$. 
Note that in this formulation $\partial_t$ 
operates at constant $z^a$, and is thus no longer the derivative 
along the trajectories. It differs from the previously
defined dot-operator ($\partial_t$ at constant $\xi^a$) just by 
${\cal C}^a \partial_a$. 
Since $H^{\rm eff}$ acts purely tangential to $\Gamma_t$ and is a 
three-covariant object, it may be rewritten in the new coordinates. 
It is thus still given by (\ref{Heff}) and (\ref{gammaab}), 
with $h_{ab}$ denoting the induced metric on $\Gamma_t$, 
${\cal N}_a$ the vector field on $\Gamma_t$ defined by (\ref{Na}), 
and $\cal N$ the scalar function (\ref{NN}). If an 
$(n-1)+1$-split of the DeWitt metric is performed in terms 
of $(t,z^a)$, the ''lapse'' is still given by $\cal N$, whereas
the ''shift'' vector field $\widehat{{\cal N}}^a$ is related to
its analogue ${\cal N}^a$ from the $(t,\xi^a)$ split 
(\ref{admdecom})
(with appropriately transformed components) by
$\widehat{{\cal N}}^a = $ ${\cal N}^a +$ ${\cal C}^a$. This 
is a consequence of the identity
$d z^a =$ $(\partial z^a/\partial \xi^b)d \xi^b-$
${\cal C}^a dt$ and may serve --- together with (\ref{Na}) --- as 
an alternative definition of ${\cal C}^a$. 
The physical significance of such a coordinate transformation 
may be illustrated by performing a Galilei transformation
$\vec{x}^{\,\rm new} = \vec{x}^{\,\rm old} - \vec{v} t$ on the
variables of the non-relativistic Schr{\"o}dinger equation
of a free particle ($H^{\rm old}=$ $-\triangle/(2m)$). The
new Hamiltonian becomes $H^{\rm new} =$
$-\triangle/(2m) +$ $i \vec{v}\,\vec{\nabla}$, which agrees with
the overall structure of (\ref{eff2}). Redefining the wave 
function by extracting the phase factor 
$\exp(i m \vec{v} \vec{x})$ --- which just amounts to a unitary 
transformation --- yields the new Hamiltonian in the 
representation $-\triangle/(2m) -$ $m \vec{v}^{\,2}/2$. This 
shows that the net effect of the coordinate transformation 
is just a shift of the energy, according to the velocity
of the new ''observer''. In the case of (\ref{eff2}),
it might be necessary to perform similar redefinitions
in order to cast the result into a familiar form. 
\medskip

The standard WKB-expansion approach should be
recovered by recalling that the (positive)
contribution $({\cal D}_a S) {\cal D}^a S$ in
(\ref{NN}) --- although it may be large --- can
have small parts that correspond to the slow 
oscillations in the quantum variables.
Trying to neglecting these, one is tempted to
introduce the notation of classical and quantum variables, 
and to decompose $S$ into the rapidly changing part
$S_0(y_{\rm cl})$ and some (slowly changing) rest 
$S_{\rm q}(y_{\rm cl},y_{\rm q})$. It is 
not always obvious how to do this, but whenever a
reasonable identification of the two parts is possible, 
one should be able to recover the effective Schr{\"o}dinger
equation (\ref{eff0}), with $\widetilde{H}^{\rm eff}$
being the ''quantum'' part of the original Hamiltonian. 
In simple cases we expect $\widetilde{H}^{\rm eff}$ to be
$V^{-1}H^{\rm eff} V$ with $V=$ $D^{-1} D_0 \exp(-i S_{\rm q})$
and possibly some small contributions omitted. In general, 
some average over neighbouring WKB-branches may be involved. 
In Section 11 we have encountered a situation where the
average over two branches is necessary. 
Let us remark that
an average over {\em infinitely} many neighbouring branches
--- although this may not be necessary in our context ---
might mathematically relate to path integral ideas. 
The details of this would be an interesting subject to pursue. 
\medskip

Let us comment once more on the advantages and disadvantages 
of the modified formalism, as presented here.
In contrast to the standard approach, we have retained in
$S$ and $H^{\rm eff}$ also small and slow contributions. 
Also, there may be situations in which the
split into $S=$ $S_0+$ $S_{\rm q}$ is not an 
obvious task (cf. also the discussion of the semiclassical
approximation in Ref.
\cite{Isham}, p. 245). 
Moreover, in the standard approach, the separation
into classical and quantum variables is usually {\it prior} 
to considering states. Only those wave functions that
fit the pattern (\ref{WKB0}) may emerge. 
In our approach, we can say that the action $S$ is
''the most classical'' coordinate on $\cal M$, and this is
determined only by the particular WKB-branch. 
We are mainly motivated by adjusting this WKB-branch according 
to the oscillations in a {\it given} state, and this can be 
regarded as a slightly more general and fundamental point 
of view (although --- or rather {\em therefore} --- 
in a larger distance from actual observations). 
Our version of local quantum structures thus 
demonstrates that ''a clean division of variables into
classical and quantum''
\cite{Vilenkininter} 
is not a necessary starting point for recovering an approximate 
unitary evolution from the Wheeler-DeWitt equation. 
Maybe one could state that {\it approximate unitarity shows up 
at a slightly more fundamental level than the split between 
classical and quantum}. 
Another advantage of our formalism becomes clear by 
comparing (\ref{WKB0}) and (\ref{WKB2}). Since $S$ solves
the {\it exact} Hamilton-Jacobi equation, in 
many situations the effective Schr{\"o}dinger
equation of our formalism will come closer to full
solutions of the Wheeler-DeWitt equation than the 
semiclassical approach. In this sense it is ''more
accurate'' than the latter. 
\medskip

The disadvantage of our formalism for many purposes
might be that some enlarged degree of ambiguity appears. 
Whereas in the standard approach the classical
''background'' action $S_0(y_{\rm cl})$ has some definite
''macroscopic'' significance, our exact action
$S(y)$ is just one among many that are
macroscopically (classically) equivalent. 
In the (rather trivial) example of the flat Klein-Gordon 
equation, one may use several actions of the type 
(with $\hbar=1$ but $c$ displayed explicitly)
$S=$ $\vec{k} \vec{x} -$ $c\, t (\vec{k}^2 + m^2 c^2)^{1/2}$
with $\vec{k}$ fixed and $\vec{k}^2\ll m^2 c^2$ in order
to find identical results: non-relativistic quantum mechanics 
of a free particle. 
We leave it open to what extent it is possible to 
''factor out'' this ambiguity in a generic situation 
by means of geometric concepts based on $\cal M$
and its foliation(s) by equal-time
hypersurfaces. The notation of classical and quantum 
variables should emerge from such an attempt rather 
than being assumed. (Vilenkin's condition 
\cite{Vilenkininter} 
that the size of the coefficients of the DeWitt metric
$\gamma_{\alpha\beta}$ between classical and
quantum variables are small --- symbolically 
$\gamma(y_{\rm cl},y_{\rm q}) \ll \gamma(y_{\rm cl},y_{\rm cl})$ 
--- is likely to emerge as a special case, but there should be 
some dependence on the wave function as well). 
On the other hand, if a division into classical and
quantum variables is {\it assumed}, 
it should be possible to make contact with
the standard semiclassical formalism by 
using coordinates analogous to our $\xi^a$ for the classical
sector alone (cf. Ref. \cite{Cosandey}). 
\medskip

To sum up, if an {\it exact} wave function of the form
$\psi=\chi D \exp(i S)$ (with $S$ and $D$ being defined as
above) has the property that in some domain of $\cal M$
the phase factor $\exp(i S)$ oscillates much faster than 
$D$ and $\chi$,
we can reasonably assume that the coefficients of
the $\epsilon^2$- and $\epsilon$-terms in (\ref{wdw3})
are negligible, and $\chi$ is subject to the 
Schr{\"o}dinger equation (\ref{eff}) with
Hamiltonian (\ref{Heff}). This generates a unitary
evolution with respect to the scalar product
(\ref{QQQ1}), or equivalently (\ref{surff}). 
\\
\\
\\
{\Large {\bf Acknowledgments}}
I would like to thank Jerome Martin, Heinz-Dieter Conradi and
Claus Kiefer for literature hints and extremely helpful remarks 
on the manuscript.

\end{document}